\documentclass[natbib]{svjour3}                     
\smartqed  

\usepackage{graphicx}
\usepackage{amsmath}

 \journalname{Space Science Reviews}

\newcommand{\aap}{{Astron. Astrophys.}}

\usepackage[backref,breaklinks,colorlinks,citecolor=blue]{hyperref}

\usepackage{aas_macros}

\begin{document}

\title{A Primer on Focused Solar Energetic Particle Transport}

\subtitle{Basic physics and recent modelling results}

\author{Jabus van den Berg \and Du Toit Strauss \and Frederic Effenberger}

\authorrunning{van den Berg et al.} 

\institute{J.P. van den Berg \at
Centre for Space Research, North-West University, Potchefstroom, South Africa\\
South African National Space Agency, Hermanus, South Africa\\
\email{24182869@nwu.ac.za}\\
https://orcid.org/0000-0003-1170-1470\\
\and
R.D. Strauss \at
Centre for Space Research, North-West University, Potchefstroom, South Africa\\
\email{dutoit.strauss@nwu.ac.za}\\
https://orcid.org/0000-0002-0205-0808\\
\and
F. Effenberger \at
GFZ German Research Centre For Geosciences, Potsdam, Germany\\
Bay Area Environmental Research Institute, NASA Research Park, Moffett Field, CA, USA\\
\email{feffen@gfz-potsdam.de}\\
}

\date{Received: \today / Accepted: date}

\maketitle

\begin{abstract}
The basics of focused transport as applied to solar energetic particles are reviewed, paying special attention to areas of common misconception. The micro-physics of charged particles interacting with slab turbulence are investigated to illustrate the concept of pitch-angle scattering, where after the distribution function and focused transport equation are introduced as theoretical tools to describe the transport processes and it is discussed how observable quantities can be calculated from the distribution function. In particular, two approximations, the diffusion-advection and the telegraph equation, are compared in simplified situations to the full solution of the focused transport equation describing particle motion along a magnetic field line. It is shown that these approximations are insufficient to capture the complexity of the physical processes involved. To overcome such limitations, a finite-difference model, which is open for use by the public, is introduced to solve the focused transport equation. The use of the model is briefly discussed and it is shown how the model can be applied to reproduce an observed solar energetic electron event, providing insights into the acceleration and transport processes involved. Past work and literature on the application of these concepts are also reviewed, starting with the most basic models and building up to more complex models.

\end{abstract}

\vspace{-0.3cm}
\tableofcontents
\vspace{-0.1cm}


\newpage

\section{Introduction}

Solar energetic particles (SEPs) are one of the key subjects in heliospheric physics, receiving even more interest in the last couple of years, mostly due to space missions focusing on the Sun, such as the Parker Solar Probe \citep[launched on 11 August 2018; \url{http://parkersolarprobe.jhuapl.edu/index.php};][]{foxetal2016} and the Solar Orbiter mission \citep[launched on 10 February 2020; \url{http://sci.esa.int/solar-orbiter};][]{mulleretal2013}. Their importance is not only related to their character as highly energetic test particles, tracing the heliospheric plasma environment between their source close to the Sun and the observer, but also to their potential impact on space hardware and interplanetary travel by humans.

A number of excellent and mostly up to date reviews on general SEP properties and their observational basis \citep{reames1999, reames2013, reames2017, ryanetal2000, mewaldt2006, kleindalla2017} including specific topics such as scattering theories and perpendicular diffusion \citep{shalchi2009, shalchi2020} exist. However, a review focusing on the various aspects of the transport of SEPs, in a broad context and with a view towards applications and common misconceptions, is still somewhat lacking. To our knowledge, the last review with a similar scope dates back to \citet{droge2000a}, with a focus on pitch-angle scattering, so a fresh look appears justified. Focus will fall in particular on simulation work in the last decade or so, and how recent results from these can be reconciled with each other and the observational basis that has been established over the last years. Although we aim for a comprehensive view of the subject, there will be important works falling through the cracks or which will be left out due to space constraints and considerations of readability. We apologize to all colleagues in advance, who feel that their favorite study is missing.

The review of the subject begins by establishing the basics of focused transport for SEPs. Common misconceptions will be highlighted throughout and the correct interpretations will be explained. To build a conceptual understanding of the processes on a pitch-angle level, Section~\ref{sec:Microphysics} will consider the microscopic physics of a single charged particle interacting with electromagnetic slab turbulence. The concept of a distribution function, to model the macroscopic physics, will be introduced in Section~\ref{sec:Macroscopic}, together with the focused transport equation. This section will also investigate the applicability of analytical approximations to the full solution of the focused transport equation. Here it will be emphasised that a numerical scheme is needed to solve the focused transport equation and in order to do so, a finite difference numerical scheme is presented in Appendix~\ref{apndx:FDSolver} with a link to the source code. Many processes, requiring at least a 2D spatial geometry to be correctly described, e.g. drift and perpendicular diffusion, are reviewed in Section~\ref{sec:Review}. The review focuses specifically on modelling work, starting form the basic 1D models and building up to the fully 3D models. Additional information are presented in further appendices, which also provide a reference to more technical aspects not fully discussed in the main text.

We hope that this review encourages scientists, especially new to the subject, to investigate and apply the theory to actual problems in SEP research. The numerical tools, as described in the appendix with their source code freely available\footnote{\url{https://github.com/RDStrauss/SEP_propagator}}, can be a starting point for such endeavours.


\newpage

\section{The Micro-physics of Charged Particles in Turbulent Electromagnetic Fields}
\label{sec:Microphysics}

Cosmic ray (CR) research usually deals with the isotropic limit (this refers to isotropy in momentum, a concept which will be clarified in this section), allowing CRs to be considered as a function of position, energy, and time. SEP transport is inherently time dependent, although certain event integrated distributions can be considered to be in a steady state. Anisotropy, however, is ubiquitous in SEP transport and isotropy is reached only during the decay phase of an SEP event. This is probably one of the most complicated aspects of SEP transport, as pitch-angle dependent transport must be considered and the processes must be described on a more fundamental level than in the isotropic limit. Focused transport is, for this reason, not well understood in general, as concepts well established in isotropic transport cannot be applied to anisotropic transport. It is, of course, possible to extend the anisotropic processes to the isotropic limit, but the reverse cannot be done. In this section the Newton-Lorentz equation, some basic definitions, and the process of magnetic focusing will be introduced. A slab turbulence model is introduced in Appendix~\ref{apndx:ModelSlab} and a particle is simulated in this turbulence field to illustrate the concept of pitch-angle scattering. The section concludes with a summary of the introduced concepts.


\subsection{The Newton-Lorentz Equation}
\label{subsec:NwtnLrntz}

The motion of a non-relativistic particle, with mass $m$ and charge $q$, moving with a velocity $\vec{v}$ in an electric $\vec{E}$ and magnetic $\vec{B}$ field, is governed by the Newton-Lorentz equation \citep{rossiolbert1970, chen1984}
\begin{equation}
\label{eq:NewtonLorentz}
\frac{{\rm d} \vec{p}}{{\rm d} t} = q ( \vec{E} + \vec{v} \times \vec{B} ) ,
\end{equation}
where $\vec{p} = m \vec{v}$ is the particle's momentum, which is the most fundamental description of charged particle transport in magnetized plasma, and the basis of all transport equations. A non-relativistic description will be used here as an approximation just to illustrate the basic concepts. Some analytical solutions of this equation can be found in any plasma physics textbook \citep[see e.g.][]{rossiolbert1970, chen1984, choudhuri1998}. For electric and magnetic fields with spatial and temporal dependencies, it is relatively easily solvable with various numerical methods \citep[see e.g.][]{boris1970, birdsalllangdon1991}. The effect of large scale electric fields will not be considered here and is only included to emphasize that a turbulent electric field will exert a force on the particle. Notice that since the magnetic force is perpendicular to the direction of motion, the magnetic field does no work on the particle and cannot change its energy \citep{rossiolbert1970, chen1984, choudhuri1998}.

For a particle moving in a constant and uniform magnetic field, with strength $B_0$, in the absence of electric fields, the vector product in Eq.~\ref{eq:NewtonLorentz} implies that the particle experiences a centripetal acceleration and will gyrate around the magnetic field, with positive and negative particles gyrating in a left- and right-hand manner, respectively. The particle will gyrate around the magnetic field at the cyclotron frequency
\begin{equation}
\label{eq:CyclotronFrequency}
\omega_c = \frac{|q| B_0}{m} ,
\end{equation}
while tracing a circle with the Larmor radius (or gyro-radius)
\begin{equation}
\label{eq:LarmorRadius}
r_L = \frac{m v_{\perp}}{|q| B_0} ,
\end{equation}
where $v_{\perp}$ is the speed of the particle perpendicular to the magnetic field (the maximal Larmor radius is defined as $R_L = m v / |q| B_0$). The particle's velocity component parallel to the magnetic field, $v_{\parallel}$, will cause the gyrating particle to trace a spiral trajectory \citep{rossiolbert1970, chen1984, choudhuri1998}.

The particle's pitch-angle is defined as the angle between the particle's velocity vector and the magnetic field vector,
\begin{equation}
\label{eq:PitchAngle}
\alpha = \arccos \left( \frac{\vec{v} \cdot \vec{B}_0}{v B_0} \right) = \arcsin \left( \frac{v_{\perp}}{v} \right) = \arccos \left( \frac{v_{\parallel}}{v} \right) = \arctan \left( \frac{v_{\perp}}{v_{\parallel}} \right) ,
\end{equation}
while the parallel and perpendicular speeds can be calculated from the pitch-angle by
\begin{subequations}
\begin{align}
\label{eq:ParallelSpeed}
v_{\parallel} & = v \cos \alpha = v \mu \\
\label{eq:PerpSpeed}
v_{\perp} & = v \sin \alpha = v \sqrt{1 - \mu^2} ,
\end{align}
\end{subequations}
respectively, where the so called pitch-cosine
\begin{equation}
\label{eq:PitchCosine}
\mu = \cos \alpha
\end{equation}
is a quantity normally used in transport equations. Since the parallel and perpendicular speed is constant in a constant and uniform magnetic field, the pitch-angle will also be constant in such a field \citep{rossiolbert1970, chen1984, choudhuri1998}.

\begin{figure*}[!t]
\centering
\includegraphics[page=1, clip, trim=10mm 210mm 150mm 13mm, scale=0.9]{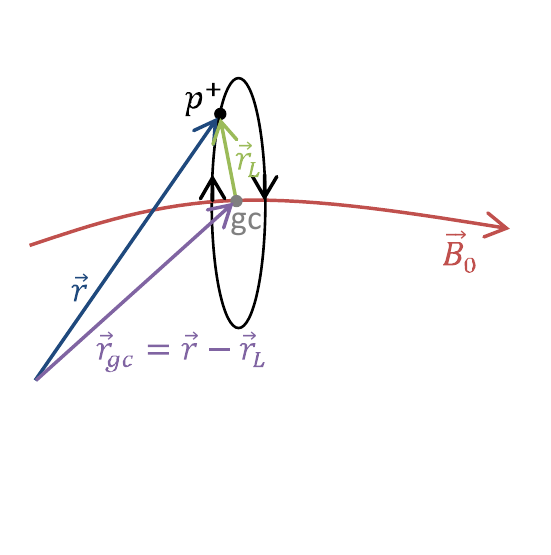}
\includegraphics[scale=0.35]{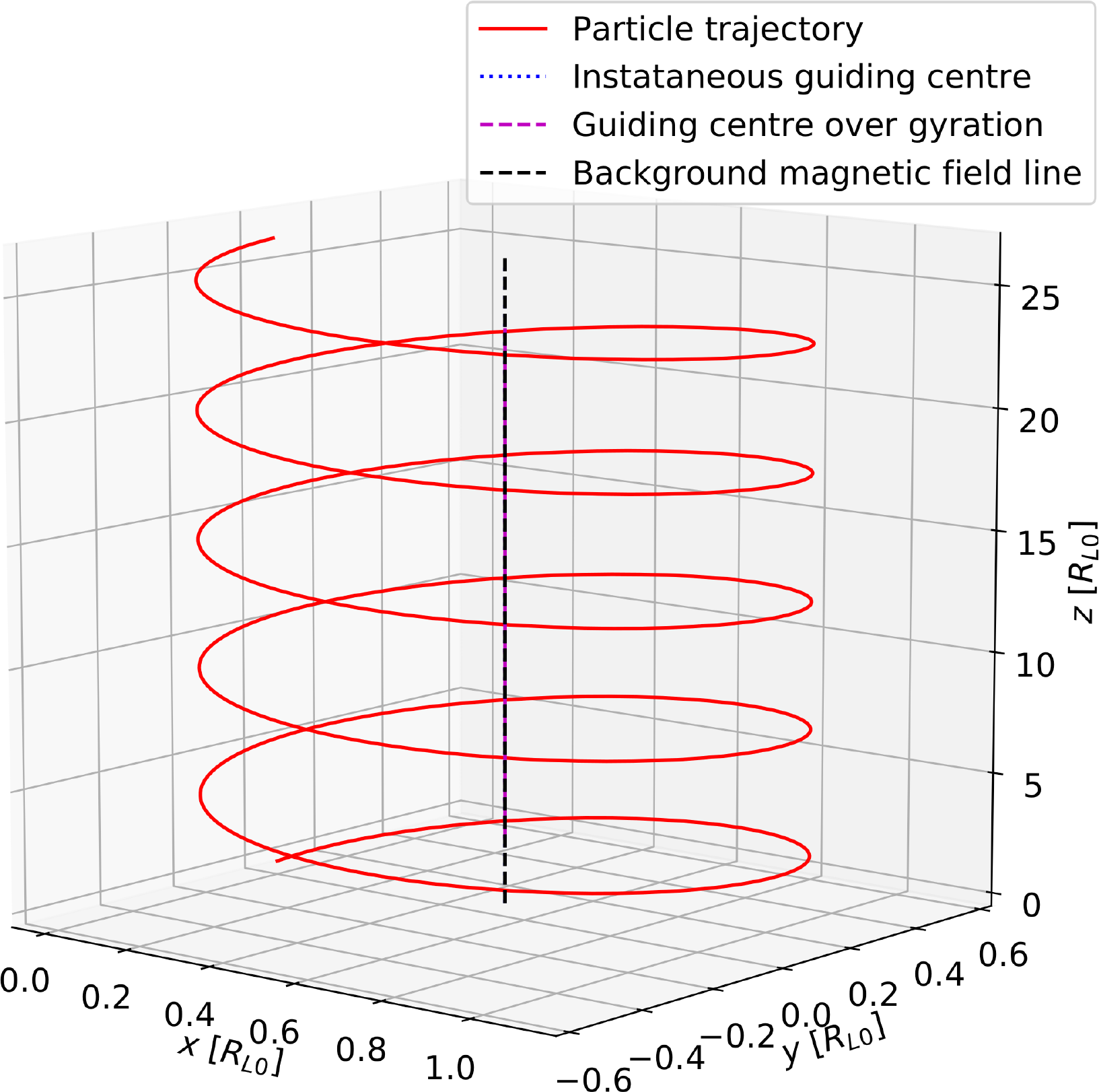}
\textbf{\caption[]{\label{fig:GCGyrate}{\textnormal{\textit{Left:} Illustration of a proton's position (blue vector), guiding centre (purple vector), and directional Larmor radius (green vector) during its gyration (black circle, with arrows indicating the direction of rotation) around the background magnetic field line (red vector). This figure was adapted from \citet{northrop1961}. \textit{Right:} Simulation of a proton in a constant and uniform magnetic field performed with a fourth-order Runge-Kutta scheme. The trajectories of the particle (solid red) and its guiding centre (dotted blue: Eq.~\ref{eq:GCPos}; dashed purple: running average of particle's position over a gyration) are shown, together with a single background magnetic field line (dashed black; coinciding with the guiding centre).}}}}
\end{figure*}

The particle gyrate around an imaginary point called the guiding centre (GC) and its position can be found by subtracting a directional Larmor radius from the particle's position. The directional Larmor radius can be interpreted as the instantaneous radius of curvature projected onto the plane perpendicular to the magnetic field. This is illustrated in the left panel of Fig.~\ref{fig:GCGyrate} and can be written as
\begin{equation}
\label{eq:GCPos}
\vec{r}_{\rm gc} = \vec{r} + \frac{m}{q B_0^2} \vec{v} \times \vec{B}_0 ,
\end{equation}
where $\vec{r}$ is the particle's position. The particle's helical path can be decomposed, as a first approximation, into a gyration around the GC and the movement of the GC along the magnetic field \citep{northrop1961, rossiolbert1970, burgeretal1985}, as illustrated in the right panel of Fig.~\ref{fig:GCGyrate}. Notice however that \emph{the GC is a mathematical construct} which is introduced as a tool to help describe the particle's motion. It is imperative to realise that \emph{the particle does not know that it has a GC and is not affected by what happens to the GC.} Furthermore, \citet{burger1987} points out that the concept of a GC is only well defined over a complete gyration. If the magnetic field change over a length (time) scale shorter than the Larmor radius (gyroperiod), the GC will be ill defined and might behave in an unexpected manner.

\subsection{Magnetic Focusing}
\label{subsec:Focus}

The theoretical background and derivations of this section is well documented in plasma physic textbooks and will only be summarised. When the magnetic field has a gradient along it, the particle will experience a force parallel to the magnetic field which will be in the opposite direction of the gradient, $\vec{F}_{\parallel} = - M (\partial B_0 / \partial \vec{s}) = - M \vec{\nabla}_{\parallel} B_0$, where ${\rm d} \vec{s}$ is a line segment parallel to the magnetic field, $\vec{\nabla}_{\parallel}$ denotes the gradient along the magnetic field, and $M = m v_{\perp}^2 / 2 B_0$ is the particle's magnetic moment. Due to the invariance of the magnetic moment (${\rm d} M / {\rm d} t = 0$) {in the absence of magnetic turbulence} and the conservation of kinetic energy, this force is accompanied by an interchange between parallel and perpendicular energy: as the particle moves into a region of larger magnetic field strength, its perpendicular speed increases, with the effect that its parallel speed decreases. Ultimately this causes the particle's motion to be reversed and the particle is mirrored. Not all particles, however, will be mirrored. It can be shown that a particle starting out in a region with field strength $B$ with
\begin{equation}
\label{eq:MirrorCondition}
|\mu| > \mu_m = \sqrt{1 - \frac{B}{B_m}} ,
\end{equation}
will not be able to penetrate a region of magnetic field strength $B_m$ \citep{rossiolbert1970, chen1984, choudhuri1998}.

Due to the decrease of the heliospheric magnetic field (HMF) strength with heliocentric radius \citep[][see also Appendix~\ref{apndx:HMF}]{parker1958}, SEPs will experience magnetic focusing. As a particle moves into regions of weaker parallel magnetic fields, the particle's perpendicular speed will decrease while its parallel speed will increase, causing the particle's motion to become increasingly ballistic. Focusing is the reason why SEP events are anisotropic, excluding the fact that the particles are propagating away from their release at the Sun \citep{roelof1969}. Since focusing causes the perpendicular speed to decrease, \emph{it might be incorrectly expected that the Larmor radius} (which is dependent on the perpendicular speed) \emph{would also decrease}. The Larmor radius, however, is inversely proportional to the magnetic field strength, which decreases as $1/r$ for the \citet{parker1958} HMF in the equatorial plane. From the definition of the magnetic moment and its invariance, it can be seen that the perpendicular speed does not change at the same rate as the magnetic field, since the magnetic moment is dependent on the square of the perpendicular speed. The HMF strength therefore decreases faster than the perpendicular speed close to the Sun and this would cause \emph{the Larmor radius to increase as SEPs move away from the Sun}. Indeed a simple calculation of the maximal Larmor radius ($v_{\perp}$ replaced by $v$ in Eq.~\ref{eq:LarmorRadius}) for a $100$ $\mathrm{keV}$ ($v \approx 0.548 \, c$) electron close to the Sun ($0.1$ $\mathrm{AU}$) and at the Earth ($1$ $\mathrm{AU}$) with a HMF field strength of $500$ $\mathrm{nT}$ and $5$ $\mathrm{nT}$, respectively, yields $\sim 1.868$ $\mathrm{km}$ and $\sim 186$ $\mathrm{km}$, respectively. Even if $99\%$ of the electron's speed is converted to parallel speed ($v_{\perp} = 0.01 \, v$) by focusing, then the electron's Larmor radius will still be equal to it's initial Larmor radius in this example.


\subsection{The Effects of Slab Turbulence}
\label{subsec:SlabTurbulence}

Charged particles in the heliosphere would have followed the smooth motions described thus far, were it not for turbulence. Turbulence can be described as \emph{seemingly random fluctuations containing some level of correlations or structures} \citep{goldsteinetal1995, brunocarbone2013}. For the following discussion a non-relativistic proton is simulated in the turbulence model presented in Appendix~\ref{apndx:ModelSlab}, with $N_{\rm RH}^+ = N_{\rm LH}^+ = 1000$, using a fourth-order Runge-Kutta method, with $21600$ steps per gyration for $5$ gyrations (with respect to the background magnetic field; $\vec{B}_0 = B_0 \, \hat{\vec{z}}$ with $B_0 = 1 \times 10^{-12}$ $\mathrm{T}$). Keep in mind that these simulations will be similar, except for the sense of gyration, for an electron with the same momentum. Unless otherwise stated, the particle was initialised at the origin with a velocity of $\vec{v}_0 = (2 \, \hat{\vec{y}} + 3 \, \hat{\vec{z}})$ $\mathrm{m} \cdot \mathrm{s}^{-1}$ and an Alfv\'{e}n speed of $V_A = 0.001 \, v_0$ was used since particles normally move much faster than the waves. All quantities of interest were calculated with respect to the background magnetic field, as this provides a natural and unchanging `reference' field.


\subsubsection{Pitch-angle Scattering}
\label{subsubsec:PitchAngleScattering}

To build a systematic understanding of the influence of turbulent fluctuations, consider a magnetostatic wave field with a single wavelength (not shown). If the wavelength is long enough, the magnetic field is changing slowly enough with position such that the particle is able to follow the perturbed magnetic field line. If the wavelength is short enough, the particle moves so quickly over the fluctuations that it does not have time to react to it and its trajectory is only slightly perturbed. If the wavelength is on the order of the Larmor radius, the particle can resonate with the wave and the particle's trajectory is perturbed from the normal helix.

\begin{figure*}[!t]
\centering
\includegraphics[scale=0.33]{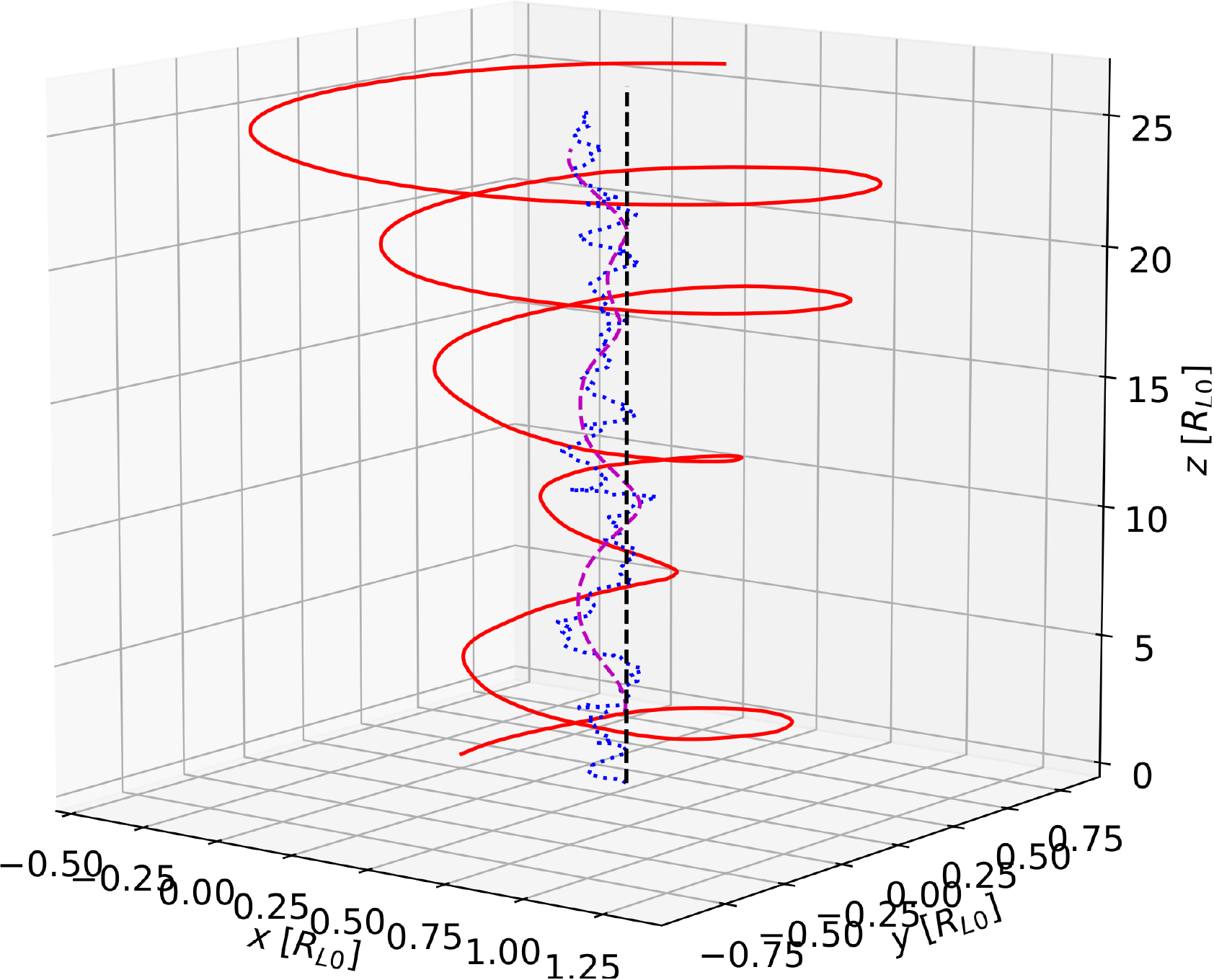}
\includegraphics[trim=4mm 10mm 24mm 28mm, clip, scale=0.26]{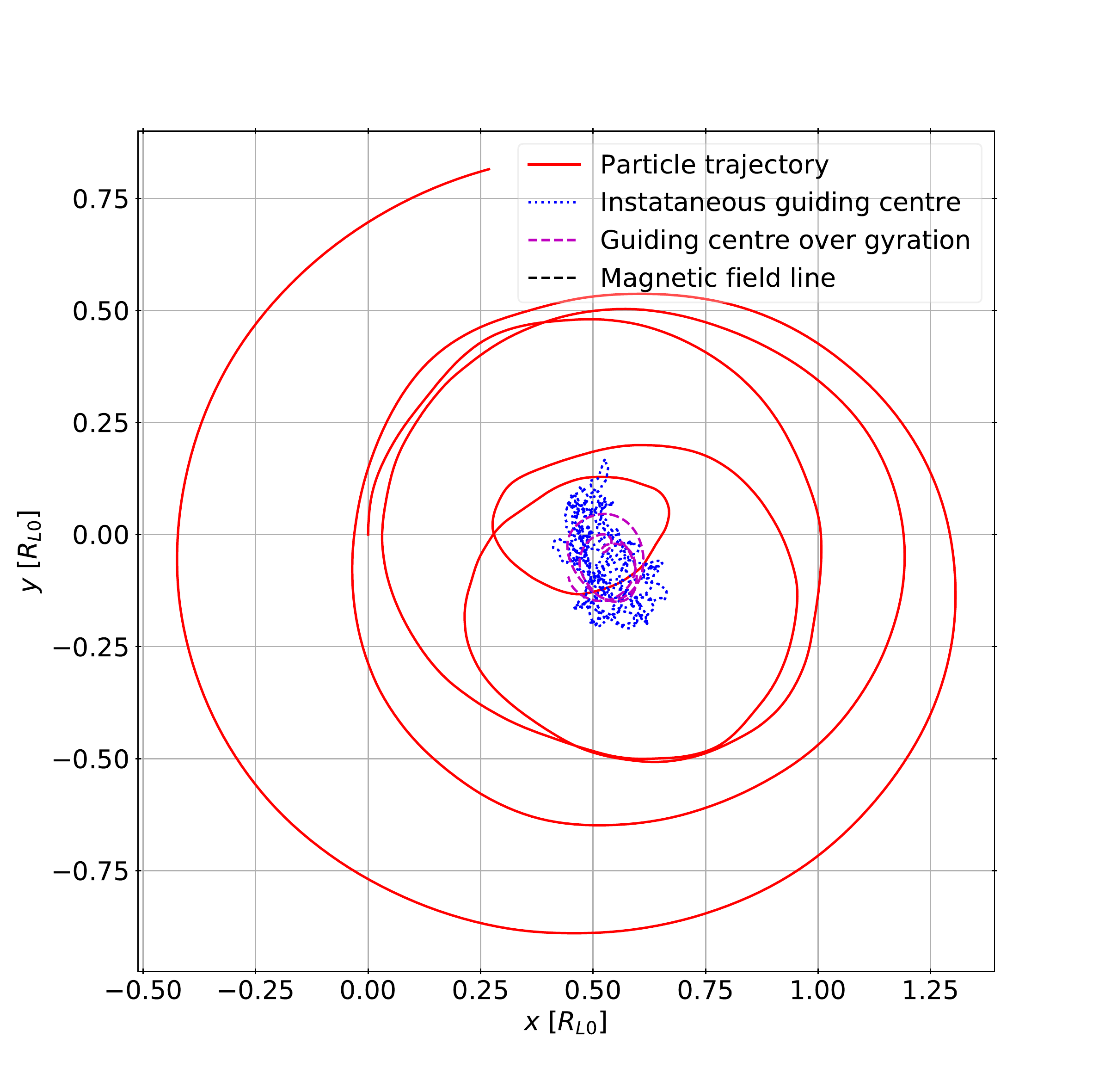} \\
\includegraphics[trim=40mm 10mm 45mm 20mm, clip, scale=0.29]{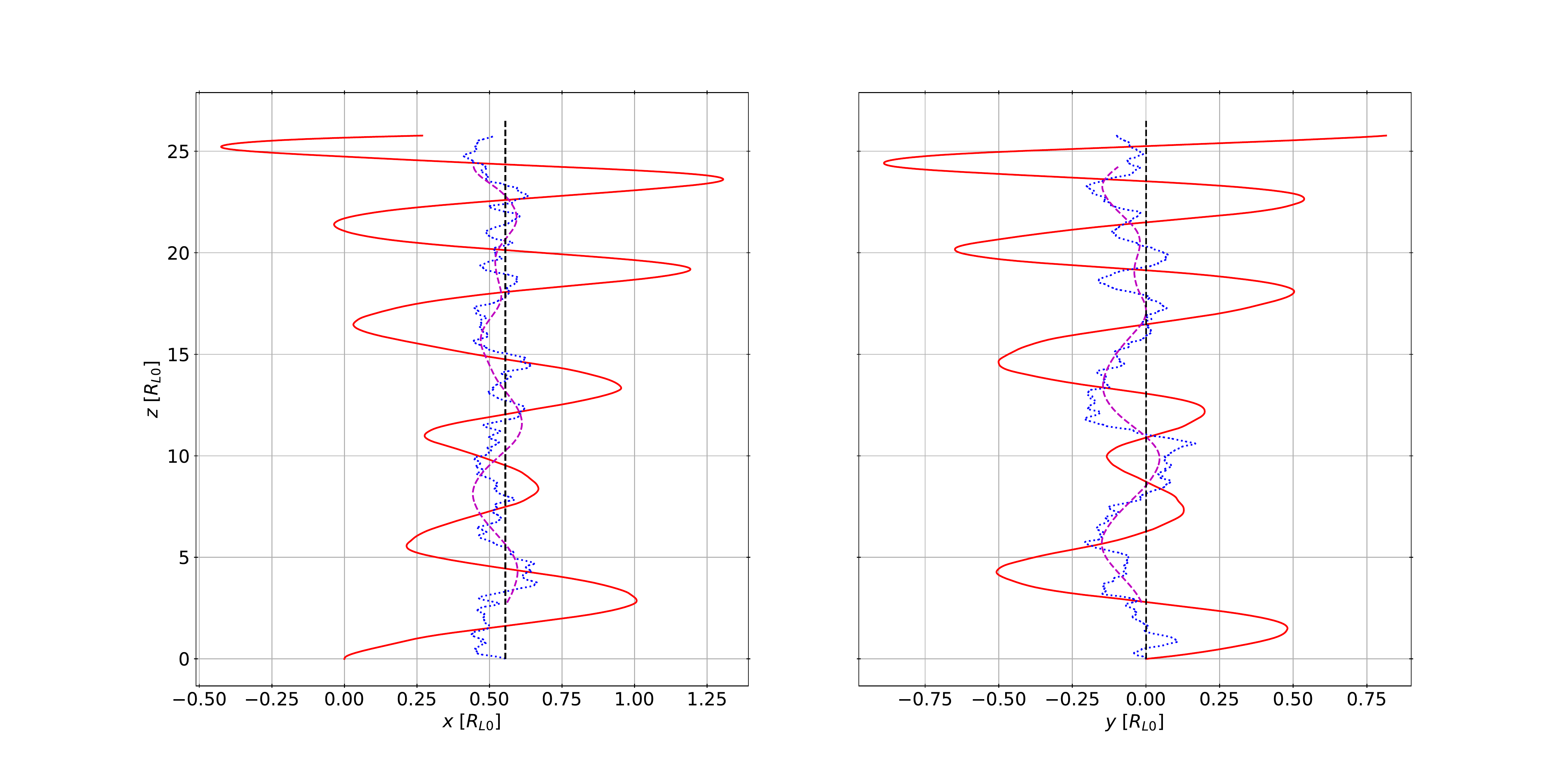}
\textbf{\caption[]{\label{fig:3DProjections}{\textnormal{Simulation of a proton in a constant and uniform background magnetic field with a spectrum of slab turbulence. The three-dimensional view (\textit{top left}) is projected onto the $xy$- (\textit{top right}), $xz$- (\textit{bottom left}), and $yz$-plane (\textit{bottom right}). The trajectories of the particle (solid red) and its guiding centre (dotted blue: Eq.~\ref{eq:GCPos}; dashed purple: running average of particle's position over a gyration) are shown, together with a single background magnetic field line.}}}}
\end{figure*}

Consider a moving resonant magnetic wave with a single wavelength, i.e. a parallel wavelength in the order of the particle's Larmor radius, $\lambda_{\parallel} \sim r_L$, but with no induced electric field. If the wave is moving much faster than the particle, the wave results in very large changes in both the perpendicular and parallel speeds and hence, in the pitch-angle. If the wave is moving much slower than the particle, the GC seems to jump to different regions of the slowly propagating magnetic field line over which it is moving. If the wave speed is equal to the particle's parallel speed (the Landau or Cherenkov resonance), a very strong resonance occur and the GC seems to be bouncing between two turning points, reminiscent of classical hard-sphere collisions. If the fluctuating electric field is also included, the only significant result is that the particle's energy then changes \citep[see][for illustrations of these discussions]{vandenberg2018}.

\begin{figure*}[!t]
\centering
\includegraphics[trim = 33mm 25mm 5mm 25mm, clip, scale=0.29]{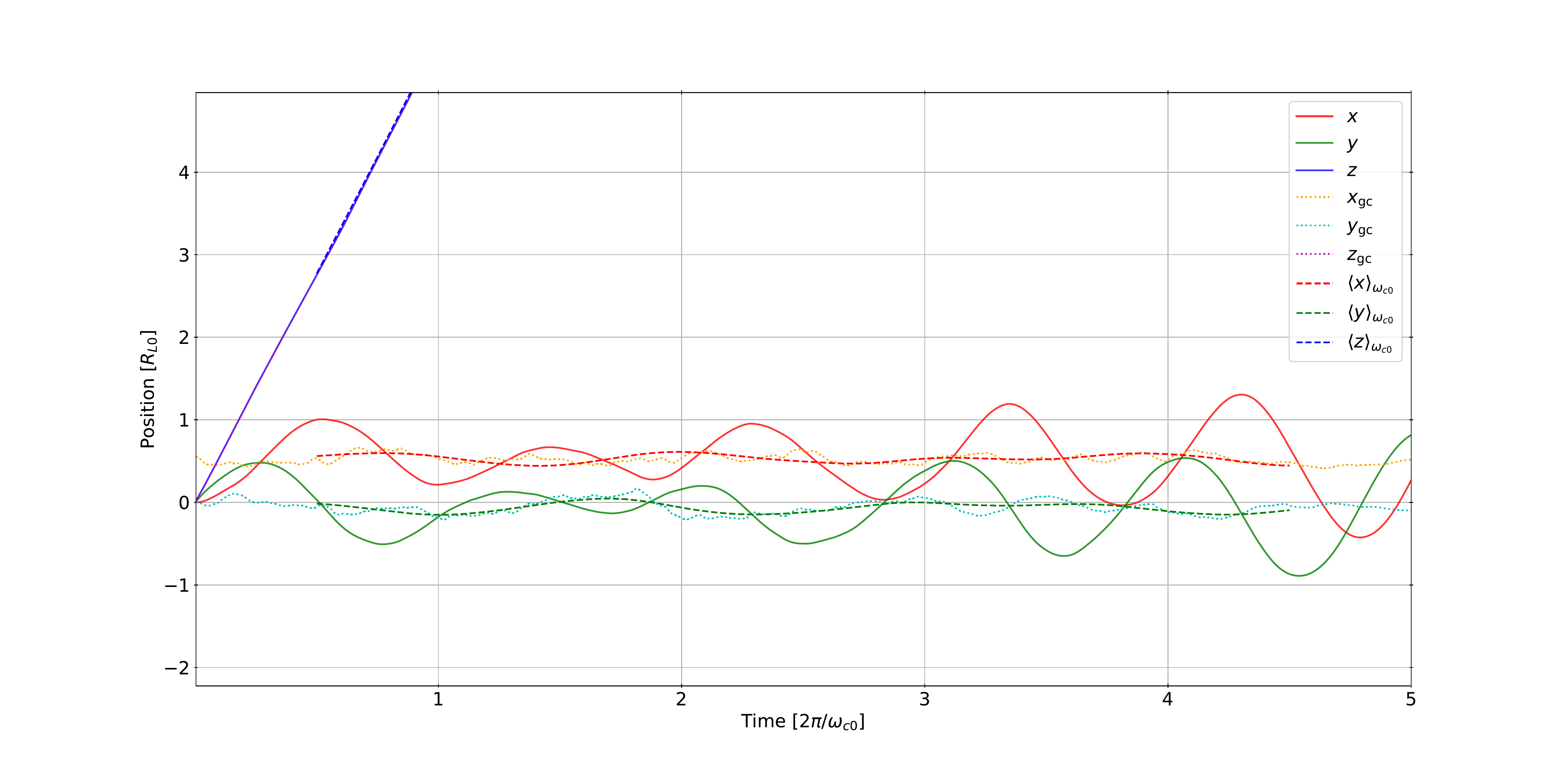}
\includegraphics[trim = 33mm 10mm 5mm 20mm, clip, scale=0.29]{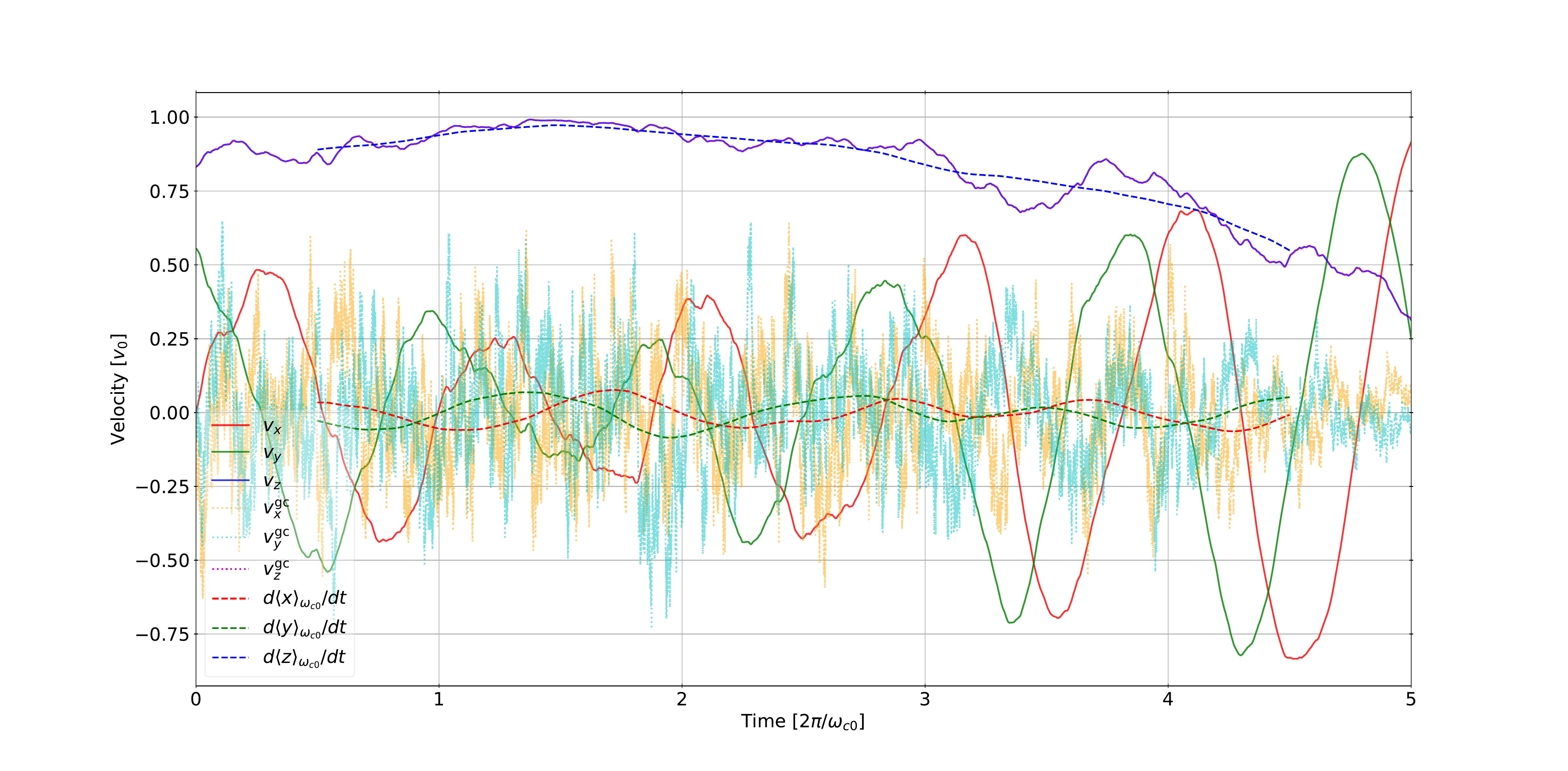}
\textbf{\caption[]{\label{fig:PosVel}{\textnormal{\textit{Top:} Cartesian components of the position (solid) and instantaneous (dotted; Eq.~\ref{eq:GCPos}) and gyro-averaged (dashed) guiding centre's position vectors for the proton in Fig.~\ref{fig:3DProjections}. \textit{Bottom:} Similar to the top panel, but for the velocity components.}}}}
\end{figure*}

Fig.~\ref{fig:3DProjections} shows the trajectories of a proton and its GC when interacting with a spectrum of slab turbulence. The GC was calculated here in two different ways, firstly the `instantaneous GC' was calculated from Eq.~\ref{eq:GCPos} and secondly the `average GC over a gyration' was calculated by performing a running average of the particle's position over a gyroperiod in the background magnetic field. Notice that although the spiral trajectory of the particle is highly perturbed, it is still smooth and continuous. The motion of the instantaneous GC, however, is more irregular and reminiscent of classical hard-sphere collisions. The different behaviour of the particle and the instantaneous GC can be understood if it is realized that any small changes in the particle's velocity would be amplified when `projecting the directional Larmor radius to the distant position of the GC'. In contrast to this, the gyro-averaged GC follow a smoother trajectory.

This qualitatively different behaviour between the particle and its GC can also be seen in Fig.~\ref{fig:PosVel} where the Cartesian components of the particle's and its GC's position and velocity vectors are shown. This behaviour is clearly seen in the velocity components: the particle's $x$- and $y$-velocity components still exhibit a fairly regular oscillation, while the $z$-velocity component have irregular features; both the position and velocity components of the gyro-averaged GC are smooth averages of the particle's components; the instantaneous GC's $z$-velocity component coincides with the particle's $z$-velocity component, but its $x$- and $y$-velocity components have discontinuous changes reminiscent of collisions.

Although not shown here, the changes in the parallel (equal to the velocity's $z$-component) and perpendicular speed components will cause the particle's pitch-angle to change continuously in an irregular way. This is then pitch-angle scattering and its effect can be seen as the particle is not moving at a constant speed along the magnetic field. \emph{Pitch-angle diffusion in velocity space therefore leads to parallel spatial diffusion in configuration space} \citep{shalchi2009}. It is also important to realise that pitch-angle scattering is a continuous process and that the pitch-angle {should not} simply be changed randomly according to some probability in simulations which integrate the Newton-Lorentz equation. Lastly notice that the GC stays close to the background magnetic field line on which it started. It is expected, both from theoretical considerations and simulations \citep[see][for a review]{shalchi2009}, that slab turbulence will lead to little or no perpendicular diffusion (mostly described as a random movement of the GC perpendicular to the background magnetic field). 


\subsubsection{Energy Conservation in the Wave Frame, but not in the Observer's Frame}
\label{subsubsec:EnergyConserv}

In a reference frame moving with the wave, where the fluctuations are magnetostatic with no induced electric field fluctuations, it is expected that the particle's energy should stay constant since the magnetic field alone cannot do any work on the particle. \citet{tsurutanilakhina1997} gives the following proof: Consider only the magnetic forces exerted on the particle by the wave, assume that the particle gains a quantum of energy $\Delta K = \hbar \omega$ from the wave during an interaction, and that the change in parallel momentum is $m \Delta v_{\parallel} = \hbar k_{\parallel}$, where $\hbar$ is Planck's constant divided by $2 \pi$. If the energy change is small compared to the particle's kinetic energy ($K = m v_{\parallel}^2 / 2 + m v_{\perp}^2 / 2$), then it would hold that $\Delta K = \omega m \Delta v_{\parallel} / k_{\parallel} \approx m (v_{\parallel} \Delta v_{\parallel} + v_{\perp} \Delta v_{\perp} )$, which gives
\begin{equation*}
\frac{1}{2} m \left( v_{\parallel} - V_A \right)^2 + \frac{1}{2} m v_{\perp}^2 = {\rm constant}
\end{equation*}
upon integration. This shows that the particle's energy in the wave frame is conserved. In the observer's frame, however, there exists an induced fluctuating electric field, which can change the particle's energy. Thus, \emph{the particle's energy is conserved in the wave frame, but not in the observer's frame}.

\begin{figure}[!t]
\centering
\includegraphics[trim = 35mm 9mm 5mm 25mm, clip, scale=0.29]{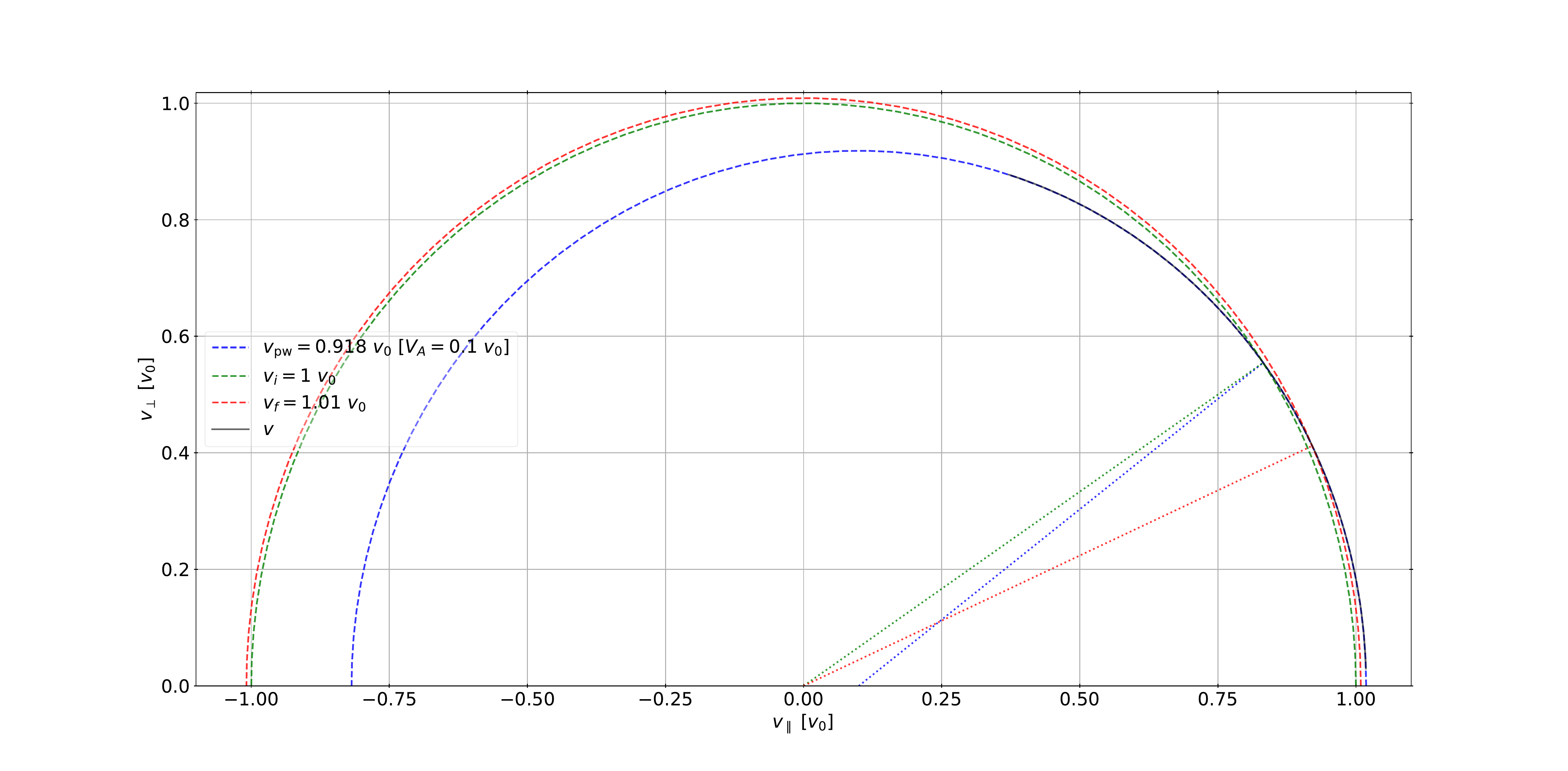}
\textbf{\caption[]{\label{fig:VelSpace}{\textnormal{Trajectory (solid black) of the proton in Fig.~\ref{fig:3DProjections} in velocity space, where the perpendicular speed (Eq.~\ref{eq:PerpSpeed}) is plotted as a function of the parallel speed (Eq.~\ref{eq:ParallelSpeed}). The dashed semi-circles indicate constant speed and the pitch-angle is the angle between the positive $v_{\parallel}$-axis and the velocity vector (dotted lines). The dashed blue semi-circle indicates the particle's initial speed in the wave frame.}}}}
\end{figure}

The trajectory of the simulated particle in velocity space ($v_{\perp}$ as a function of $v_{\parallel}$) is shown in Fig.~\ref{fig:VelSpace}. The dashed semi-circles indicates constant speed, with the green and red vectors representing the particle's initial and final velocity vectors, respectively. The dashed blue semi-circle indicates the particle's initial speed in the wave frame ($v_{\rm pw} = \sqrt{(v_{\parallel 0} - V_A)^2 + v_{\perp 0}^2}$) and the blue dotted line is its initial velocity vector in the wave frame. This figure clearly illustrates that the particle's energy is conserved in the wave frame since the trajectory lies on the blue semi-circle, but that the particle's energy is continuously changing in the observer's frame. In this graph, the pitch-angle is the angle between the positive $v_{\parallel}$-axis and the velocity vector. Pitch-angle scattering can therefore be seen here as the trajectory moves on the semi-circle. Although this is the velocity space trajectory for only a single particle, the extent of the trajectory towards both $0^{\circ}$ and $90^{\circ}$ pitch-angles are indicative of turbulence trying to isotropise the distribution of particles (in this case a single particle) in the wave frame.

\subsection{Summary of Micro-physic Concepts}
\label{subsec:MicroSummary}

\begin{figure*}[!t]
\centering
\includegraphics[page=2, clip, trim=18mm 174mm 100mm 18mm, scale=0.7]{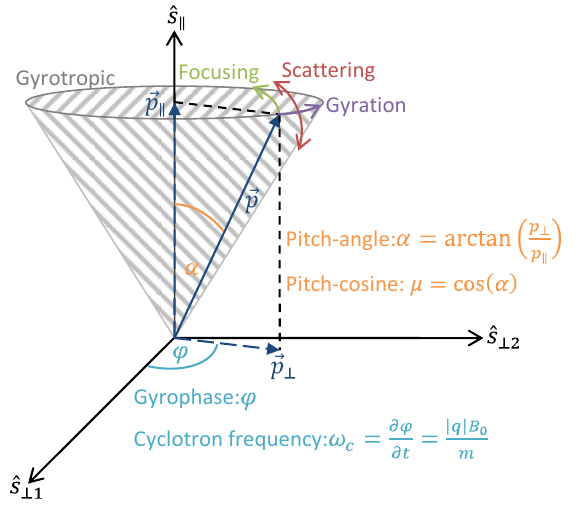}
\textbf{\caption[]{\label{fig:MicroSummary}{\textnormal{Illustration of the various processes and definitions introduced. Shown is the particle's momentum space in a field-aligned coordinate system. See Section~\ref{subsec:MicroSummary} for details. This picture was inspired by \citet{prinslooetal2019}.}}}}
\end{figure*}

All of the concepts and processes discussed in this section can be connected conceptually to the next section with the illustration given in Fig.~\ref{fig:MicroSummary}. The particle's momentum space in a field-aligned reference frame is shown. The polar angle, the angle between the $s_{\parallel}$-axis and the momentum vector $\vec{p}$, is the pitch-angle $\alpha$ (Eq.~\ref{eq:PitchAngle}). Using the definition of the pitch-cosine $\mu$ (Eq.~\ref{eq:PitchCosine}), the particle's momentum can be decomposed into a parallel $\vec{p}_{\parallel} = \mu p \, \hat{\vec{s}}_{\parallel}$ ($\hat{\vec{s}}_{\parallel}$ is a unit vector in the direction of the background magnetic field) and perpendicular $\vec{p}_{\perp} = p \sqrt{1 - \mu^2} \left( \cos \varphi \; \hat{\vec{s}}_{\perp 1} + \sin \varphi \; \hat{\vec{s}}_{\perp 2} \right)$ component ($\hat{\vec{s}}_{\perp 1}$ and $\hat{\vec{s}}_{\perp 2}$ are two mutually perpendicular unit vectors lying in the plane perpendicular to the background magnetic field), similar to Eq.~\ref{eq:ParallelSpeed} and Eq.~\ref{eq:PerpSpeed}, respectively. Here $p$ is the magnitude of the momentum and can be thought of as a radius in momentum space. The parallel momentum is the projection of the momentum vector onto the background magnetic field direction, while the perpendicular momentum is the projection of the momentum vector onto the plane perpendicular to the background magnetic field. The azimuthal angle, the angle between the $s_{\perp 1}$-axis and the perpendicular momentum, is the particle's gyrophase $\varphi$ and its rate of change is the cyclotron frequency $\omega_c$ (Eq.~\ref{eq:CyclotronFrequency}). The gyration of the particle around the magnetic field causes the momentum vector to precess around the $s_{\parallel}$-axis at the cyclotron frequency.

Focusing will decrease the particle's pitch-angle, while scattering will either increase or decrease it, as indicated by the green and red arrows, respectively. As the number of particles under consideration in a real event is so large that there is, for all practical purposes, a particle in every phase of gyration, a gyrotropic distribution of particles are normally considered. This can be illustrated as a collection of particles having the same pitch-angle, but different gyrophases and is indicated by the grey circle (also referred to as a ring-distribution). This also represents a gyrotropic distribution of mono-energetic particles, since all of their momentum vectors have the same magnitude. If the particles were to have different energies and gyrophases, but the same pitch-angle, then their momentum vectors will form the shaded cone. The cone will then represent a possible anisotropic distribution as \emph{the particles have a preferred direction of motion along the background magnetic field}. For a gyrotropic distribution of mono-energetic particles, pitch-angle scattering will cause the circle to change into a spherical shell (also referred to as a shell-distribution), assuming that the scattering does not change the particles' energy and that enough time has elapsed. Similarly, pitch-angle scattering will cause the cone of an anisotropic distribution to become a filled sphere. In such a case, the distribution will be called isotropic with particles of all energies moving in all directions. Turbulence can therefore drastically change the characteristics of the original particle distribution and will mostly act to isotropise an anisotropic distribution.


\newpage

\section{The Distribution Function and Focused Transport Equation}
\label{sec:Macroscopic}

A macro-physical or ensemble averaged description of SEPs is needed for most modelling purposes and since SEPs are highly anisotropic, the \citet{parker1965} transport equation, used to model CRs, is inadequate for this purpose. The evolution of the anisotropic SEP distribution function can be described by a so called focused transport equation (FTE). The concept of a distribution function and the simplest form of the FTE will be introduced here.


\subsection{The Distribution Function}
\label{subsec:DistFunc}

The distribution function $f(\vec{x}; \vec{p}; t) = {\rm d}N / {\rm d}^3x \, {\rm d}^3p$ of a system is defined as the number density ${\rm d}N$ in a volume element ${\rm d}^3x \, {\rm d}^3p$ of the $6$-dimensional phase-space spanned by the three spatial $\vec{x}$ and momentum $\vec{p}$ coordinates. It can be interpreted as the number of particles at time $t$ having position vectors between $\vec{x}$ and $\vec{x} + {\rm d}\vec{x}$ with momentum vectors between $\vec{p}$ and $\vec{p} + {\rm d}\vec{p}$. Integrating the distribution function over all space and momentum would give the total number of particles in the system. Dividing the distribution function by the total number of particles, results in a probability distribution to find particles in the phase-space volume ${\rm d}^3x \, {\rm d}^3p$ around $(\vec{x}; \vec{p})$ at time $t$ \citep{choudhuri1998, moraal2013, zank2014}.

Plasma physics or transport theory textbooks \citep[see e.g.][]{chen1984, choudhuri1998} usually defines the distribution function in terms of velocity and not momentum. Such a distribution function is fine for non-relativistic particles, but for relativistic particles, a distribution function defined in terms of momentum is preferred. Consider an observer frame, where quantities are unprimed, and a frame moving with respect to the observer frame, where quantities are primed. It can be proven that the phase-space volume element is invariant, ${\rm d}^3x \, {\rm d}^3p = {\rm d}^3x' {\rm d}^3p'$ \citep[see e.g.][]{zank2014}, with the implication that the distribution function would also be invariant, $f(\vec{x}; \vec{p}; t) = f'(\vec{x}'; \vec{p}'; t')$. This is expected since the distribution function is related to the particle number density which is invariant between different reference frames. {It is here implicitly assumed that a non-relativistic transformation can be made between the stationary and solar wind (SW) frames, so that $t = t'$ may be assumed.} However, if the distribution function is defined in terms of velocity, then the phase-space volume element is not invariant and the distribution function also not \citep{moraal2013, zank2014}.

For a plasma with a stationary background or large scale average magnetic field, the magnetic field can be used as a reference point. The distribution function can then be defined in a field aligned coordinate system and a transformation from Cartesian to spherical coordinates can be made in momentum space (see Fig.~\ref{fig:MicroSummary}), such that ${\rm d}N = f (s_{\parallel}; \vec{s}_{\perp}; p; \mu; \varphi; t) {\rm d}s_{\parallel} {\rm d}^2s_{\perp} {\rm d}p \, {\rm d}\mu \, {\rm d}\varphi$. The dependence of the distribution function on $\varphi$ can be averaged out to yield the gyrotropic distribution function $f (s; p; \mu; t) = \int_0^{2 \pi} f (s_{\parallel}; \vec{s}_{\perp}; p; \mu; \varphi; t) {\rm d}\varphi / 2 \pi$. By performing such an average, transport perpendicular to the magnetic field is removed \citep[see e.g.][]{zank2014}, hence the dependence on $\vec{s}_{\perp}$ was neglected and $s$ was written for $s_{\parallel}$. Drifts or diffusion perpendicular to the magnetic field is therefore not described here and this distribution function can be thought of as describing the number of particles per phase space volume in a given flux tube \citep{ngwong1979}. Notice that the neglect of perpendicular transport implies that the intensity of an SEP event might be overestimated.

The distribution function is a quantity of theoretical interest, but it can give a complete description of a system's state and various useful quantities can be calculated from it \citep[][e.g. show how the hydrodynamic equations can be derived from the distribution function and its governing equation]{chen1984, choudhuri1998}. The omni-directional intensity (ODI)
\begin{equation*}
F_0(s; p; t) = \frac{1}{2} \int_{-1}^1 f (s; p; \mu'; t) {\rm d}\mu' ,
\end{equation*}
is essentially the distribution function without a pitch-angle dependence and represents the number of particles at time $t$ within ${\rm d}s$ from $s$ with a momentum between $p$ and $p + {\rm d}p$. It is related to the measured differential intensity in terms of kinetic energy by $j = p^2 F_0 / 2$ for protons or electrons, which has the dimensions of particles per unit area, per unit time, per unit solid angle, per unit kinetic energy \citep[][discusses in detail the relation among different observable quantities]{moraal2013} \citep[see also the summary of][]{prinslooetal2019}. The first order anisotropy
\begin{equation*}
A(s; p; t) = 3 \frac{\int_{-1}^1 \mu' f (s; p; \mu'; t) {\rm d}\mu'}{\int_{-1}^1 f (s; p; \mu'; t) {\rm d}\mu'} ,
\end{equation*}
is a measure of how anisotropic the distribution is at a certain phase-space point $(s;p)$ at a time $t$. Notice that the distribution function (phase-space density) is changed to a probability by dividing with $\int_{-1}^1 f (s; p; \mu'; t) {\rm d}\mu'$, and the anisotropy can therefore be interpreted as essentially three times the average or expected pitch-cosine. It has a value of $3$ ($-3$) if all particles are moving along (in the opposite direction of) the magnetic field and a value of zero if there are equal number of particles moving in opposite directions (isotropic) or if all the particles have no parallel speed (an unlikely case). The anisotropy is usually calculated in observations from the pitch-angle distribution (PAD)
\begin{equation*}
F (s; p; \mu; t) = \frac{f (s; p; \mu; t)}{\int_{-1}^1 f (s; p; \mu'; t) {\rm d}\mu'} ,
\end{equation*}
which is a probability distribution and is normally constructed from the sectored measurements of detectors looking in different directions.


\subsection{The Focused Transport Equation}
\label{subsec:FTE}

The distribution function's evolution is in general governed by the Fokker-Planck equation, which is a generalisation of Liouville's theorem for a distribution function including the effects of random changes to the momentum coordinates by turbulence or collisions \citep{choudhuri1998, zank2014}. A transformation from Cartesian to spherical coordinates in momentum space is made and an average over gyrophase is then preformed, as described in the previous paragraphs. Additionally, a transformation can first be made from the observer's frame to a wave frame, usually assumed to be the SW frame, because momentum diffusion can be neglected in this frame. Alternatively, the Vlaslov equation, essentially the collisionless Boltzmann equation with the Lorentz force substituted, can be used as a point of departure. The distribution function and the electric and magnetic field must then be written as the sum of a large scale average and a rapid fluctuating part, with the fluctuating part acting as a perturbation on the average part. Such derivations, as given by \citet{zhang2006} or \citet{zank2014}, lead to the focused transport equation (FTE), but are lengthy and beyond the scope of the current discussion.

Although the name ``focused transport equation'' might be a misnomer, as it describes the evolution of any anisotropic distribution, it is appropriate in the case of SEPs since the anisotropy is caused primarily by focusing. The simplest form of the FTE, is that of \citet{roelof1969} without advection or energy losses
\begin{equation}
\label{eq:FTPE}
\frac{\partial f}{\partial t} + \frac{\partial}{\partial s} \left[ \mu v f \right] + \frac{\partial}{\partial \mu} \left[ \frac{(1 - \mu^2) v}{2L(s)} f \right] = \frac{\partial}{\partial \mu} \left[ D_{\mu \mu} \frac{\partial f}{\partial \mu} \right] ,
\end{equation}
where $L(s)$ is the focusing length of the magnetic field given by Eq.~\ref{eq:FocusingLength} and $D_{\mu \mu}$ is the pitch-angle diffusion coefficient (PADC) describing the random changes of the pitch-angle due to turbulence. This equation describes the evolution of the distribution function $f(s;\mu;t)$ for a constant particle speed $v$. The various terms, from left to right, describe temporal, spatial (the streaming of particles along the magnetic field, since $\mu v$ is their parallel speed), and pitch-angle changes (discussed in Appendix~\ref{apndx:FocusPAD}) on the left hand side, and pitch-angle diffusion on the right hand side. It should be noticed that the FTE is a highly non-linear, second order, parabolic partial differential equation. The different processes' effects cannot be added linearly because each process is dependent on quantities which are affected by the other processes. The various terms therefore affect one another and the dominating process is ultimately determined by its relative strength. This non-linearity and competition between terms imply that none of the terms can be neglected to model SEPs realistically.

\begin{figure*}[!t]
\centering
\includegraphics[trim=10mm 10mm 15mm 20mm, clip, scale=0.4]{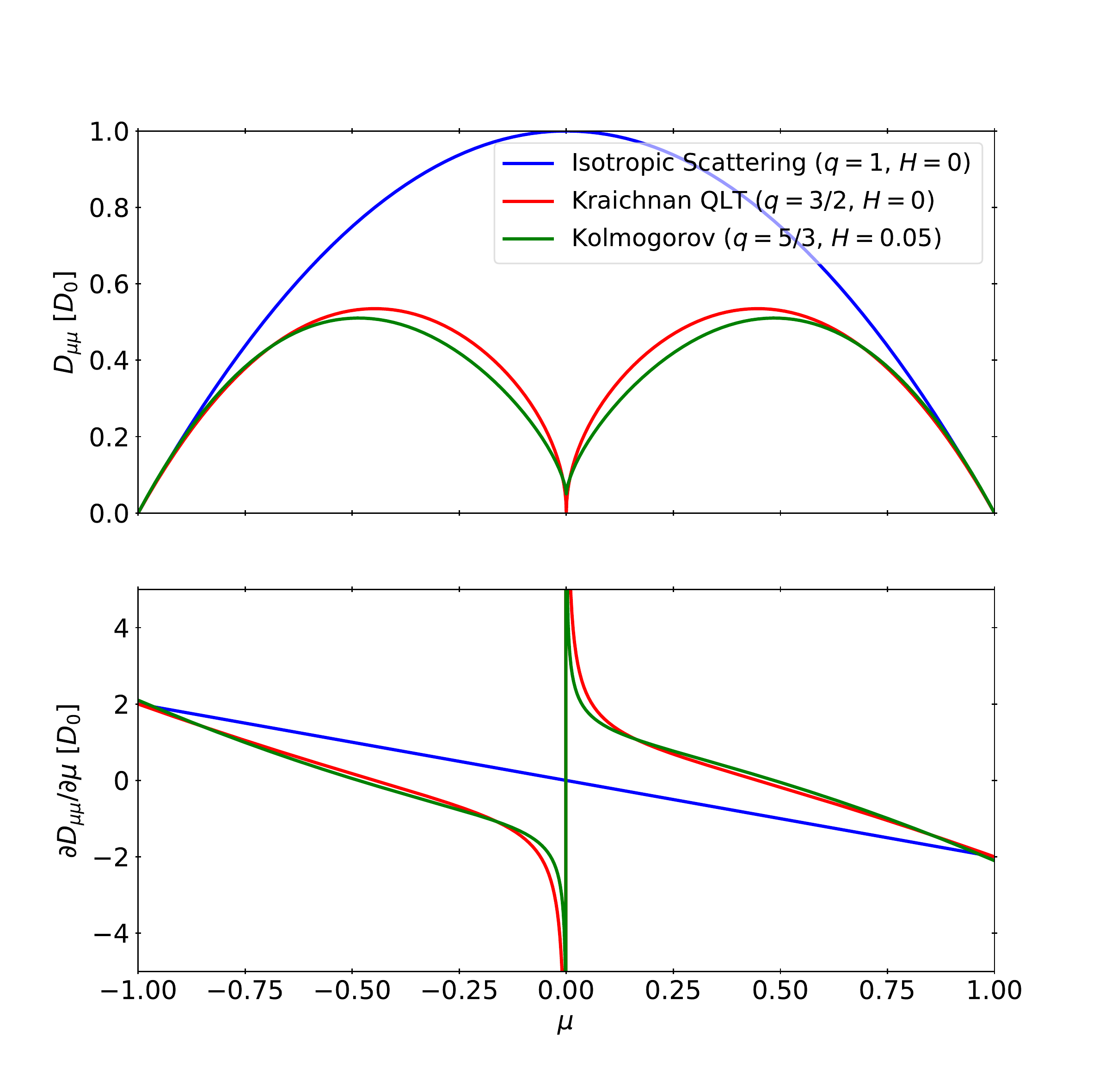}
\textbf{\caption[]{\label{fig:DiffCoefs}{\textnormal{Pitch-angle diffusion coefficients (\textit{top}) and their derivatives (\textit{bottom}) of isotropic scattering (blue; Eq.~\ref{eq:IsotropicScattering}), quasi-linear theory with a Kraichnan inertial range (red; Eq.~\ref{eq:QLT} with $q = 3/2$), and \citet{beeckwibberenz1986} with a Kolmogorov inertial range (green; Eq.~\ref{eq:FixedQLT} with $q = 5/3$ and $H = 0.05$).}}}}
\end{figure*}

The PADC must be specified and a variety of options are available from different theories. Three rather simple forms will be used here for illustrative proposes. A widely used PADC is that of \citet{beeckwibberenz1986},
\begin{equation}
\label{eq:FixedQLT}
D_{\mu \mu}^{\rm BW} = D_0 (1 - \mu^2) (|\mu|^{q-1} + H) ,
\end{equation}
based on quasi-linear theory \citep[QLT;][]{jokipii1966, shalchi2009}. Here $D_0$ is the scattering amplitude, $q$ is the spectral index of the magnetic turbulence's inertial range, and $H$ is an arbitrary (in terms of its value) correction to describe the inclusion of dynamical effects. If $q = 1$ and $H = 0$, then
\begin{equation}
\label{eq:IsotropicScattering}
D_{\mu \mu}^{\rm iso}(\mu) = D_0 (1 - \mu^2)
\end{equation}
is called isotropic scattering. This PADC can be used in the presence of very strong turbulence, but if the turbulence is weaker and pitch-angle scattering is caused by resonances with a spectrum of waves, then anisotropic scattering must be used. If dynamical effects are neglected ($H = 0$), then
\begin{equation}
\label{eq:QLT}
D_{\mu \mu}^{\rm QLT}(\mu) = D_0 (1 - \mu^2) |\mu|^{q-1}
\end{equation}
has the known problem of a resonance gap at $\mu = 0$ ($D_{\mu \mu}^{\rm QLT}(0) = 0$) \citep{droge2000a}. Fig.~\ref{fig:DiffCoefs} shows the different PADCs and their derivatives. Care should be taken here not to confuse isotropic or anisotropic scattering with an isotropic or anisotropic distribution.

The scattering amplitude is usually calculated from the parallel mean free path (MFP). The MFP can be generally defined as \emph{the average distance moved by a particle before its velocity is uncorrelated with its initial velocity}. Based, however, on the results of the previous section, the parallel MFP might be better interpreted as \emph{the average distance a particle would move in a turbulent plasma, being continuously subjected to small pitch-angle changes, before the pitch-angle is changed significantly and the particle's GC reverses its direction of motion parallel to the background magnetic field}. The connecting formula between $D_0$ and the parallel MFP, is
\begin{equation}
\label{eq:ParallelMFP}
\lambda_{\parallel}^0 = \frac{3}{8} v \int_{-1}^1 \frac{(1 - \mu'^2)^2}{D_{\mu \mu} (\mu')} {\rm d}\mu'
\end{equation}
for an isotropic distribution. Notice that this is not a formal definition, but rather a consequence of averaging Eq.~\ref{eq:FTPE} over pitch-cosine in the absence of focusing for an isotorpic distribution \citep{jokipii1966, hasselmannwibberenz1970, shalchi2009}. In keeping the focusing term (as is done with the diffusion-advection and telegraph equations in Appendix~\ref{apndx:DiffTel}), the parallel MFP for an anisotropic distribution becomes
\begin{equation}
\label{eq:ParalMFPFocus}
\lambda_{\parallel} = 3 L \frac{\int_{-1}^1 \mu' e^{G(\mu')} {\rm d}\mu'}{\int_{-1}^1 e^{G(\mu')} {\rm d}\mu'} ,
\end{equation}
where $G(\mu)$ is given by Eq.~\ref{eq:PADg}. This expression reduces to the former in the absence of focusing \citep{beeckwibberenz1986, heschlickeiser2014}. From this it can be seen that \emph{the interpretation of the parallel MFP is modified in the presence of focusing}. It was already stated by \citet{earl1981} that the MFP would change due to the focusing length, because focusing causes the distribution to have pitch-angles close to $\mu \sim 1$ where particles would experience less scattering and have a larger parallel MFP. Additionally, in the heliosphere where $L(s)$ is position dependent, \emph{the parallel MFP would also change with position}. Due to these reasons, it might be better to calculate $D_0$ from observable turbulence properties instead \citep[][presents a spatially varying MFP based on these considerations]{hewan2012}.


\subsection{Comparison of the Diffusion and Telegraph Approximations to Describe Focused Transport}
\label{subsec:Compare}

In the isotropic limit, the transport could be well described by a diffusion equation \citep[see][]{parker1965}. The force field approximation could successfully be applied to galactic CR spectra, even though all the complicated modulation processes (such as advection, diffusion, energy losses, and drifts) were absorbed into a single parameter \citep[i.e. the modulation potential; see][]{moraal2013}. Analytical approximations also exist for the propagation time and average energy losses of CRs \citep[see again][]{parker1965}. Within focused transport there is unfortunately no simplistic approximation which give satisfactory results. The advection-diffusion and telegraph approximations are introduced in Appendix~\ref{apndx:DiffTel} and it will be shown to what extend these approximations can be used. The analytical approximations will be compared to two numerical solutions of the FTE. The first solution, revered to as `the model', uses a finite difference scheme and is given in Appendix~\ref{apndx:FDSolver} (including a link to the source code). The second solution, used as synthetic data, uses a stochastic differential equation approach and is discussed in Appendix~\ref{apndx:SDESolver}.

Energy losses can be considered to be negligible for $100$ $\mathrm{keV}$ electrons and will be used here as an example. A constant parallel MFP and focusing length of $\lambda_{\parallel}^0 = 0.3$ $\mathrm{AU}$ and $L = 0.9$ $\mathrm{AU}$, respectively, will be used. These choices are informed by Section~\ref{subsubsec:SEPevent} (the focusing length used here is the average value within the first $2$ $\mathrm{AU}$ from the Sun) and yields $\xi = \lambda_{\parallel}^0 / L = 1/3$, which is in the weak focusing limit necessary for the anisotropic case. The injection is located at $s_0 = 0$ $\mathrm{AU}$ and an observer is assumed to be located at $s = 1.2$ $\mathrm{AU}$ (roughly the position of Earth). The coefficients used in the analytical approximations are given in Appendix~\ref{subsec:Coefficients}. See \citet{fiskaxford1969}, \citet{earl1981}, \citet{litvinenkonoble2013}, \citet{litvinenkoschlickeiser2013}, and \citet{effenbergerlitvinenko2014} for similar or further discussions, including results in the absence of focusing.


\subsubsection{Isotropic Scattering with a Constant Focusing Length}
\label{subsubsec:Isotropic}

\begin{figure*}[!t]
\centering
\includegraphics[trim=10mm 0mm 20mm 12mm, clip, scale=0.5]{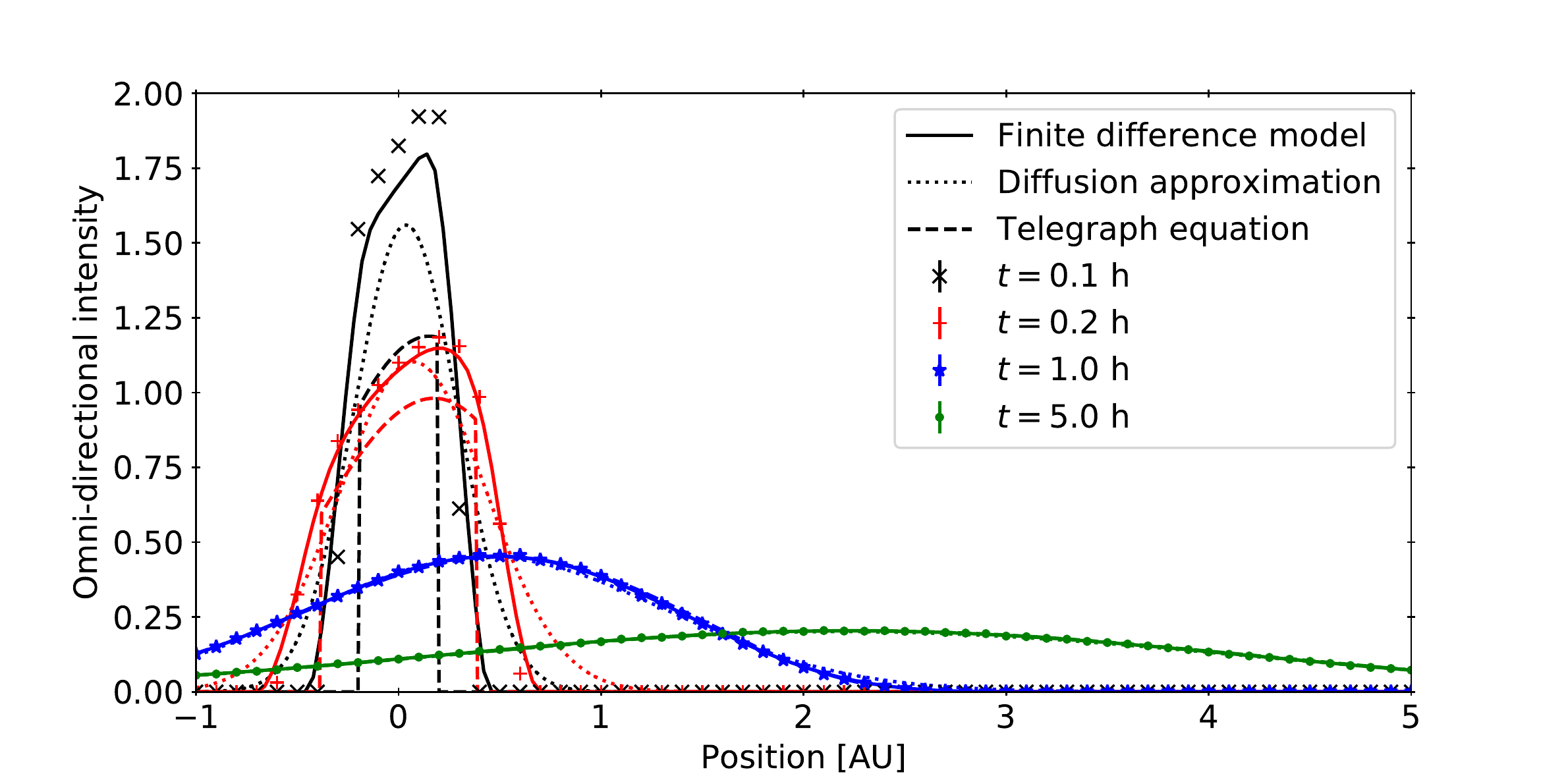}
\includegraphics[trim=10mm 0mm 20mm 12mm, clip, scale=0.5]{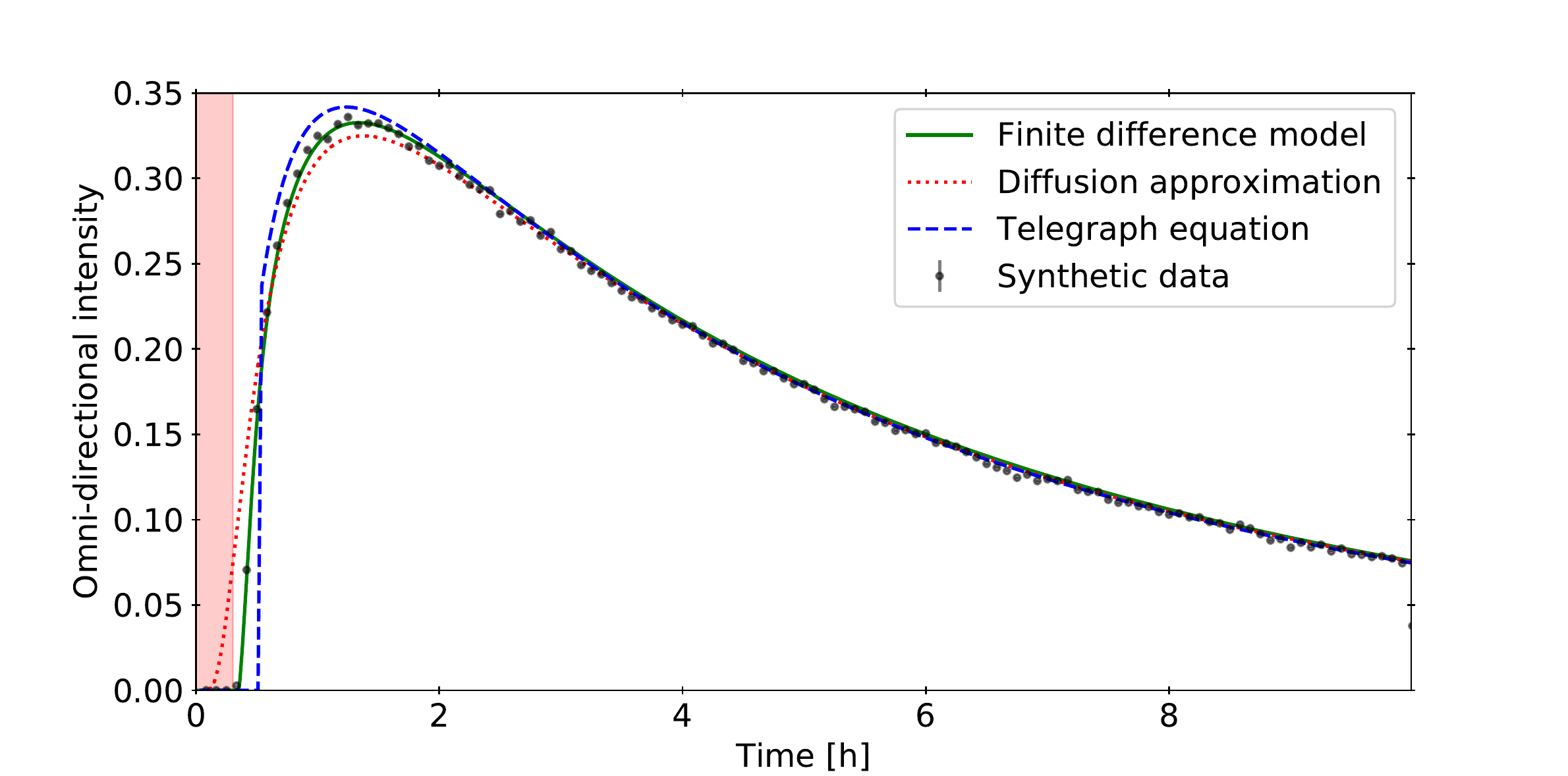}
\textbf{\caption[]{\label{fig:IsoODI}{\textnormal{\textit{Top:} Temporal evolution of the normalised omni-directional intensity as a function of position for $100$ $\mathrm{keV}$ electrons with isotropic pitch-angle scattering (Eq.~\ref{eq:IsotropicScattering}), $\lambda_{\parallel}^0 = 0.3$ $\mathrm{AU}$, and $L = 0.9$ $\mathrm{AU}$. The synthetic data (symbols; calculated from a stochastic differential equation model) and finite difference model (solid lines) are compared to the diffusion approximation (dash dotted lines; Eq.~\ref{eq:DiffusionApprox}) and telegraph equation (dashed lines; Eq.~\ref{eq:TelegraphApprox}). \textit{Bottom:} Normalised omni-directional intensity as a function of time as seen by an observer at $s = 1.2$ $\mathrm{AU}$. The red shaded period indicates how long it would take particles to reach the observer if they propagate in a ballistic fashion along the magnetic field.}}}}
\end{figure*}

\begin{figure*}[!t]
\centering
\includegraphics[trim=10mm 0mm 18mm 12mm, clip, scale=0.5]{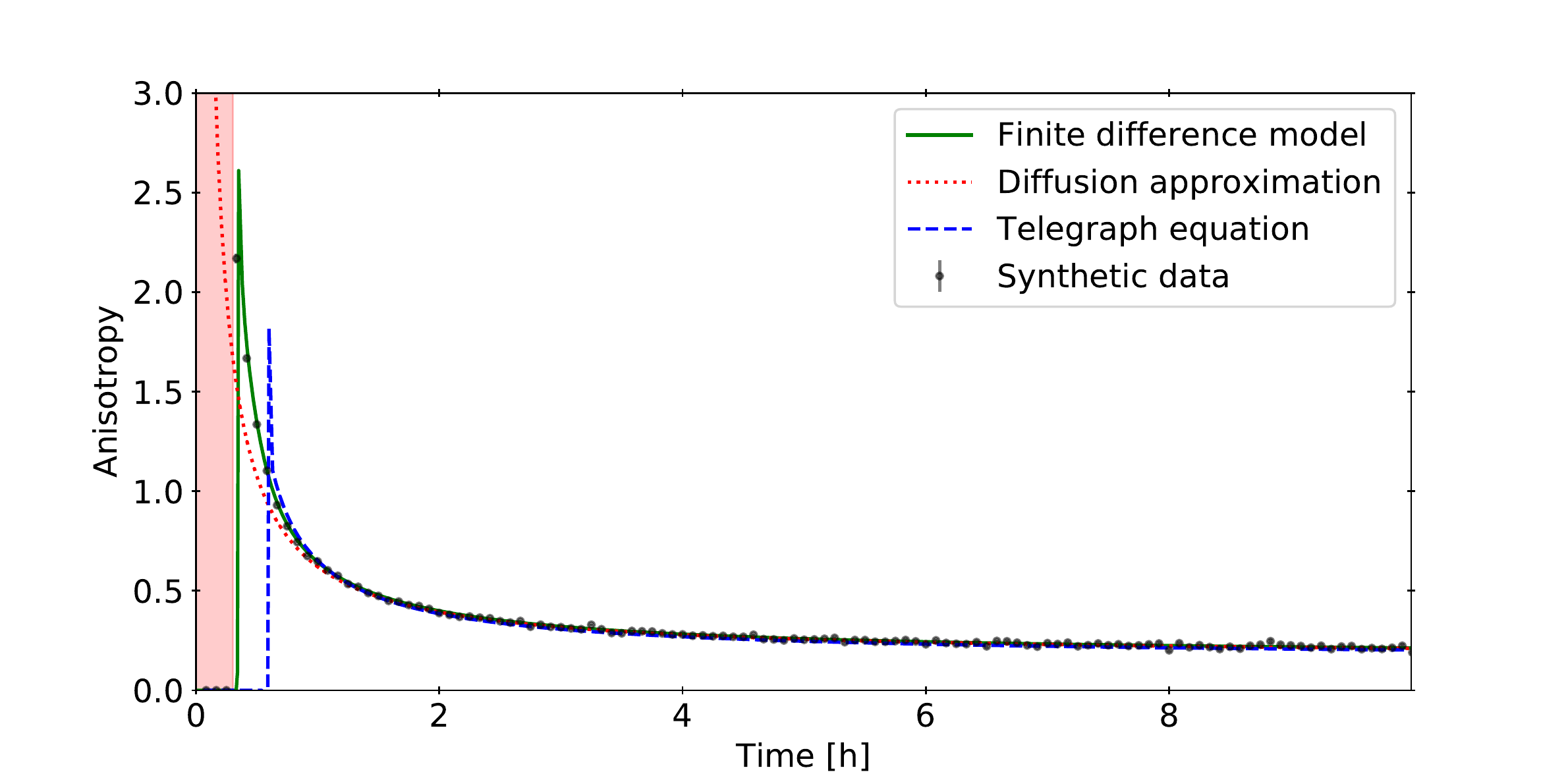}
\includegraphics[trim=10mm 0mm 18mm 12mm, clip, scale=0.5]{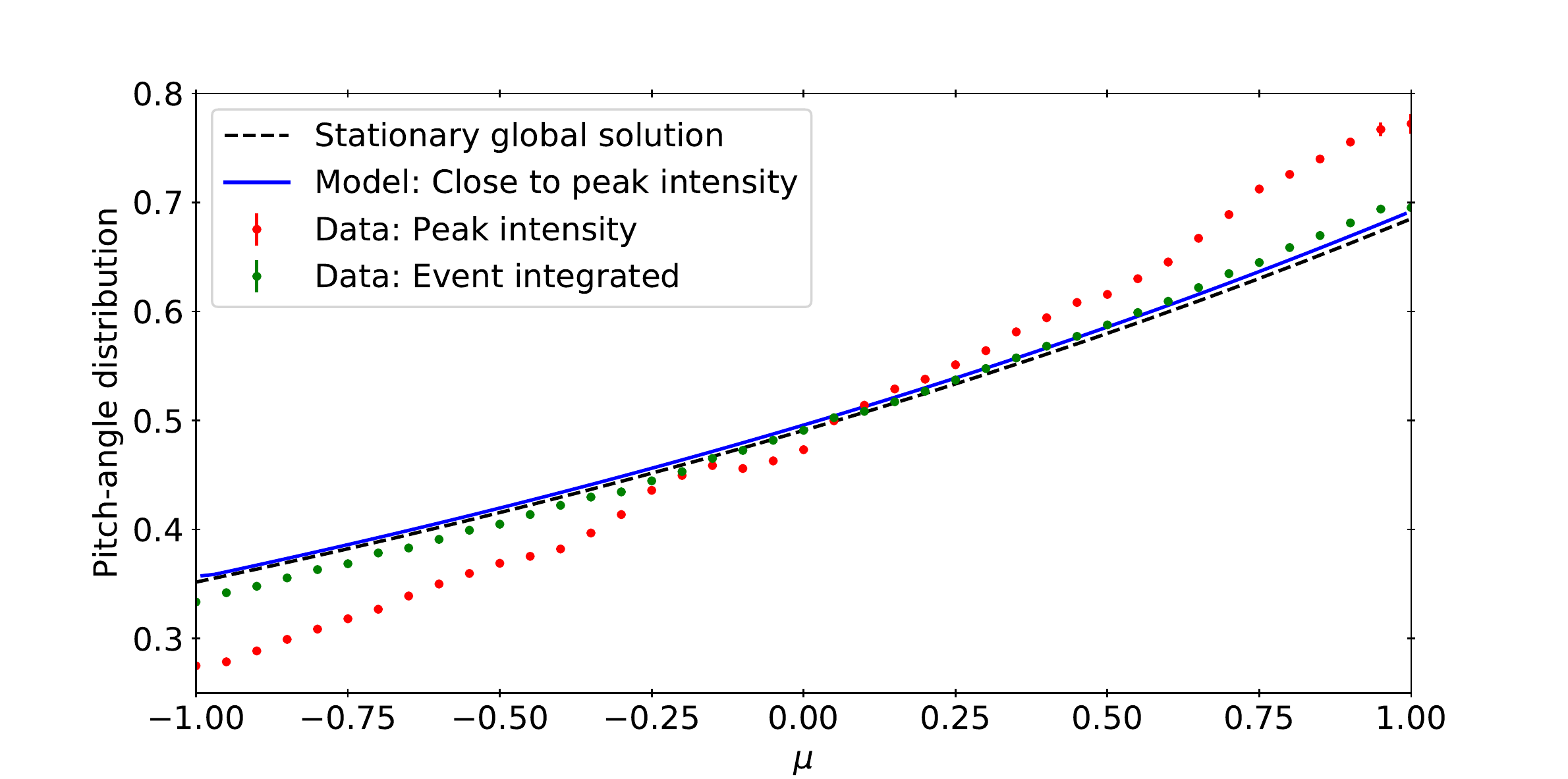}
\textbf{\caption[]{\label{fig:IsoAnisotropyPAD}{\textnormal{\textit{Top:} Anisotropy as a function of time as seen by the observer in Fig.~\ref{fig:IsoODI}. \textit{Bottom:} Pitch-angle distribution at peak intensity and time integrated pitch-angle distribution at the observation point in Fig.~\ref{fig:IsoODI} compared to the analytical stationary pitch-angle distribution (Eq.~\ref{eq:StationaryPAD}; black dashed line) and the model pitch-angle distribution at $t = 1$ $\mathrm{h}$.}}}}
\end{figure*}

The temporal evolution of the probability density (comparable to the ODI through a proper scaling constant) as a function of position is shown in the top panel of Fig.~\ref{fig:IsoODI}, where the model, the diffusion approximation, and telegraph equation are compared to the synthetic data. Focusing causes a coherent pulse to form, propagating with speed $\sim u$ (see Appendix~\ref{apndx:DiffTel}) and composed mainly of particles which have not yet undergone significant scattering. The pulse spreads out with time due to scattering, while the scattered particles, having smaller parallel speeds than the focused particles in the pulse, form a wake behind the pulse. Both the diffusion approximation and telegraph equation is in good agreement with the synthetic data at late times, while at early times the diffusion approximation is too diffusive and the telegraph equation predicts a very sharp propagation front. The Gaussian shape of the diffusion approximation is clearly inadequate to describe the non-symmetric density, except in the wake, and the causality violation is clearly visible ahead of the pulse. The model is in best agreement with the synthetic data, although it is a little too diffusive at very early times (flux limiters are used to reduce numerical diffusion which is, to some extend, always present in finite difference models).

The ODI as a function of time at the observer is shown in the bottom panel of Fig.~\ref{fig:IsoODI}. The intensity has a quick rise time up to a peak intensity, after which the flux decreases slowly, characteristic of impulsive SEP events. This behaviour can be understood by looking at the top panel: as the pulse propagates past the observer, the event onset is seen followed by the peak intensity and the intensity decrease in the wake of the pulse or the decay phase. Although both the diffusion approximation and telegraph equation are comparable at late times, initially these two solutions are too diffusive or restrictive, respectively, and it seems as if the true solution (i.e. the model) is an interpolation between the two approximations.

The anisotropy as a function of time at the observer is shown in the top panel of Fig.~\ref{fig:IsoAnisotropyPAD}. The distribution is initially highly anisotropic after which it becomes more isotropic. Physically this is because the first particles to arrive at the observation point are particles with small pitch-angles that was focused and have experienced little scattering, while the particles in the wake have experienced more scattering and are approaching diffusive behaviour. {Some text refer to these first arriving particles, associated with large anisotropies, as particles undergoing {\it scatter-free propagation}. The phrase {\it scatter free} is however a misnomer when considering particle propagating in magnetic turbulence: all charged particles will experience these turbulent fluctuations and will, to some extent, have their smooth gyro-motion disturbed.} The causality violation of the diffusion approximation can be seen as infinite anisotropies before the event onset, while predicting a lower anisotropy during the event's onset. The telegraph equation is generally better at predicting the anisotropy, but has a significantly delayed onset time.

The bottom panel of Fig.~\ref{fig:IsoAnisotropyPAD} shows the event integrated PAD at the observer compared to the analytical stationary solution. The event integrated PAD corresponds very well to the analytical solutions of the stationary PAD. It is more interesting to note the temporal behaviour of the PAD at the observer (a three point average in time and pitch-cosine was taken in the synthetic data to smooth out fluctuations). The distribution is beam-like at the event onset (not shown) with the PAD coinciding with the stationary solution just after the peak intensity (the model result), after which the distribution slowly approach isotropy (not shown). This seems to imply that the pulse has a quasi-stationary distribution set up by a balance between focusing and scattering, as suggested by Eq.~\ref{eq:PADstationaryODE} \citep{beeckwibberenz1986}.

\begin{figure*}[!t]
\centering
\includegraphics[trim=10mm 0mm 20mm 12mm, clip, scale=0.5]{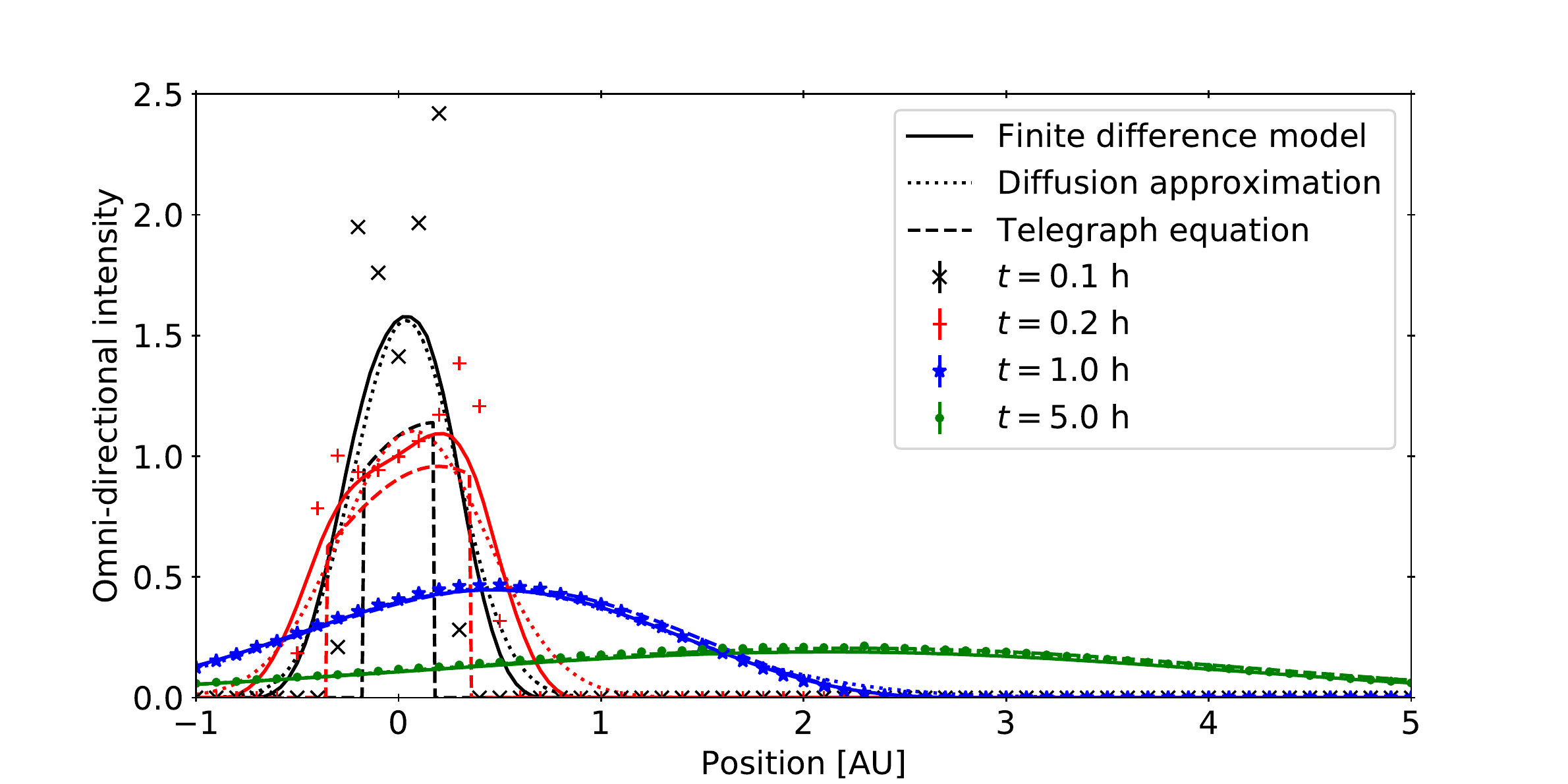}
\includegraphics[trim=10mm 0mm 20mm 12mm, clip, scale=0.5]{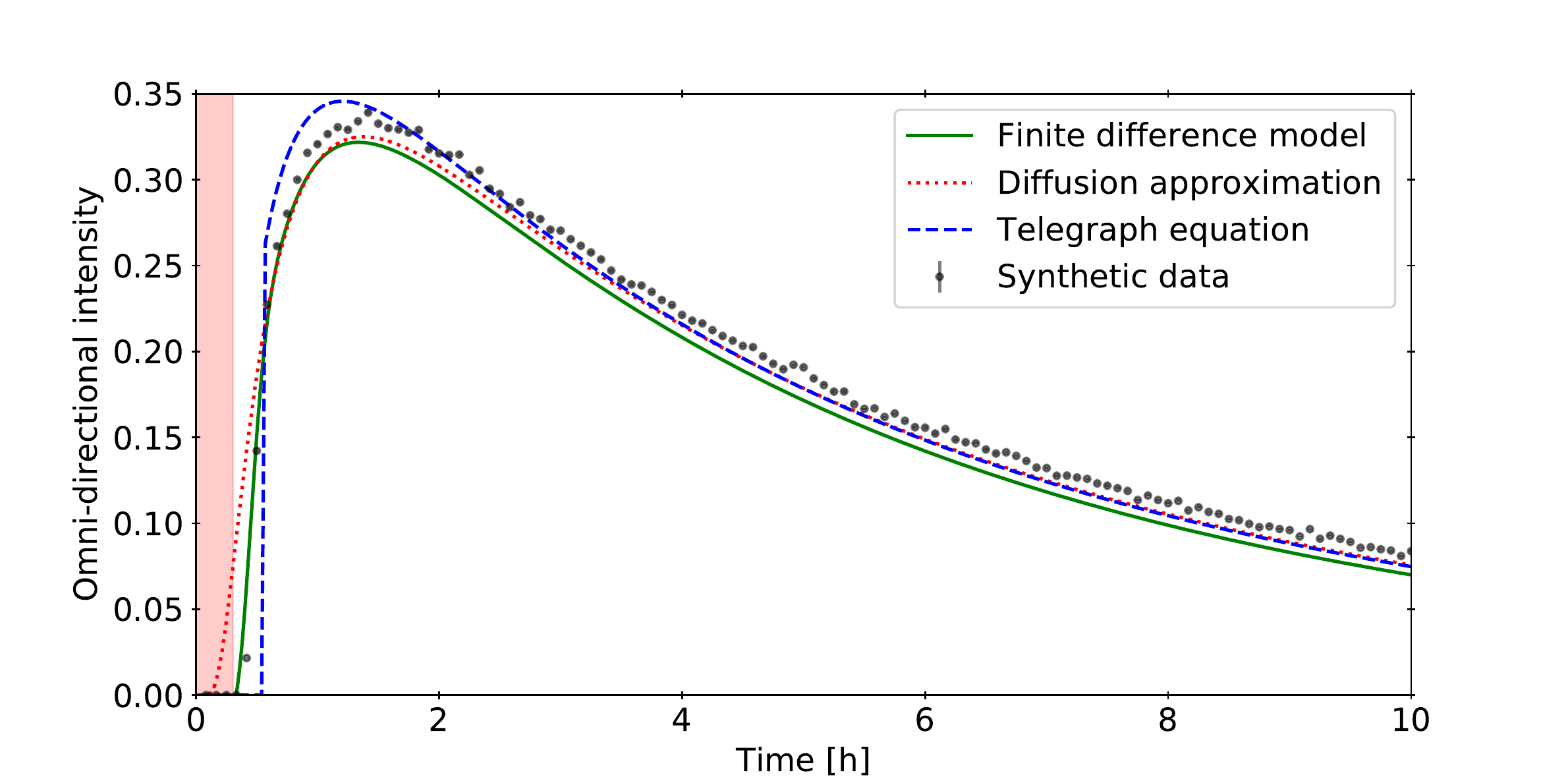}
\textbf{\caption[]{\label{fig:AnisoODI}{\textnormal{Similar to Fig.~\ref{fig:IsoODI}, but for anisotropic scattering (Eq.~\ref{eq:QLT}).}}}}
\end{figure*}


\subsubsection{Anisotropic Scattering with a Constant Focusing Length}
\label{subsubsec:Anisotropic}

\begin{figure*}[!t]
\centering
\includegraphics[trim=10mm 0mm 18mm 12mm, clip, scale=0.5]{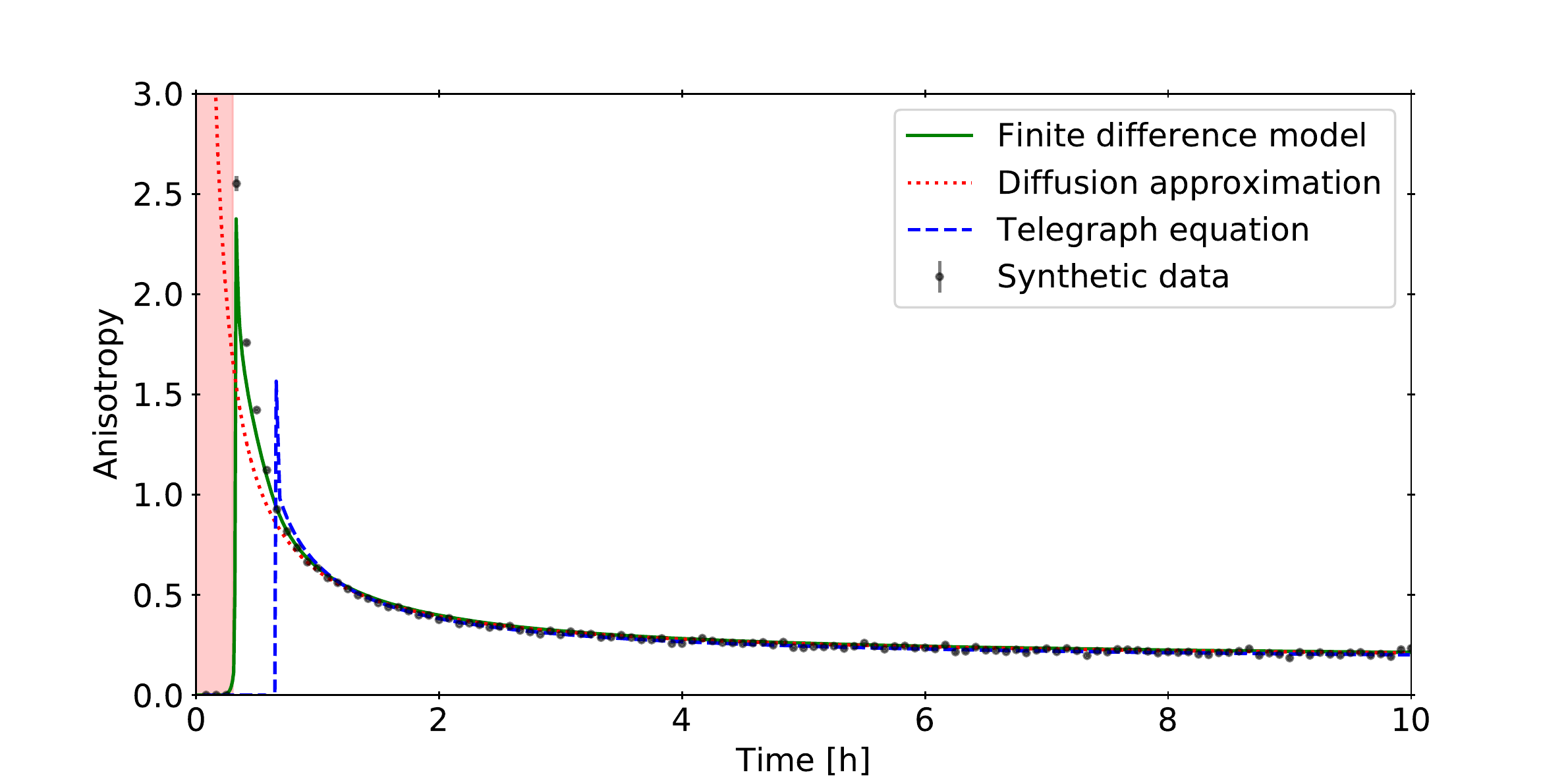}
\includegraphics[trim=10mm 0mm 18mm 12mm, clip, scale=0.5]{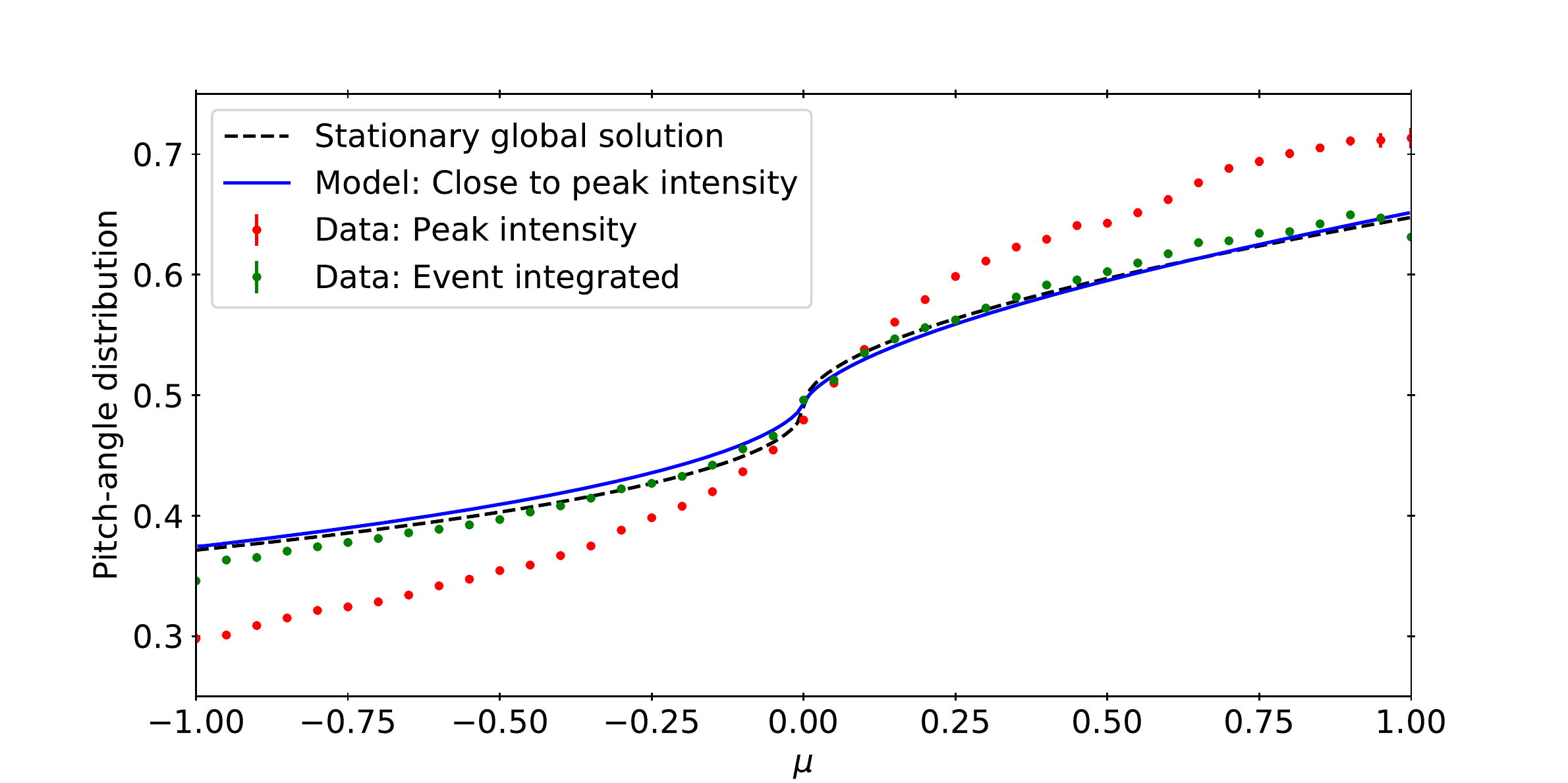}
\textbf{\caption[]{\label{fig:AnisoAnisotropyPAD}{\textnormal{Similar to Fig.~\ref{fig:IsoAnisotropyPAD}, but for the anisotropic scattering (Eq.~\ref{eq:QLT}) in Fig.~\ref{fig:AnisoODI}}}}}
\end{figure*}

The temporal evolution of the probability density as a function of position is shown in the top panel of Fig.~\ref{fig:AnisoODI}. The effect of anisotropic pitch-angle scattering is interesting and visible in the initial phase of the event. The delta injection seem to split into two coherent pulses propagating away from one another. This behaviour is due to ineffective scattering across $\mu = 0$ and the use of an isotropic injection. The effect of focusing can be seen as there are more particles in the pulse propagating towards weaker magnetic fields. These two pulses, however, are combined into one pulse by scattering some time after the injection. The diffusion approximation and telegraph equation is again in good agreement with the synthetic data at late times, but at early times even the model is too diffusive to replicate these two pulses (it is well known that finite difference models do not handle steep gradients well).

The ODI as a function of time at the observer is shown in the bottom panel of Fig.~\ref{fig:AnisoODI}, while the anisotropy is shown in the top panel of Fig.~\ref{fig:AnisoAnisotropyPAD}. These results are comparable to the isotropic scattering scenario. It should be kept in mind that the coefficients presented in Appendix~\ref{apndx:DiffTel} for the telegraph equation, are only appropriate if the weak focusing limit is considered. The model predicted intensity is a bit lower, probably due to numerical diffusion of the initial pulse. The event integrated PAD and PAD at peak intensity at the observer are shown in the bottom panel of Fig.~\ref{fig:AnisoAnisotropyPAD}. The effect of anisotropic scattering can be seen in the PAD as a decrease in crossing $\mu = 0$ from positive to negative values. Also note here that the PAD at peak intensity (the `synthetic data') does not coincide exactly with the stationary solution, but only shortly before or after the peak (the `model'). This should be kept in mind if the PAD is used to extract parameters from data and it might be tempting to use the PAD at peak intensity because it is easier than calculating the event integrated PAD.


\subsubsection{Application towards a Solar Energetic Particle Event}
\label{subsubsec:SEPevent}

\begin{figure*}[!t]
\centering
\includegraphics[trim=5mm 10mm 15mm 20mm, clip, scale=0.5]{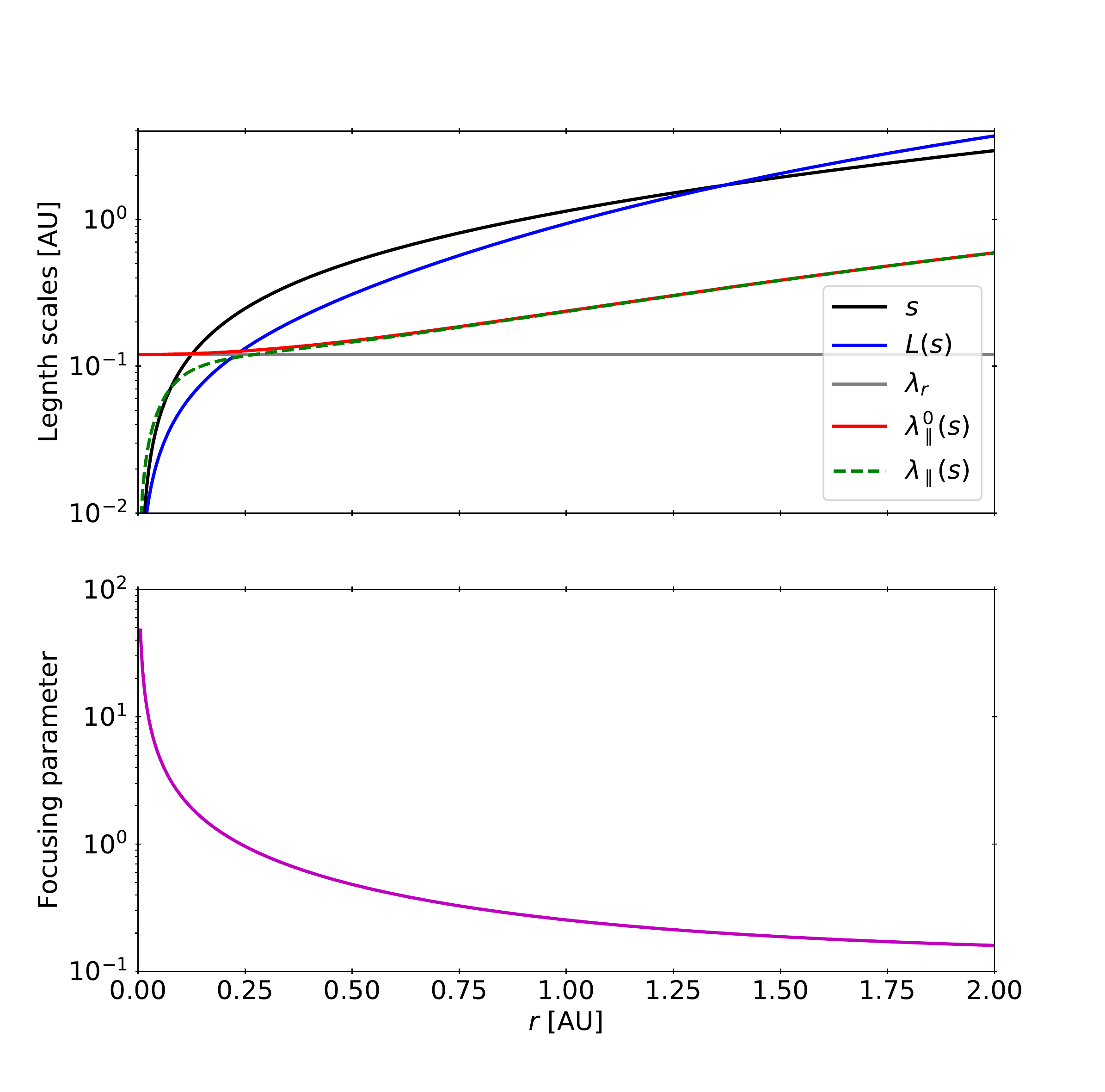}
\textbf{\caption[]{\label{fig:LengthScales}{\textnormal{\textit{Top:} Magnetic field arch length (black; Eq.~\ref{eq:ArcLength}), focusing length (blue; Eq.~\ref{eq:L(s)}), and constant radial (grey), `isotropic' parallel (red, $\lambda_{\parallel}^0 = \lambda_r / \cos^2 \psi$), and `focusing included' parallel (dashed green; Eq.~\ref{eq:ParalMFPFocus}) mean free paths as a function of heliocentric radius. \textit{Bottom:} Focusing parameter (ratio of the parallel mean free path to the focusing length) as a function of radius.}}}}
\end{figure*}

The $65-105$ $\mathrm{keV}$ solar energetic electron event of 7 February 2010 observed by STEREO-B \citep[see Figure 9a in][]{drogeetal2014} will be considered in this section. Electrons are injected at $s_0 = 0.05$ $\mathrm{AU}$ with an energy of $80$ $\mathrm{keV}$ and a reflective inner boundary assumed at $s = 0$ $\mathrm{AU}$, supposedly caused by mirroring in the HMF. A constant radial MFP of $0.12$ $\mathrm{AU}$ will be used with $q = 5/3$ and $H = 0.05$ in Eq.~\ref{eq:FixedQLT}, similar to \citet{drogeetal2014}. The parallel and radial MFPs are related by $\lambda_{\parallel}^0 = \lambda_r / \cos^2 \psi$. The arc length, focusing length, radial and parallel MFP, and focusing parameter are shown in Fig.~\ref{fig:LengthScales}. These parameters have values of $s = 1.139$ $\mathrm{AU}$, $L = 0.936$ $\mathrm{AU}$, $\lambda_{\parallel}^0 = 0.238$ $\mathrm{AU}$, and $\xi = 0.253$ at Earth. The parallel MFP formulation, which directly incorporates focusing, (Eq.~\ref{eq:ParalMFPFocus}) is also shown, assuming that $D_0$ is calculated from Eq.~\ref{eq:ParallelMFP}. From the focusing parameter it can be seen that focusing will have the largest effect within the first $\sim 1$ $\mathrm{AU}$ from the Sun.

\begin{figure*}[!t]
\centering
\includegraphics[scale=0.5]{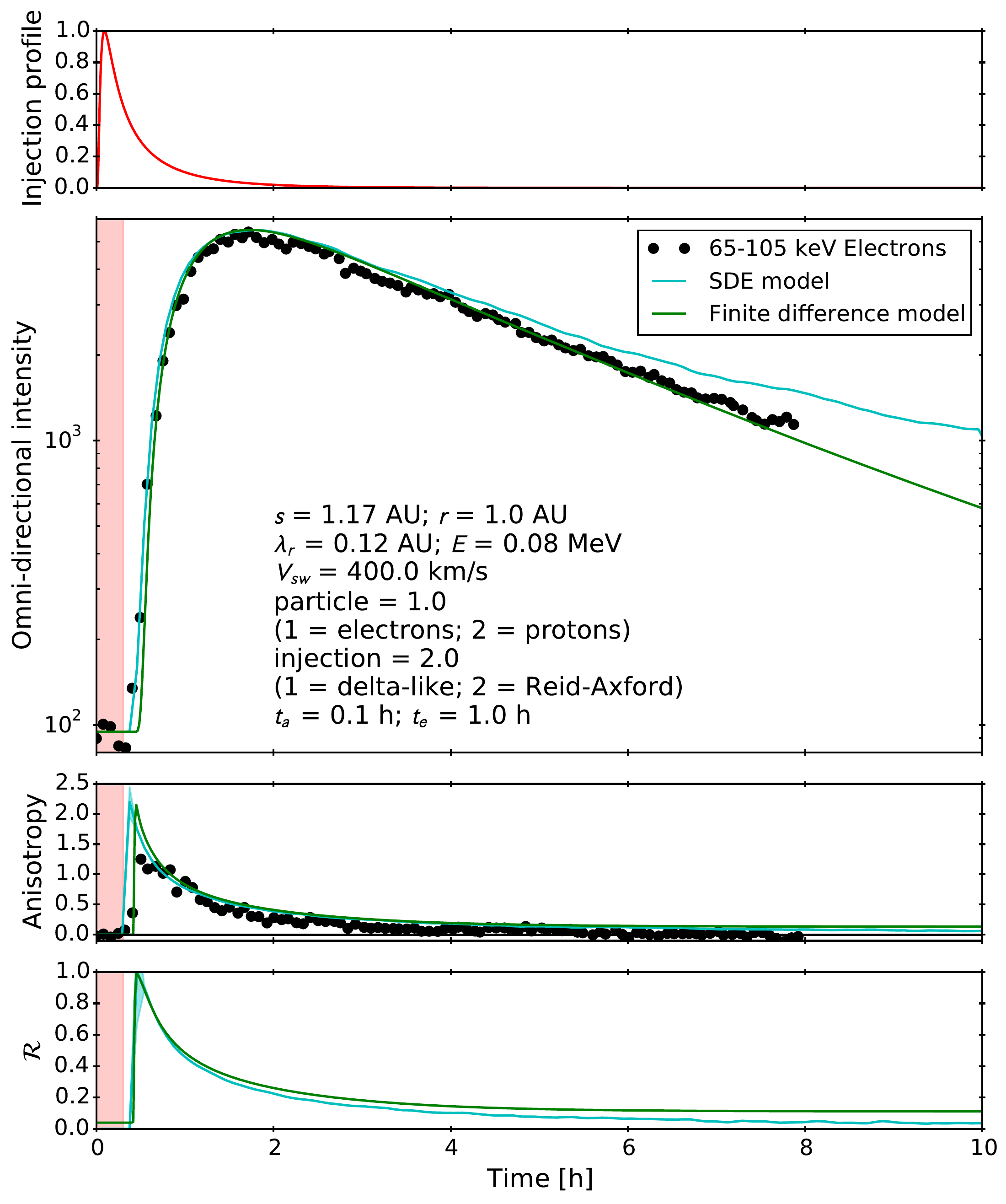}
\textbf{\caption[]{\label{fig:BestFitSEP}{\textnormal{Best fit results of the finite difference model (solid green), together with the stochastic differential equation model (solid cyan), to the electron data from the 7 February 2010 event (black dots). \textit{Top:} Injection function (Eq.~\ref{eq:ReidAxford}) normalised to its peak value. \textit{Middle:} Omni-directional intensity and anisotropy. \textit{Bottom:} The ratio of forward to backward propagating particles as calculated by Eq. \ref{eq:FwBwratio}.}}}}
\end{figure*}

Unlike \citet{drogeetal2014}, who assumed a piece-wise linear injection function and inferred its form from fitting the data, a Reid-Axford \citep{reid1964} injection function,
\begin{equation}
\label{eq:ReidAxford}
f(s=s_0, t) = \frac{C}{t} e^{-\tau_a / t - t/\tau_e} 
\end{equation}
with $C$ a normalisation constant and $\tau_a=0.1$ hr and $\tau_e = 1$ hr the acceleration and escape time, respectively, will be assumed here. These best-fit model results are compared to observations in Fig.~\ref{fig:BestFitSEP}. The top panel shows the assumed injection function as a function of time, the two middle panels the calculated ODI and the anisotropy. Appendix \ref{apndx:FDSolver} illustrates how sensitive these results are to changing transport parameters. Notice that there is a significant discrepancy between the finite difference and stochastic differential equation (synthetic data) model during the decay phase of the event, even though the two models were run with the same parameters. This is due to an implicitly assumed absorbing outer boundary condition at $s = 3$ $\mathrm{AU}$ in the finite difference model.

\paragraph{Calculating the Number of Particles Injected}
\label{prgrph:CountParticles}

The distribution function can also be used to calculate the average propagation time or energy losses suffered by particles. Some analytical diffusion approximations can be found in \citet{parker1965}, for example, while \citet{straussetal2011} show how easily a stochastic differential equation model can be used to calculate these quantities for CRs and Jovian electrons, for example. \citet{litvinenkoetal2015} gives an approximation for the average propagation time of SEPs using the telegraph equation. Average energy losses have been investigated by \citet{kocharovetal1998} and \citet{zhangetal2009}, for example, the latter of which also briefly investigated propagation times. As a less obvious application, the problem of calculating the total number of particles released in an SEP event will be considered here.

\citet{denolfoetal2019} compared the number of particles released from a long duration solar flare to the number of particles needed to produce the gamma-rays observed by Fermi-LAT in order to access the possible acceleration mechanisms. These authors used PAMELA and STEREO A and B data to calculate the total number of $> 200$ $\mathrm{MeV}$ protons observed in the heliosphere $N_{\rm obs}$ during such flares. This number, however, is larger than the number of protons released in the flare $N_{\rm inj}$ due to particle scattering causing particles to move past the observation point multiple times. A correction is therefore necessary as the PAMELA detector cannot discriminate between Sunwards and anti-Sunwards propagating particles, i.e. the pitch-angle dependence cannot be observed. To correct for this, $N_{\rm obs}$ was divided by the average number of times a particle would cross the observation point $\bar{N}_{\rm cross}$ ($N_{\rm inj} = N_{\rm obs} / \bar{N}_{\rm cross}$). This number $\bar{N}_{\rm cross}$, of course, depends on the underlying turbulence and, as such, cannot be determined experimentally and must be estimated from simulations. In \citet{denolfoetal2019}, this number was calculated from two test particle simulations using an unspecified plasma turbulence field for the scattering. These simulations yielded numbers varying between $3.6$ and $31$, while a constant $\bar{N}_{\rm cross} = 8$ was used for all fourteen events considered.

Information regarding the propagation direction of the SEPs are, however, contained in the distribution function and if the pitch-angle dependence of the distribution function can be determined (either experimentally or through simulations), the use of any ad-hoc corrections, such as implementing a rather arbitrary $\bar{N}_{\rm cross}$ factor, is unnecessary. Formally, the ratio of forward (outwards) to backward (inwards) propagating particles can be calculated as
\begin{equation}
\label{eq:FwBwratio}
\mathcal{R} = \frac{f_{\rm out} - f_{\rm in}}{f_{\rm out} + f_{\rm in}} = \frac{\int_0^{+1} f {\rm d}\mu - \int_{-1}^0 f {\rm d}\mu}{\int_0^{+1} f {\rm d}\mu + \int_{-1}^0 f {\rm d}\mu} = \frac{\int_0^{+1} f {\rm d}\mu - \int_{-1}^0 f {\rm d}\mu}{\int_{-1}^{+1} f {\rm d}\mu}.
\end{equation}
This calculation was performed with the numerical model discussed in Section~\ref{apndx:FDSolver}, using the same parameters as in Fig.~\ref{fig:BestFitSEP}, and shown in the bottom panel of Fig.~\ref{fig:BestFitSEP}. $\mathcal{R}$ is, as expected, not constant for the entire duration of the event and roughly follows the temporal evolution of the anisotropy, $A$. There is, however, not a simple linear relationship between these two quantities, with the ratio being
\begin{equation*}
\frac{A}{\mathcal{R}} = \frac{3 \int_{-1}^{+1} \mu f {\rm d}\mu}{\int_0^{+1} f {\rm d}\mu - \int_{-1}^0 f {\rm d}\mu},
\end{equation*}
which should be integrated numerically. In order to estimate the total number of SEPs passing e.g. a spacecraft position where the distribution function is not known, a simple numerical model can be used, tuning the transport parameters to fit the ODI, and using the computed distribution function to calculate $\mathcal{R}$ and ultimately use $\mathcal{R}$ to calculate the particle flux from the omni-directional particle intensity.


\newpage

\section{A Brief Review of Contemporary Models and Simulation Results}
\label{sec:Review}

In this section different models and/or applications of the FTE that were applied to SEP transport will be reviewed. Models with increasing levels of complexity will be discussed, starting from the spatial 1D version, the so-called \citet{roelof1969} equation, and extending to 3D full-orbit simulations. 


\subsection{1D Simulations}
\label{subsec:1D}


\subsubsection{Velocity Dispersion Analysis and the Effect of Scattering}
\label{subsubsec:VDA}

\begin{figure*}[!t]
\centering
\includegraphics[width=0.39\textwidth]{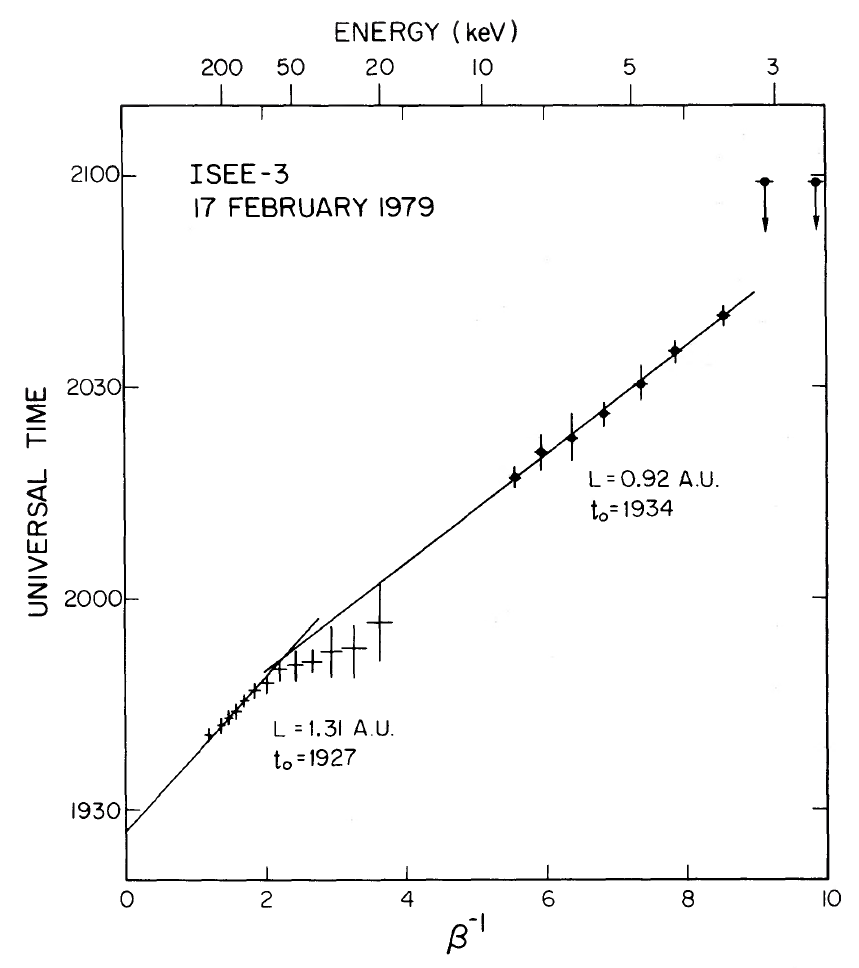}
\includegraphics[width=0.59\textwidth]{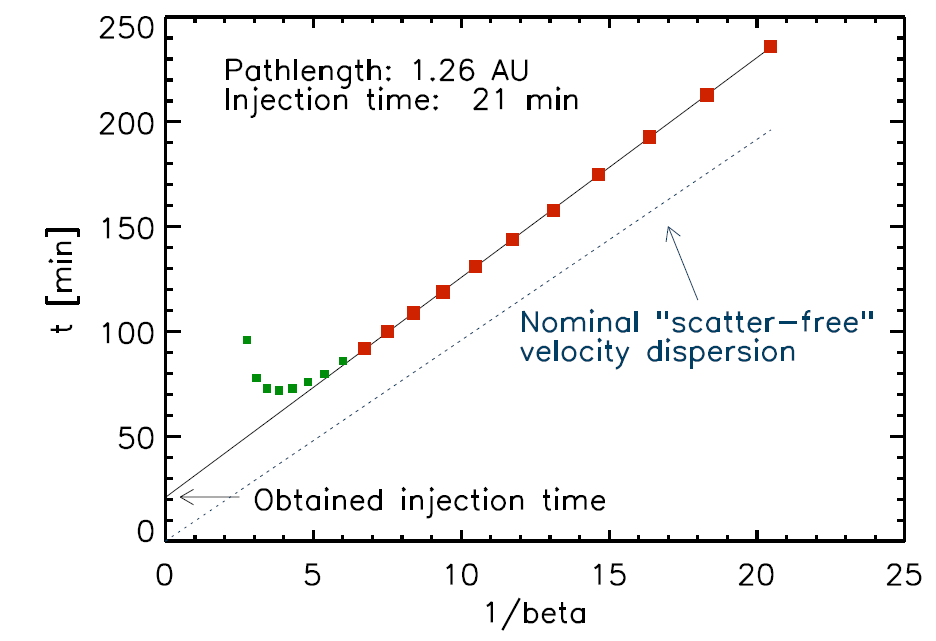}
\textbf{\caption[]{\label{fig:VDA}{\textnormal{\textit{Left panel:} An early velocity dispersion analysis from \citet{linetal1981}. \textit{Right panel:} Simulation results of a velocity dispersion analysis by \citet{laitinenetal2015}.}}}}
\end{figure*}

The most simplistic view of SEP transport is that of ballistic motion along a single smooth and unperturbed \citet{parker1958} magnetic field line. Under this unrealistic assumption, the so-called onset time, $t_o$, i.e. the time that a detector will start measuring an increase in SEP intensity in a given energy channel, is given by
\begin{equation*}
t_o (v) = t_i + \frac{s}{v},
\end{equation*}
where $t_i$ is the injection time (i.e. when the particles are released from their acceleration site) and $s$ is their (magnetic) propagation length. This is generally referred to as a velocity dispersion analysis (VDA). The left panel of Fig.~\ref{fig:VDA} shows such an analysis from \citet{linetal1981}, where a linear fit of $t_o$ against the inverse of $\beta = v/c$, gives an estimation of $s$. However, these analysis lead to seemingly contradictory results, e.g. the results presented in Fig.~\ref{fig:VDA} indicate, for some energies, $s<1$ $\mathrm{AU}$. While instrumental (including the natural background of the detector) effects can play a role in leading to such discrepancies, most notable, this analysis ignores particle scattering by magnetic turbulence. The effect of scattering on simulated particle dispersion relations are presented by e.g. \citet{laitinenetal2015}, with a selection of their results presented in the right panel of Fig.~\ref{fig:VDA}. Here, the symbols are the onset times as calculated from the simulation results, solving the 1D FTE, while the straight lines give the dispersion relationship either obtained from the results (solid line), or what would be expected from scatter-free ballistic motion. It is clear that VDA, for this simulation, over-estimates both the injection time and the particle propagation length.

As already discussed in Section~\ref{subsubsec:Isotropic}, the assumption of scatter free particle events is clearly an over-simplification, and any results obtained from the VDA approach should be used with care \citep[see the discussion by][]{laitinenetal2015}. Similar conclusions are reached, again based on 1D simulation results, by \citet{lintunenvainio2004} and \citet{saizetal2005}, where the latter also included adiabatic energy losses. \citet{wangqin2015a} also show how VDA can be affected by cross-field transport, discussed in more detail in Section~\ref{Subsec:2D3D}. Recently, there has also been analytical approximations to the initial, highly anisotropic, phase of an SEP events, with results presented by \citet{bianemslie2019} and \citet{lilee2019}, showing that the particle onset is delayed due to (weak) scattering. An improvement over the traditional VDA, using a so-called fractional VDA, which mitigates the uncertainties in detector onset determination, was recently established by \citet{zhao2019}.

\begin{figure*}[!t]
\centering
\includegraphics[width=0.43\textwidth]{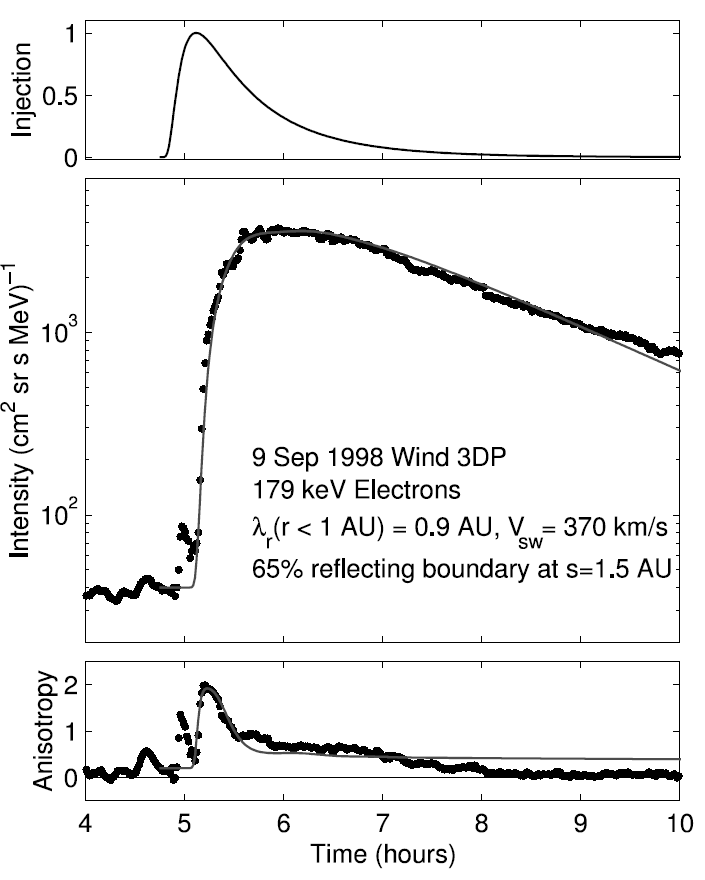}
\includegraphics[width=0.55\textwidth]{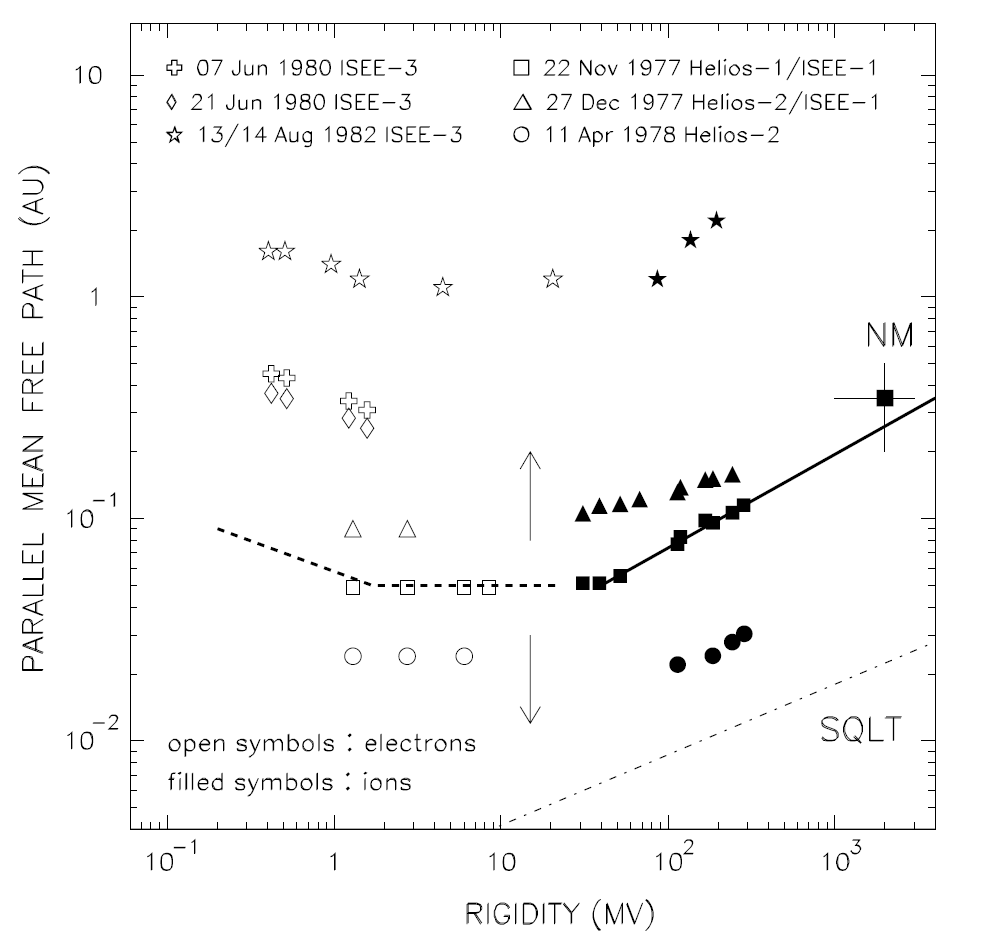}
\textbf{\caption[]{\label{fig:Duus}{\textnormal{\textit{Left panel:} A comparison between simulated and observed solar energetic electron fluxes, taken from \citet{drogeetal2006}. \textit{Right panel:} A selection of transport parameters obtained by comparing simulated and observed solar energetic particle fluxes and anisotropies, from \citet{droge2000b}.}}}}
\end{figure*}


\subsubsection{Phenomenological Approach to Determining Transport Parameters}
\label{subsubsec:MakeParametersFit}

Although simulations of the VDA can lead to insight regarding particle scattering, a more thorough approach is to reproduce the observed temporal profiles of both the observed SEP intensity and anisotropy with simulation results. The left panel of Fig.~\ref{fig:Duus} shows an example of such a data comparison from \citet{drogeetal2006}: The top panel shows the assumed injection function (discussed more in the next section), the middle panel the ODI, and the bottom panel the anisotropy. A careful comparison between simulations and observations, adjusting the correct combination of parameters (predominantly $D_{\mu \mu}$ and the injection function in the 1D approach), can lead to accurate estimations of e.g. the parallel MFP. These estimations, for a large number of events and different energy channels, are presented by \citet{droge2000b} and shown in the right panel of Fig.~\ref{fig:Duus}. Here, open and filled symbols show results for electrons and protons, respectively, which are compared to a theoretical prediction for protons. These results show that the SEP transport parameters can have very large (almost two orders of magnitude) inter-event variability, most likely related to the changing plasma conditions and levels of magnetic turbulence between the Sun and the observer. In addition, these results show different behaviour for low energy protons and electrons, with the MFP for low energy electrons increasing for decreasing energies \citep[see also][]{drogekartavyhk2009}. Such dependencies are expected from scattering theory \citep[][]{teufelschlickeiser2002} if a dissipation range is included in the assumed slab turbulence spectrum: low energy electrons resonate in this weak-turbulence regime, experiencing very little pitch-angle scattering, leading to large MFPs. However, recent simulations of very low energy electron transport by \citet{kartavykhetal2013} have shown rather large discrepancies with predicted theoretical results, and this is yet to be explained.

Most of the 1D simulation results neglect SW effects, including SW convection and adiabatic energy losses. Results from \citet{ruffolo1995} indicate that both of these effects could potentially be negligible for relativistic electrons while becoming increasingly important for low energy particles. When both SW convection and energy losses are considered,  the event onset and peak time are slightly earlier, due to convection, while the peak intensity is lower and the decay is quicker, due to energy losses \citep{ruffolo1995, kocharovetal1998}. Moreover, \citet{qinetal2006} found that a model without energy losses would generally overestimate the derived parallel MFP.

\begin{figure*}[!t]
\centering
\includegraphics[width=0.50\textwidth]{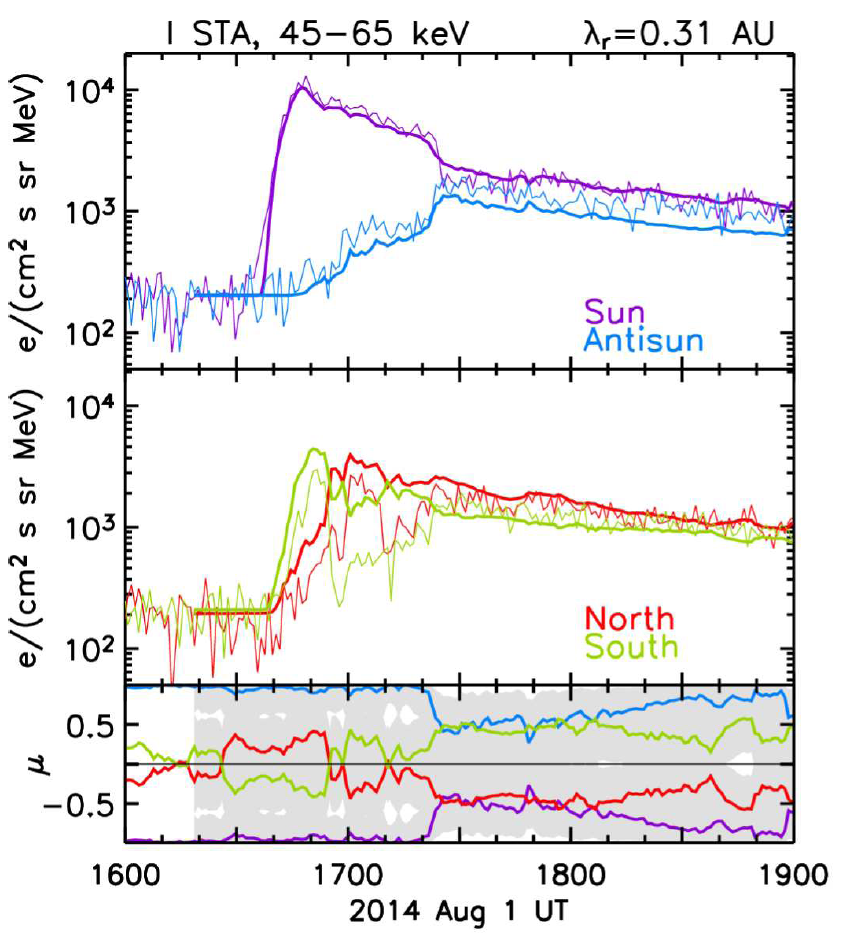}
\includegraphics[width=0.48\textwidth]{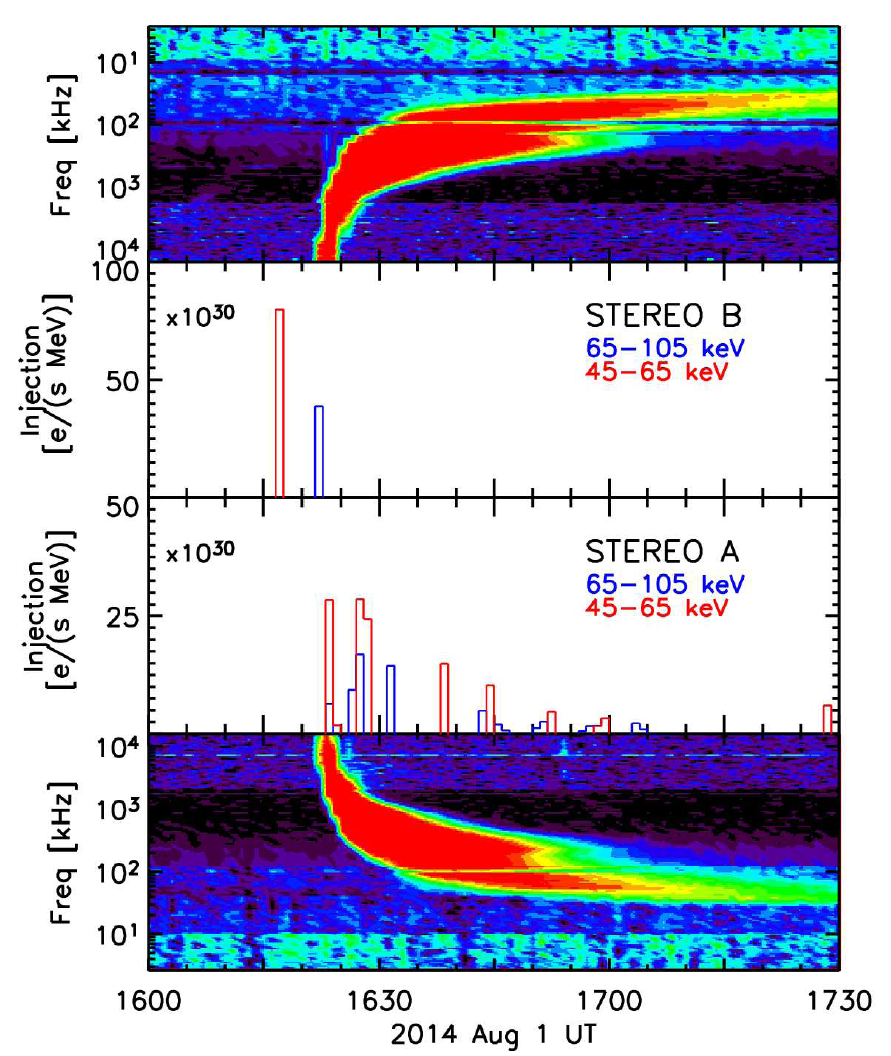}
\textbf{\caption[]{\label{fig:inversion}{\textnormal{\textit{Left panel:} A comparison between pitch-angle dependent simulation results (thick lines) and STEREO B spacecraft observations (thin lines) for four different detector viewing directions (Sun, Anti-sun, North, and South). \textit{Right panel:} The derived injection function, this time for both STEREO A and B, for two different energy channels, along with radio-observations. Both figures are taken from \citet{paschecoetal2017}.}}}}
\end{figure*}


\subsubsection{Deriving the Injection Function}
\label{subsubsec:Injection}

In the previously discussed modelling work, the focus was on simulating the effect of, specifically, pitch-angle scattering on the particle intensity observed at some point far away from the SEP source. The acceleration process near the Sun (through flares and/or coronal mass ejections) are therefore mostly neglected and reduced to a so-called injection function: The temporal profile of SEPs released into the interplanetary medium, which could also be energy, spatially, and pitch-angle dependent. \citet{aguedaetal2008} and \citet{aguedaetal2009} implement a 1D SEP model, where a series of short, impulsive bursts of particles are injected, allowing for a deconvolution of the transport and injection processes. The calculation of these Green's functions allow the modelling inversion to be done for a large number of events, forming a database of particle injection histories \citep[e.g.][]{aguedaetal2012, vainioetal2013}. By comparing the simulated injection profiles to remote-sensing observations (e.g. soft- and hard-X-rays and radio-observations) can lead to insight regarding the source of these SEP electrons \citep[][]{aguedaetal2014, aguedalario2016, pachecoetal2019}. An example of this inversion approach, taken from \citet{paschecoetal2017}, is shown in Fig.~\ref{fig:inversion}: The left panel compares observations (thin lines) and modelling results (thick lines) for different pitch-angles (i.e. detector viewing directions). The best fit radial MFP, for 45--65 keV electrons, is given at the top of the left panel. The right panel shows the derived injection function (release history) of these electrons near the Sun, as compared with radio-observations. The inversion results presented here are consistent with the acceleration of electrons in a solar flare, with low-energy escaping electrons producing the observed radio beams.

Most modelling results, however, do not apply such a detailed deconvolution method, and the standard approach is to adopt a Reid-Axford profile \citep[Eq.~\ref{eq:ReidAxford};][]{reid1964} for the injection function, characterized by an acceleration and a decay timescale. Even these parameters can, of course, be constrained by observations \citep[][]{ruffoloetal1998}. Moreover, all of the injection functions discussed above assume an isotropic injection of particles, which might be an over-simplification. Results from \citet{kocharovetal1998} suggest that the pitch-angle dependence of the injection function may affect the simulated intensities, especially close to the SEP source.


\subsubsection{Other 1D Applications}
\label{subsubsec:Other1D}

In addition to the results discussed above, \citet{kartavykhetal2016} presented 1D simulation results where pitch-angle dependent shock acceleration at a moving shock is included, and hence also an energy coordinate. They show that particle acceleration occurs predominantly near the Sun ($r< 0.05$ $\mathrm{AU}$), leading to a so-called ``prompt" phase of the SEP event being observed at Earth, where after the propagation of the shock modulated the observed ``gradual" phase of the event, potentially explaining the observation of so-called ``mixed" particle events \citep[e.g.][]{canetal2003} showing a combination of both gradual and impulsive SEP characteristics. As a last application, \citet{straussetal2017b} compared simulated and observed temporal profiles of high-energy SEP events as measured by ground-based neutron monitors (so-called ground level enhancements) to characterise the rise and decay times of these events. These authors showed that the effect of pitch-angle scattering can be significant, even for high-energy ($>100$ $\mathrm{MeV}$ protons) ground level enhancement events.


\begin{figure*}[!t]
\centering
\includegraphics[width=0.48\textwidth]{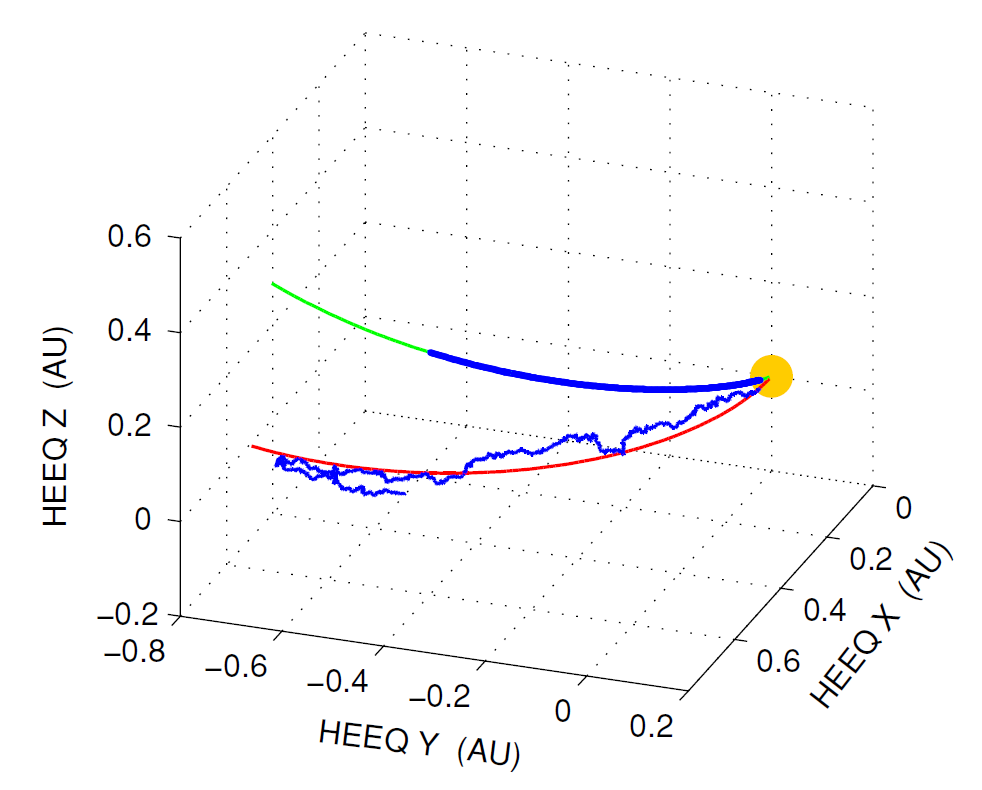}
\includegraphics[width=0.50\textwidth]{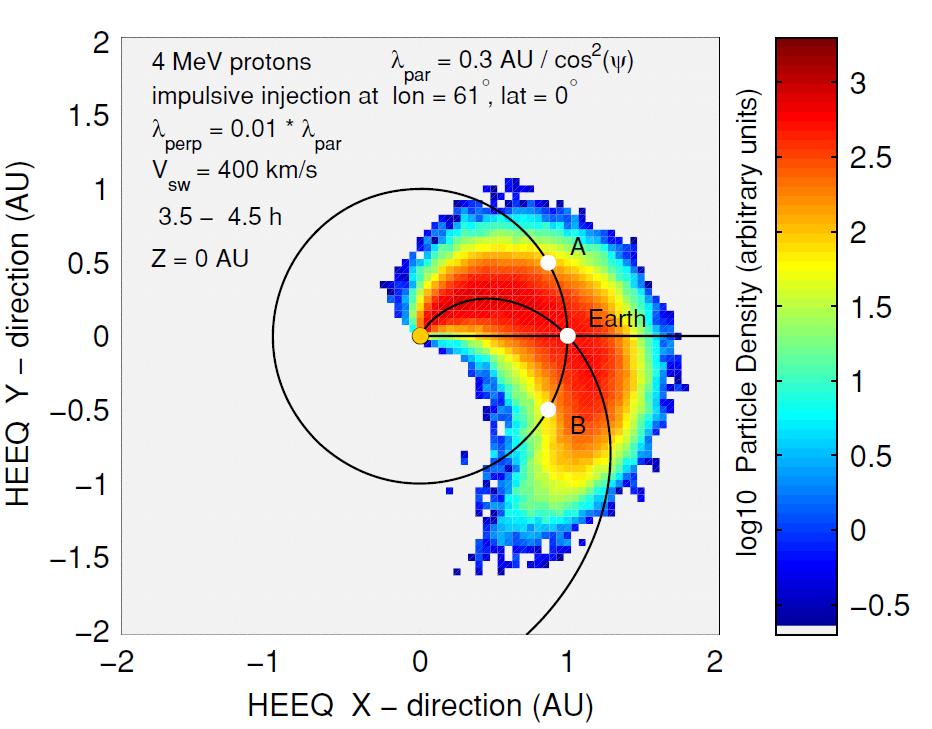}
\textbf{\caption[]{\label{fig:perpDiff}{\textnormal{\textit{Left panel:} An illustration of particle propagation without the inclusion of perpendicular diffusion (blue particle tied to the green magnetic field line) and with perpendicular diffusion (blue particle able to decouple from the red magnetic field line). \textit{Right panel:} Simulated SEP intensities in the equatorial plane, illustrating how particles diffuse perpendicular to the mean field. Both figures are taken from \citet{drogeetal2010}.}}}}
\end{figure*}

\subsection{2D and 3D Simulations}
\label{Subsec:2D3D}

The observations of so-called widespread SEP events \citep[see e.g.][amongst others]{dresingetal2012} have shown that impulsively accelerated SEP electrons can be observed up to $\sim 180^{\circ}$ away from their source. Possible mechanisms invoked to explain this seemingly efficient cross-field transport are perpendicular diffusion and drift effects, in addition to a wide injection region. These effects cannot be included in a spatially 1D model, resulting in the development of SEP models with higher dimensionality. These models are briefly discussed in this section.

So far, in this work, the focus was only on discussing field-aligned transport and the perpendicular diffusion process was not touched on, for the most part. However, the discussion in Sec. \ref{subsubsec:FLMeandering} presents an illuminating picture of this process. Perpendicular transport can be retained in the FTE by first transforming to the GC position before changing to spherical coordinates in momentum space and preforming a gyro-averaging. Details of such derivations can be found in \citet{zhang2006}, \citet{lerouxwebb2007}, and \citet{wijsen2020}. The FTE then includes particle drifts and diffusion perpendicular to the magnetic field, and is given by
\begin{align}
\label{eq:FTE}
& \frac{\partial f}{\partial t} + \frac{\partial}{\partial x_i} \left[ \frac{{\rm d}x_i}{{\rm d}t} f \right] + \frac{\partial}{\partial p} \left[ \frac{{\rm d}p}{{\rm d}t} f \right] + \frac{\partial}{\partial \mu} \left[ \frac{{\rm d}\mu}{{\rm d}t} f \right] \nonumber \\
= \; & \frac{\partial}{\partial \mu} \left[ D_{\mu \mu} \frac{\partial f}{\partial \mu} + D_{\mu j}^{\perp} \frac{\partial f}{\partial x_j} \right] + \frac{\partial}{\partial x_i} \left[ D_{i \mu}^{\perp} \frac{\partial f}{\partial \mu} + D_{ij}^{\perp} \frac{\partial f}{\partial x_j} \right] ,
\end{align}
where $D_{ij}^{\perp}$ are the perpendicular diffusion coefficients, with the mixed terms $D_{i \mu}^{\perp} = D_{\mu j}^{\perp}$ usually neglected. Note that the distribution function $f(\vec{x}; p; \mu; t)$ is written in a mixed coordinate system where the GC position $\vec{x}$ and time are measured in the observer's frame, and the momentum $p$ and pitch-cosine $\mu$ are measured in the SW frame.

In Eq.~\ref{eq:FTE},
\begin{align*}
\frac{{\rm d}\vec{x}}{{\rm d}t} & = \mu v \; \hat{\vec{b}} + \vec{v}_{\rm sw} + \vec{v}_d \\
\frac{{\rm d}p}{{\rm d}t} & = p \left\{ \frac{1 - 3 \mu^2 }{2} \; \hat{\vec{b}} \hat{\vec{b}} : \vec{\nabla} \vec{v}_{\rm sw} - \frac{1 - \mu^2}{2} \vec{\nabla} \cdot \vec{v}_{\rm sw} - \frac{\mu}{v} \; \hat{\vec{b}} \cdot \left[ \frac{\partial \vec{v}_{\rm sw}}{\partial t} + (\vec{v}_{\rm sw} \cdot \vec{\nabla}) \vec{v}_{\rm sw} \right] \right\} \\
\frac{{\rm d}\mu}{{\rm d}t} & = \frac{1 - \mu^2}{2} \left\{ v \; \vec{\nabla} \cdot \hat{\vec{b}} + \mu \; \vec{\nabla} \cdot \vec{v}_{\rm sw} - 3 \mu \; \hat{\vec{b}} \hat{\vec{b}} : \vec{\nabla} \vec{v}_{\rm sw} - \frac{2}{v} \; \hat{\vec{b}} \cdot \left[ \frac{\partial \vec{v}_{\rm sw}}{\partial t} + (\vec{v}_{\rm sw} \cdot \vec{\nabla}) \vec{v}_{\rm sw} \right] \right\} ,
\end{align*}
with $\hat{\vec{b}} = \vec{B} / B$ a unit vector in the direction of the background magnetic field, $\vec{v}_{\rm sw}$ the SW velocity,
\begin{align}
\label{eq:AvgGCDriftVel}
\vec{v}_d & = \frac{\mu p}{qB} \hat{\vec{b}} \times \left[ \frac{\partial \hat{\vec{b}}}{\partial t} + (\vec{v}_{\rm sw} \cdot \vec{\nabla}) \hat{\vec{b}} \right] + \frac{pv}{q B} \left[ \mu^2 (\vec{\nabla} \times \hat{\vec{b}})_{\perp} + \frac{1 - \mu^2}{2} \frac{\hat{\vec{b}} \times \vec{\nabla} B}{B} \right] + \nonumber \\
 & \;\;\;\;\; \frac{m}{qB} \vec{v}_{\rm sw} \times \left[ \frac{\partial}{\partial t} \left( \frac{\hat{\vec{b}}}{B} \right) + (\mu v \,  \hat{\vec{b}} + \vec{v}_{\rm sw}) \cdot \vec{\nabla} \left( \frac{\hat{\vec{b}}}{B} \right) \right]
\end{align}
the gyrophase averaged GC drift velocity perpendicular to the magnetic field \citep{northrop1961, rossiolbert1970, burgeretal1985, wijsen2020}, and $\vec{a} \vec{b} : \vec{c} \vec{d} = a_i b_j c_j d_i$ a tensor contraction. The diffusion coefficients are also gyrophase averaged, although not explicitly indicated, and it is assumed that momentum diffusion is negligible in the SW. In the derivation it is additionally assumed  that the SW is non-relativistic, and that $v \gg v_{\rm sw}$. See \citet{skilling1971}, \citet{riffert1986}, \citet{ruffolo1995}, \citet{zhang2006}, \citet{lerouxwebb2012}, \citet{lerouxetal2014}, \citet{zank2014}, and \citet{wijsen2020} for additional details and discussions about the FTE and its derivation.

Perpendicular diffusion is still hotly debated and an aspect which is not entirely understood in SEP transport. This is mainly because the exact pitch-angle dependence of the perpendicular diffusion coefficients are not yet known. The exact amount of perpendicular diffusion (i.e. the perpendicular MFP) is also currently uncertain. However, it should be emphasized that some level of perpendicular transport {\it must} be present in SEP events: Any turbulent magnetic field with a level of transversal complexity will lead to the perpendicular transport of charged particles. The SW has a significant transverse component \citep{matthaeusetal1990}, with \citet{bieberetal1996} suggesting a ratio of 80:20 for the ratio of 2D (transversal) to slab turbulence, suggesting that, at $1$ $\mathrm{AU}$, SW turbulence is predominantly transversal. Therefore, the question should not be whether perpendicular diffusion occurs in the SW, but rather {\it how much} perpendicular diffusion do SEPs experience?


\subsubsection{Effect of Perpendicular Diffusion}
\label{subsubsec:PerpDiff}

The first model to simulate pitch-angle dependent SEP transport in a full 3D geometry, including adiabatic energy losses, was presented by \citet{zhangetal2009}, showing that the inclusion of perpendicular diffusion allows SEPs to propagate rather efficiently across magnetic field lines and thereby explaining the origin of widespread SEP events. In addition to this, they also showed that perpendicular diffusion tends to smooth out any longitudinally dependent fine-structure in the SEP source. This was also later confirmed by \citet{zhangzhao2017} and \citet{straussetal2017a}. A similar 3D model without energy losses was presented by \citet{drogeetal2010}, with selected results shown in Fig.~\ref{fig:perpDiff}. These authors pointed to one of the major unsolved problems in multi-dimensional SEP transport modelling: The inclusion of perpendicular diffusion can explain the existence of widespread SEP events, but the results are inconsistent with so-called drop-out events; SEP events where very low levels of perpendicular diffusion is present and SEPs seem tightly tied to magnetic fieldlines \citep[as inferred by e.g.][]{mazuretal2000}. This apparent dichotomy is yet to be resolved \citep[see also][]{wangetal2014}. In later work, \citet{dresingetal2012}, \citet{drogeetal2014}, and \citet{drogeetal2016} implemented a phenomenological description of the diffusion parameters, and through a detailed comparison between simulation results and SEP electron observations from the STEREO spacecraft, were able to constrain the level of perpendicular diffusion needed. Similarly to the case of pitch-angle scattering, there is a rather large inter-event variation in these parameters. 

\citet{qinetal2011}, investigating SEP transport in 3D, focusing especially on the simulated anisotropy of SEP events, and similarly to \citet{zhangetal2009}, shows that the level of the observed anisotropy depends strongly on the level of magnetic connectivity: When a virtual observer in a model is well connected to the SEP source, it will observe a high anisotropy, whereas SEP particles that undergo significant perpendicular diffusion, and therefore are not magnetically well connected to the source, show very low/insignificant anisotropies. This was later confirmed in the simulation of \citet{straussetal2017a}, using a more fundamental description of the transport coefficients, and is evident in the observations presented by \citet{dresingetal2014}. For impulsive SEP events, the observed anisotropy can therefore be used as a proxy for the level of magnetic connectivity to the SEP source, and this could assist in estimating the size of the SEP source. Results indicate that, for impulsive electron events, an extended source alone cannot explain the observed longitudinal spread of SEPs; some level of perpendicular transport must be present. 

Many additional modelling studies are examining, amongst other effects, the role of perpendicular diffusion in influencing the energy spectrum of impulsive \citep[][]{straussetal2020} and gradual \citep[][]{wangqin2015b} SEP events, the effect of perpendicular diffusion on SEPs released from a moving source \citep[i.e. propagating interplanetary shock;][]{qinetal2013, qinwang2015}, and whether there may be simulated effects that can be tested against observations, such as possible asymmetries and anisotropies in the simulated distribution \citep[][]{heetal2011, he2015, hewan2017}. Indeed, there are currently numerous simulation studies looking at the effects of perpendicular diffusion and this has opened up a very rich research sub-field.

Although the effects of perpendicular diffusion can be effectively studied in transport models, there are outstanding theoretical questions, including whether a diffusive description for perpendicular diffusion is valid (see Section~\ref{subsubsec:FullOrbit}), whether perpendicular diffusion should rather be described as a field line meandering process (see Section \ref{subsubsec:FLMeandering}), and the form of the perpendicular diffusion coefficient, which on the pitch-angle level is currently not well studied. Studies by e.g. \citet{straussfichtner2014, straussfichtner2015} has shown that the pitch-angle dependence of this coefficient is an important parameter, and implementing different forms lead to very different simulation results. This is, of course, not unexpected for SEP transport, where a highly anisotropic particle distribution is formed. This remains an ongoing topic of investigation, both theoretically \citep[][]{straussetal2016, engelbrecht2019, shalchi2020} and from numerical simulations \citep[][]{qinshalchi2009, qinshalchi2014}.


\subsubsection{Field-line Meandering}
\label{subsubsec:FLMeandering}

\begin{figure*}[!t]
\centering
\includegraphics[width=0.95\textwidth]{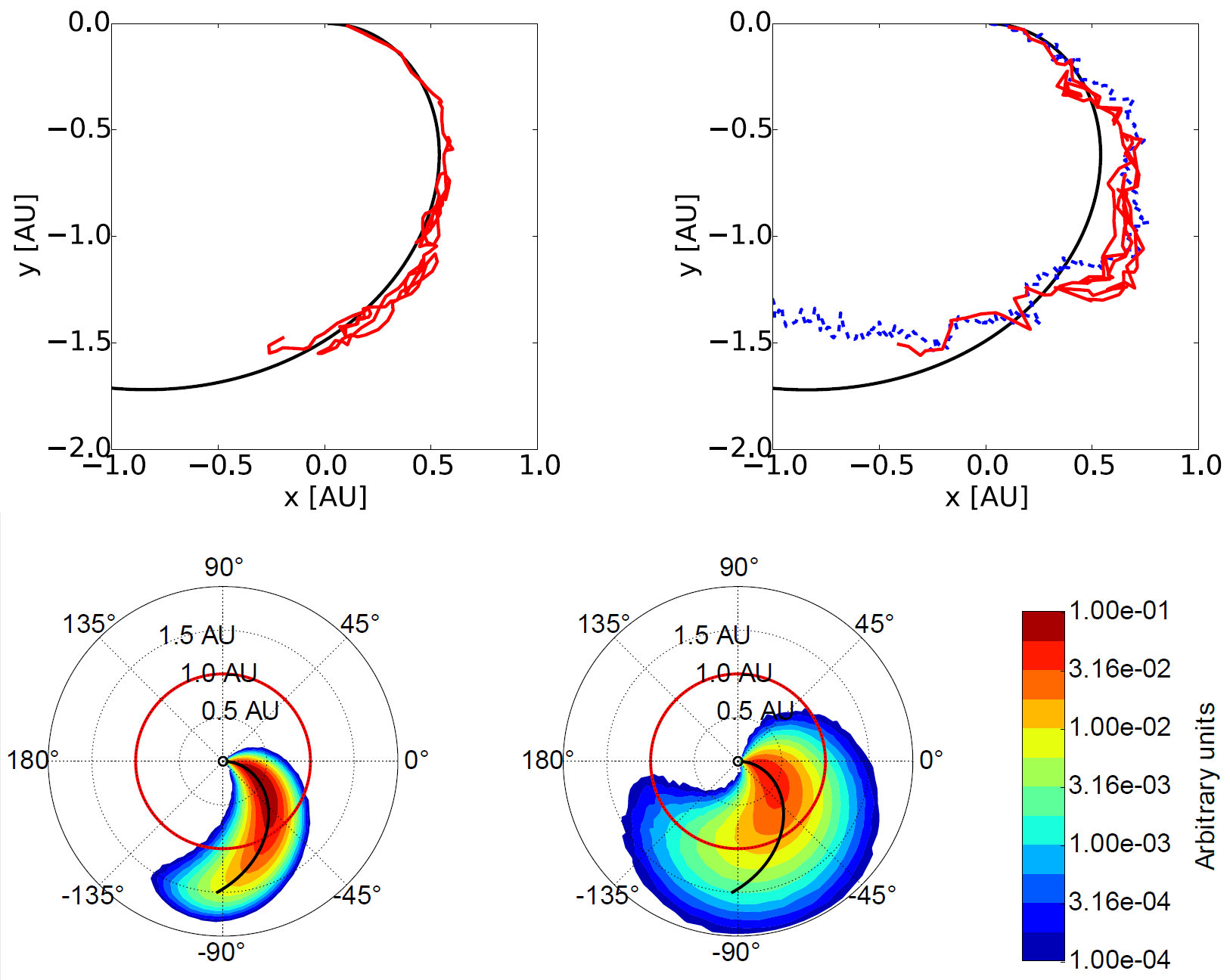}
\textbf{\caption[]{\label{fig:meandering}{\textnormal{\textit{Top:} A sample path of a 10 $\mathrm{MeV}$ proton injected at the Sun (in red) superimposed on a nominal Parker spiral (in black; \textit{left}). The dashed blue line (\textit{right}) shows a meandering field line and how the particle in this case follows the field line while scattering back and forth while decoupling slowly. \textit{Bottom:} Difference in longitudinal spread between a standard Fokker-Planck simulation with pitch-angle independent perpendicular diffusion (\textit{left}) and a combined field line random walk plus Fokker-Planck model (\textit{right}). These contours are calculated for 3h after the injection at $\phi_0=0$ at the sun. The figures are from \citet{laitinenetal2016}.}}}}
\end{figure*}

Perpendicular transport is believed to be significantly controlled by the behaviour of the large-scale background magnetic field. Commonly, in modelling, this is assumed to follow a simple Parker spiral configuration. It is clear, however, that even on larger scales, the SW is turbulent and field lines can deviate from such a simple geometry \citep[e.g.][]{shalchi2010}.
Thus, the field-line random walk (FLRW), in principle, has to be taken explicitly into account to model the perpendicular spread of particles correctly. One approach, discussed in \citet{laitinenetal2016, laitinenetal2017}, is to calculate a random ensemble of meandering field lines and follow a stochastic trajectory of particles along those lines. Fig.~\ref{fig:meandering} illustrates how particles follow the field line while slowly decoupling from it. In the bottom panel of the figure, the resulting differences for the spread of the distribution function are shown. Together with the width of the injection region, this effect can be a significant contribution to the observed width of the particle distribution at 1AU and beyond. It can also impact on the actual path-length a particle experiences \citep{laitinendalla2019}.

Note, furthermore, that this modelling approach offers an explanation for the simultaneous occurrence of wide-spread events and the observed drop-outs in particle intensity \citep{mazuretal2000}. Since in each actual event, a particular realization of meandering field lines exist, empty and filled flux tubes can be close to each other, while still spreading the particles to wide longitudes. Of course, with current modelling capabilities and limited knowledge of the interplanetary magnetic field configurations, these effects can only be understood statistically, i.e. for an ensemble of events.

The validity of the coupled FLRW and perpendicular diffusion approach has been established to some degree through MHD and full-orbit simulations in synthetic turbulence (see the additional discussion in Section~\ref{subsubsec:FullOrbit}). \citet{chuychai2007} discuss the field-line topology and trapping that can occur in a two-component turbulence model. \citet{ruffoloetal2012} show an explicit calculation of the perpendicular diffusion coefficient with a `random ballistic interpretation' that can lead to a reduction in the coefficient because of the parallel or pitch-angle scattering of the particles. These are just a few examples that illustrate the need for a detailed look at the interplay between parallel and perpendicular transport of particles and the influence of different scales of turbulent fluctuations on the resulting particle distributions.


\subsubsection{Drift Effects}
\label{subsubsec:Drifts}

Drifts can also lead to transport of particles across magnetic fields and \citet{dallaetal2013} recently derived expressions for the drift velocities of SEPs in a \citet{parker1958} HMF. The electric field drift due to the motional electric field will be in the plane containing the Parker spiral and describe the co-rotation of particles with the HMF as the Sun rotates. The gradient and curvature drift will furthermore be smaller for low energy particles and in opposite directions for positively and negatively charged particles. \citet{marshetal2013} verified these general predictions by integrating the Newton-Lorentz equation (Eq.~\ref{eq:NewtonLorentz}) in a uni-polar Parker HMF. These authors found that pitch-angle scattering and the MFP has little effect on the drifts and that drifts will be the most pronounced for high energy particles ($\sim 100$ $\mathrm{MeV}$ protons) or partially ionised heavy ions. \citet{dallaetal2017a} and \citet{dallaetal2017b} used this model to attribute the observed energy dependent charge state of iron and the temporal evolution of the iron-to-oxygen-ratio, respectively, to the mass-to-charge-ratio dependence of drifts and not the usual rigidity dependent MFP with turbulence generated by streaming protons of similar rigidity and the acceleration process \citep[see e.g.][and references therein]{reames1999}.

\begin{figure*}[!t]
\centering
\includegraphics[width=0.95\textwidth]{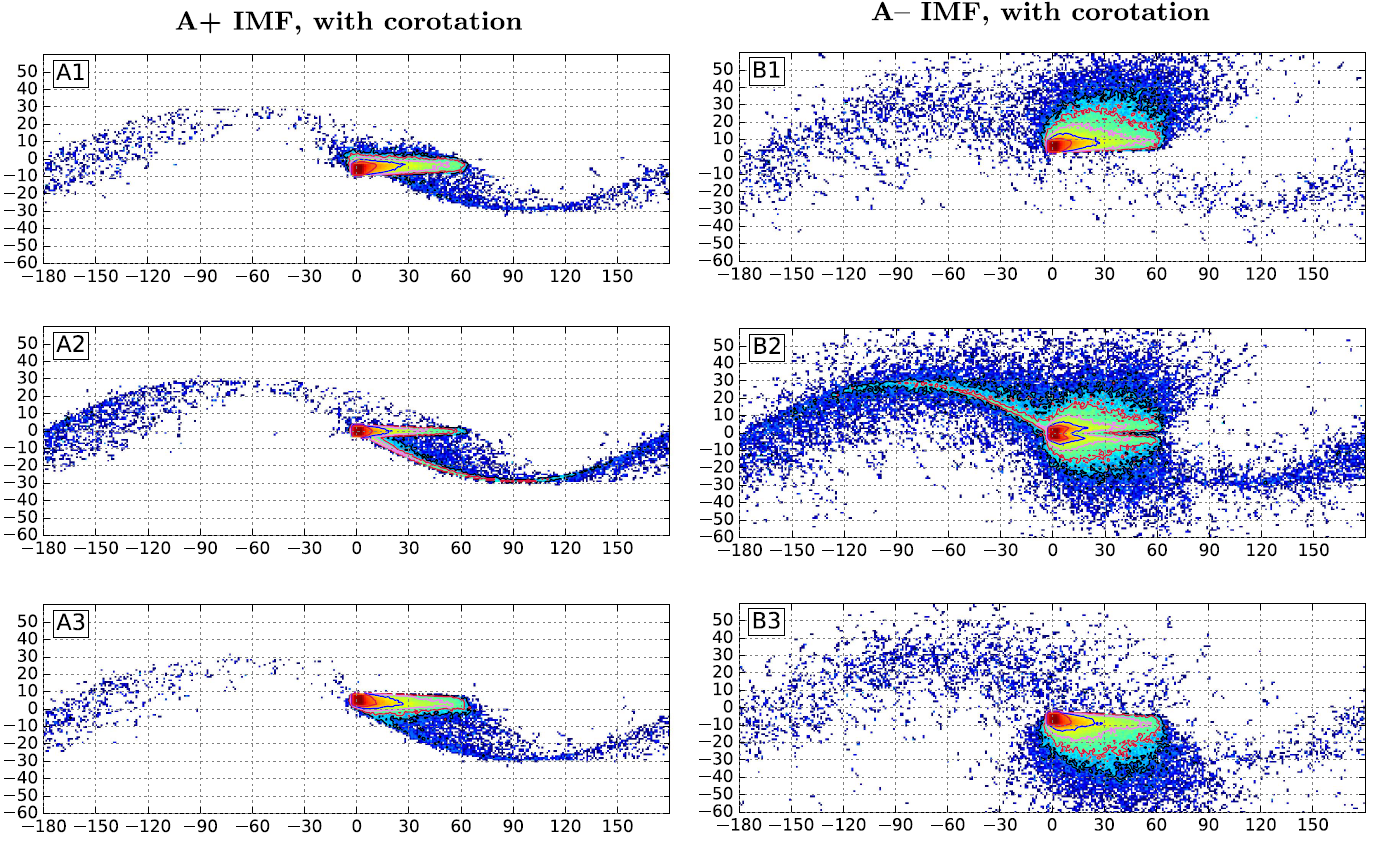}
\textbf{\caption[]{\label{fig:DriftPatterns}{\textnormal{A simulation of proton drift patterns at 1 AU along a wavy heliospheric current sheet in two different polarity cycles (magnetic field pointing outwards, \textit{left row}, or inwards, \textit{right row}, in the Northern hemisphere) for different positions of the injection (above, \textit{top row}, in, \textit{middle row}, or below, \textit{bottom row}, the current sheet). The drift of particles towards positive longitudes are due to the co-rotation. These figures are from \citet{battarbeeetal2018a}.}}}}
\end{figure*}

\begin{figure*}[!t]
\centering
\includegraphics[width=0.95\textwidth]{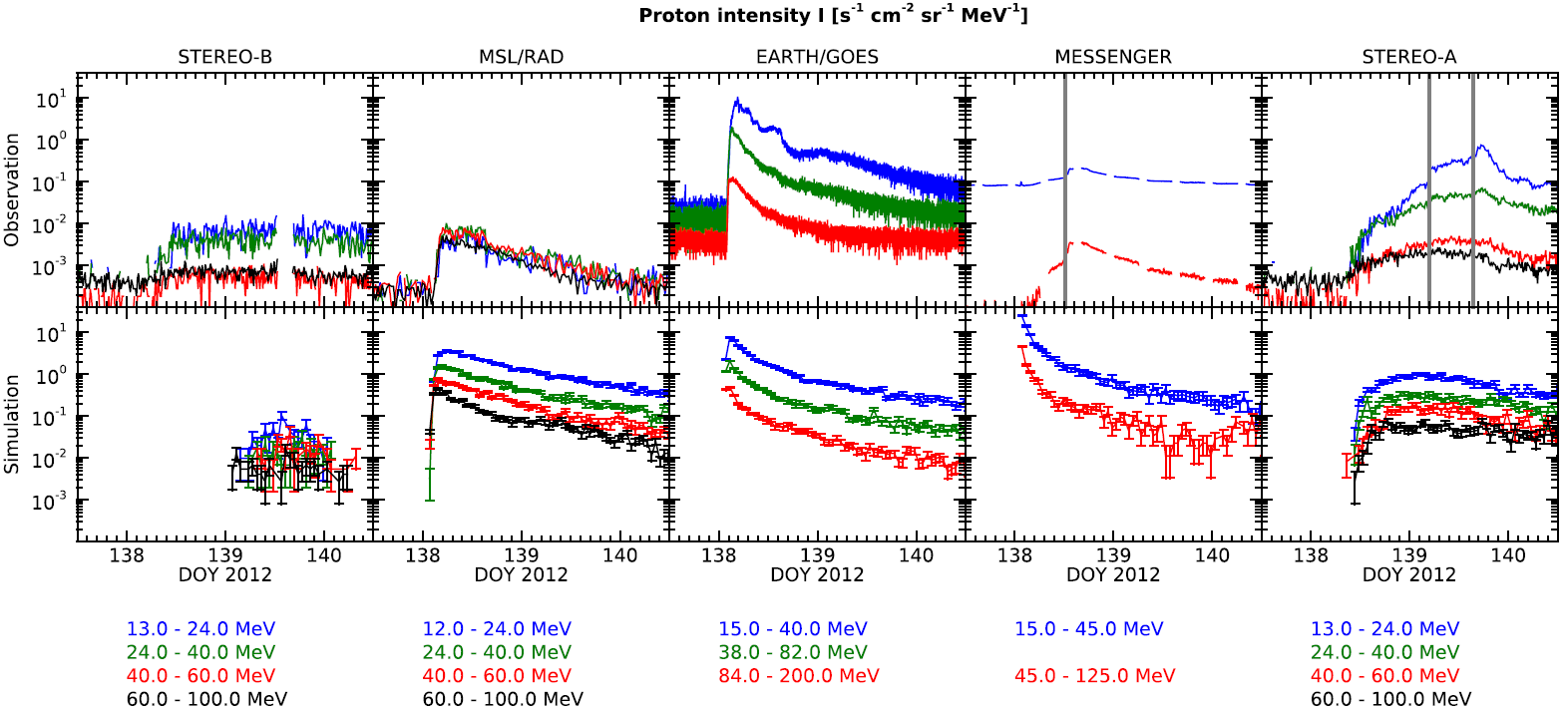}
\textbf{\caption[]{\label{fig:DriftGLE}{\textnormal{A simulation of proton intensity profiles (\textit{bottom row}) at different observers of the 17 May 2012 ground level enhancement event (observations shown in the \textit{top row}). Note that the data from the Radiation Assessment Detector on the Mars Science Laboratory (MSL/RAD) is from measurements inside the protective flight shielding and does not represent the true intensities. This figure is taken from \citet{battarbeeetal2018b}.}}}}
\end{figure*}

\citet{battarbeeetal2017} extended the investigation of \citet{marshetal2013} to include neutral sheet drifts in a flat heliospheric current sheet (HCS). They found that SEP drift patterns will be similar to galactic CR drift patters: if $A$ is the polarity of the HMF in the Northern hemisphere ($A = +1$ for outwards and $A = -1$ for inwards), then particles will drift towards the equator and in the direction of the solar rotation in the inner heliosphere or outwards in the outer heliosphere if $qA > 0$, while particles will drift towards the poles and in the opposite direction of the solar rotation in the inner heliosphere or inwards in the outer heliosphere if $qA < 0$. Particles can therefore be confined to the HCS in the $qA > 0$ polarity cycle and might have difficulty reaching an observer on the other side of the HCS. The energy dependence of drift velocities can also cause observers to see different spectra. \citet{battarbeeetal2018a} extended this investigation to a wavy HCS and found that the HCS can efficiently transport particles to the poles where they will drift more efficiently if they can escape the HCS through scattering. An example of the described drift patterns can be seen in Fig.~\ref{fig:DriftPatterns}. \citet{battarbeeetal2018b} used this model to model the ground level enhancement event of 17 May 2012 with mixed results, as can be seen in Fig.~\ref{fig:DriftGLE}. A source $80^{\circ}$ wider in longitude than the inferred coronal mass ejection was needed for particles to reach the STEREO spacecrafts and the fluxes at Mercury was overestimated.

Caution, however, should be taken with these results as strong pitch-angle scattering was implemented at Poisson-distributed scattering intervals (i.e. random adjustments of the pitch-angle and gyro-phase at random times), and the turbulent reduction of drifts has been neglected. Turbulent fluctuations disrupts the large scale drifts and therefore decrease the drift velocity in Eq.~\ref{eq:AvgGCDriftVel}. A review of the current knowledge and understanding of this subject is given by \citet{engelbrechtetal2017}. In summary, drift reduction do not seem to occur in purely magnetostatic slab turbulence, the drift coefficient is reduced by the same factor in both a homogeneous magnetic field and a magnetic field with large scale gradients for the same turbulence conditions, and drifts decrease with an increase in the turbulence strength or a decrease in the particle energy \citep{burgervisser2010, engelbrechtetal2017}. It is important to note that the exact form of the drift suppression factor is not yet known and that all studies on this have only considered isotropic distributions and do not include pitch-angle dependencies.
 
\citet{wijsenetal2020} included drifts in a uni-polar HMF into the FTE with a pitch-angle independent perpendicular diffusion coefficient for $3-36$ $\mathrm{MeV}$ protons. They verified that different observers will see different spectra and found that perpendicular diffusion will diminish the effects of drifts. \citet{richardsonetal2014} investigated the $14-24$ $\mathrm{MeV}$ proton events observed during the first seven years of the STEREO mission and compared this with the $0.7-4$ $\mathrm{MeV}$ electron events. Both the onset or peak delay and the angular position of the peak intensity, as a function of the angular separation between the flare and the spacecraft's field line footpoint, follow nearly the same trend in both the proton and electron events. If the perpendicular transport were primarily due to drifts, then it would be expected that the electrons and protons should behave differently, although the data are not presented according to the polarity of the magnetic field.


\begin{figure*}[!t]
\centering
\includegraphics[width=0.44\textwidth]{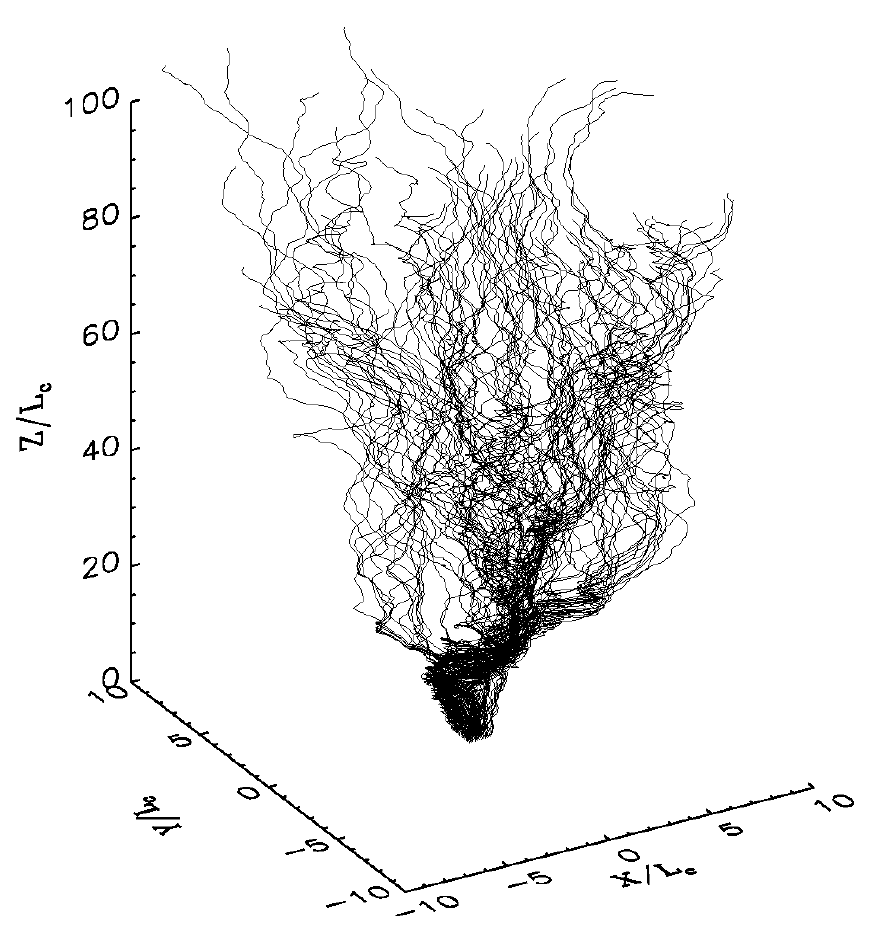}
\includegraphics[width=0.54\textwidth]{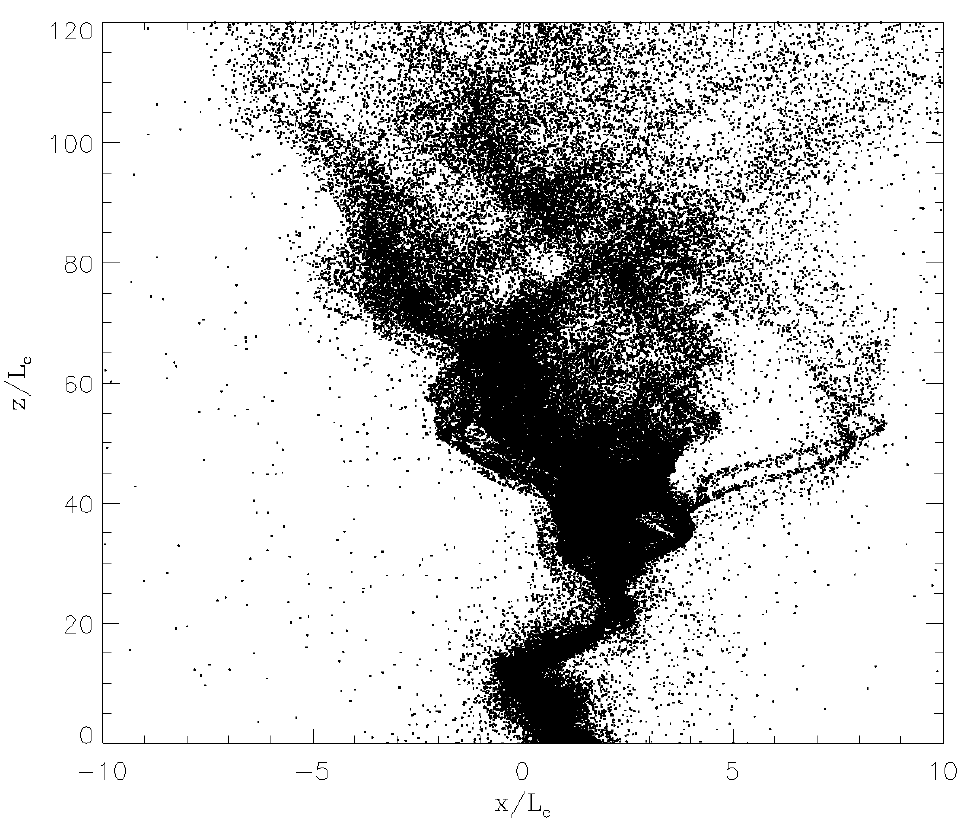}
\textbf{\caption[]{\label{fig:full_orbit}{\textnormal{A realization of a turbulent magnetic field (\textit{left panel}) and the position of SEPs released into this structure (\textit{right panel}). Both figures are taken from \citet{guogiacalone2014}.}}}}
\end{figure*}

\subsubsection{Towards Full-orbit Simulations}
\label{subsubsec:FullOrbit}

One potential disadvantage of the diffusive SEP description mostly discussed above is that these models cannot capture the initial ballistic phase of SEP transport before pitch-angle scattering results in diffusive transport \citep[see e.g.][]{laitinendalla2017, laitinenetal2017}. The time, from SEP acceleration and release, until diffusive behaviour is reached, depends on the SW turbulence characteristics near the SEP source, and as such, is not well known. SEP transport models utilizing full-orbit simulations (i.e. solving the Newton-Lorentz equations directly) does not have this limitation. However, such models are much more computationally expensive and are not always practically feasible.

\citet{kellyetal2012} present full-orbit simulations in a Parker HMF with superimposed large scale magnetic fluctuations, based on earlier work by \citet{peietal2006}. Similarly to model solutions described in Section~\ref{subsubsec:Drifts} by \citet{marshetal2013} and later co-workers, these simulations do not include small scale turbulence leading to pitch-angle scattering, with particle scattering included in an ad-hoc fashion. However, even without scattering included, simulations from \citet{kellyetal2012} show the role of field-line meandering in leading to cross-field particle transport.

Full-orbit simulations by \citet{guogiacalone2014} implemented a more detailed turbulence model, covering both small and larger scales, so that both pitch-angle scattering by small-scale turbulence, and large scale meandering is included. Fig.~\ref{fig:full_orbit} shows examples of these simulations: The left panel shows simulated magnetic fieldlines originating from a small source region, while the right panel shows the distribution of particles, a certain time after release into such as magnetic realization. These results shows that when SEP are released from compact sources, smaller than the turbulence correlations scale, particles can be confined to {\it fluxtubes}, forming a pattern of alternating empty and filled (with SEPs) regions of space. This has been put forward as an explanation for the observations of so-called {\it drop-out} events, where, presumably, the spacecraft moved through such a {\it patchy} region of filled and empty fluxtubes. These simulations do, however, depend on the structure and strength of the underlying turbulence, and continue to be an avenue of further research \citep[][]{tooprakaietal2016, ablassmayeretal2016}.


\subsection{Towards Predictive Capability}
\label{subsec:Predicting}

With the recent interest in crewed space travel, predicting, and thereby mitigating the radiation risk posed by SEPs has become an ongoing problem is space science. In some sense the development of a physics-based model with real-time SEP predictive capabilities has become the holy grail of SEP modelling studies. To understand how close we are to reaching this goal, it is useful to examine the application usability levels (AUL) as defined by \citet{halfordetal2019} which show the natural progression from a basic research question (AUL 1) to validation and approval for use (AUL 9). Most, if not all, of the SEP modelling studies presented up to here are concerned with basic research questions and trying to understand and characterize the underlying processes shaping SEP transport. Although these fundamental studies will, in future, inform the predictive models, it is clear that there is much more work to be done before SEP models will have true predictive capability. At the moment most SEP prediction algorithms are based on observed empirical relationships \citep[e.g.][]{balch2008}. See also \citet{anastasiadisetal2019} for a recent review on this topic.

Most SEP models deal only with the transport of SEPs and neglect the acceleration thereof by pre-specifying the injection function discussed in previous sections. Such a 1D transport model, including only an energy coordinate, is given by \citet{kuboetal2015}, while \citet{marshetal2015} uses a full 3D model also including drift effects. A different approach is used by \citet{luhmannetal2017} where SEPs are back-tracked in an MHD simulated heliospheric background until the shock (source) region is reached, leading to a re-weighting of the SEP intensities based on some analytical approximations for shock acceleration efficiency. While the ballistic propagation assumption is an over-simplification for particle transport, more information regarding the source region is possible. \citet{aranetal2006} present a large number of pre-computed 1D SEP modelling scenarios, including different MHD generated shock scenarios, that could be applied to observed SEP events.

Based on the relative success of the prediction models discussed above, it appears that a physics-based SEP prediction model should have the following properties: (i) In order to capture cross-field particle propagation, and the longitudinally dependent acceleration efficiency of shocks in the inner heliosphere, a spatially 2D or 3D geometry must be used. (ii) A pitch-angle dependent (i.e. focused transport) approach is needed to capture the large particle anisotropies related to SEP events. (iii) The model needs to treat the SEP source in a self-consistent manner, either through implicitly including shock acceleration for protons, or by specifying the appropriate remote-sensing observations for flare accelerated electrons. (iv) If shock acceleration is handled numerically, an energy coordinate is needed in the model. (v) To account for large non-Parkerian magnetic field variations, an MHD (or equivalent) model must be used to simulate the underlying magnetic geometry. (vi) Appropriate SEP transport parameters must be specified, although these may be based on a phenomenological description derived from the results of basic research models.

SEP models that conform to most of these requirements, although sometimes using simplifying assumptions, are EMMREM (Energetic particles, radial gradients, and coupling to MHD) model, described by \citet{kozarevetal2010} and \citet{schwadronetal2010}, the iPATH (Particle Acceleration and Transport in the Heliosphere) model, described by \citet{huetal2017}, and the PARADISE (PArticle Radiation Asset Directed at Interplanetary Space Exploration) model with initial developments described by \citet{wijsenetal2019}. These models are being constantly improved, while some open research avenues remain, such as the details of SEP seed population, whether the acceleration process is handled correctly, and whether the turbulent magnetic structures between the source and the observer are correctly described. These questions remain unanswered as we do not have, and will most likely never have, sufficient in-situ measurements of these quantities.


\newpage

\section{Summary and Conclusion}
\label{sec:Summary}

Section~\ref{sec:Microphysics} started by illustrating SEP motion in a fluctuating magnetic field by directly solving the Newton-Lorentz equation. It shows that, even with turbulent fluctuations included, the particle trajectory remain relatively smooth, forming a so-called perturbed nearly-circular orbit. The guiding center, however, behaves as one would naively expect for a particle undergoing diffusion: the velocity components show random changes reminiscent of Brownian motion. It should always be remembered that, when describing SEP transport, one deals with so-called \textit{small angle scattering} which is a slow process where particle quantities (in this case, most importantly, the pitch-angle) undergoes many \textit{small} changes, accumulating over several gyro-cycles into the particle changing it propagation direction. The particle is then \textit{scattered} after moving an avrage distance $\lambda_{\parallel}$ through the slowly (slow with respect to the particle's gyro-motion) interacting turbulence.

Also shown, by considering an anisotropic Alfv{\'e}nic turbulence wave field (modelled in Appendix~\ref{apndx:ModelSlab}), is that particle scattering conserves energy in the wave frame. In reality, however, it is not possible in most cases to define a single wave frame, as turbulence can, at best, be approximated by a large number of waves propagating both along and perpendicular to the mean field at different propagation directions. In such a scenario, the particle can be considered to scattered from one wave frame to another, leading, after many such interactions, to energy (velocity/momentum) diffusion. In the solar wind, momentum diffusion is usually slow enough to be neglected in most applications.

To specify the solar wind turbulence from the Sun to the Earth, and solving the Newton-Lorentz equation for each SEP particle, remains physically (in terms of knowing the exact turbulence structure) and computationally (in terms of the simulation) impossible. Therefore, one rather evaluates the evolution of the SEP's phase-space density, i.e. simulate the evolution of the SEP distribution function. This quantity was introduced in Section~\ref{sec:Macroscopic}, along with the so-called focused transport equation which describes its evolution. A rigorous derivation of this equation was not presented and the interested reader should refer to other, more complete works in the references. In this \textit{macroscopic} description of SEP transport, the particle-turbulence interactions are incorporated into a diffusion coefficient. For the pitch-angle diffusion coefficient, for example, there is a large number of phenomenological descriptions, although based on theoretical arguments, and contains a number of free-parameter that can be adjusted in order to reproduce observations by using a suitable model. These parameters can then be compared to theoretical quantities and our fundamental knowledge of SEP transport can be improved.

The focused transport equation cannot be solved analytically for the most general scenario, and only approximate solutions are available. In Section~\ref{subsec:Compare}, two popular analytical approximations (the Telegraph and Diffusion approximations, summarised in Appendix~\ref{apndx:DiffTel} for reference) were compared, which showed that, even for very simplistic modeling problems, these approximations do not give satisfactory solutions. Therefore, the focused transport equation must be solved numerically. This approach is still only an approximation of the true solution (some shortcomings of the finite-difference model were discussed, for instance, numerical diffusion), but allows for the incorporation of all the required processes to give a consistent description of an SEP event.

As suitable numerical models to simulate SEP transport are not widely available and difficult to construct, we have presented a finite difference model for solving the focused transport equation in the dimension along the magnetic field. The model is briefly discussed in Appendix~\ref{apndx:FDSolver} and is available for use by the community\footnote{\url{https://github.com/RDStrauss/SEP_propagator}}. We hope that this model might help scientists to have an alternative to the limited analytical approximations.

The last part of this manuscript, Section~\ref{sec:Review}, review the last $\sim 10$ years of SEP simulation studies, starting from very simplistic 1D ballistic approximations, and ending at the most complex state-of-the-art 3D numerical models currently available. Each subsection discussed the applicability of each of these approximations and modelling approaches to specific SEP transport problems, and have outlined possible pitfalls associated with each. Depending on the aim of a study, it may be possible to use a simplified approximation, as long as the user is aware of the physical processes neglected and the assumptions made when using that approximation. {All of these models have aided in our current understanding of and insight into the focused transport of SEPs, but have also raised many questions and highlighted the aspects which are still not fully understood, i.e. the research questions which should be answered by current and upcoming researchers in the fields of SEPs and turbulence.}

The attentive reader will notice that we have not reviewed the physics of perpendicular diffusion in much detail in this manuscript and have focused more on field-aligned transport and pitch-angle scattering. The reason for this is threefold: (1) A comprehensive review on the perpendicular diffusion coefficient was recently published by \citet{shalchi2020}. (2) The derivation of the correct focused transport equation that includes perpendicular diffusion is available (and presented here in Eq.~\ref{eq:FTE}), but remains poorly understood: The most complete derivation of this equation was recently published in the PhD thesis of \citet{wijsen2020} and has led to some interesting implications, including that the perpendicular diffusion terms are only retained when the position vector is first transformed to the position of the guiding center \textit{where after} averaging over gyro-phase is performed. In contrast, when specifying the SEP distribution's position in terms of the particle position, assuming the distribution to be gyrotropic, and averaging over gryo-phase, all perpendicular diffusion terms disappear. The fact that the order of operations has such large implications (which has not been discussed in the literature in detail) implies that the due diligence on the transport equation has not been performed in enough detail to review in any sense and remain a (very) active research field. (3) The physics of perpendicular diffusion, on the pitch-angle level, also remains poorly understood. It is now clear that perpendicular diffusion can be described as a combination of magnetic field wandering/meandering (where particles simply follow \textit{large scale} (i.e. larger that the particles' gyro-radius) turbulent fluctuations and \textit{small scale} (on the scale of the particles' gyro-radius) `scattering' which displaces the particle's guiding center to different field lines. The second process allows the particles to decouple from their field lines and to follow different meandering field lines. This `scattering' process can be due to perpendicular propagating Alfv\'enic fluctuations, or simply due to drift effects in a turbulent magnetic field. In addition, it is not yet clear if a diffusive (perpendicular) description for SEP transport is applicable. With these processes unknown, we are not even confident, in this manuscript, to propose a definition for the perpendicular mean free path, beyond the most generic interpretation: \textit{The perpendicular mean free path is the average distance a particle propagates, perpendicular to the mean field, before it is decoupled from it's original field line}. The ambiguity of this statement should convey the fact that perpendicular diffusion is by no means well understood and that more research needs to be done.

A brief account of physics-based SEP prediction models was also given. These models are able to deal with the transport of SEPs, from their acceleration site, to e.g. Earth, where the intensity is needed. However, the accuracy and applicability of these models are limited by our lack of understanding of \textit{where} and \textit{how} SEPs are accelerated. In future, it is expected for these models to incorporate remote-sensing observations of flares and/or CMEs to better constrain the SEP acceleration site and, ultimately, the energy-dependent time profile of SEPs released into the interplanetary medium. Once this is known, SEP transport models can propagate this \textit{injection profile} to any region in the heliosphere. We express our hope that the ongoing interest and new discoveries in SEP research will ultimately lead also to better physics-informed predictive models that will be of value to our space-faring society.


\begin{acknowledgements}

This work is based on the research supported in part by the National Research Foundation of South Africa (NRF grant numbers 120847, 120345, and 119424). Opinions expressed and conclusions arrived at are those of the authors and are not necessarily to be attributed to the NRF. JPvdB acknowledge support from the South African National Space Agency. FE acknowledges support from NASA grant NNX17AK25G. Additional support from an Alexander von Humboldt group linkage program is appreciated.  We thank the International Space Science Institute (ISSI) for hosting our team on `Solar flare acceleration signatures and their connection to solar energetic particles'. We appreciate, as always, constructive research discussions with our colleagues, in particular, we would like to thank Nicolas Wijsen, Timo Laitinen, Nina Dresing, and Kobus le Roux. Figures prepared with Matplotlib \citep{hunter2007}.

\end{acknowledgements}


\newpage
\appendix

\section{A Finite Difference Solver}
\label{apndx:FDSolver}

As shown in this work, analytical approximations of Eq. \ref{eq:FTPE} have very severe limitations, and therefore, it has to be integrated (solved) numerically to capture the transport processes involved. Such a numerical implementation, for this spatially 1D version of the transport equation, is discussed by \citet{straussetal2017b}, which is based on the numerical techniques discussed in \citet{straussfichtner2015}. Details are also given in the dissertation of \citet{heita2018}. This model has subsequently been developed to be more user-friendly, and the source-code thereof can be found at \url{https://github.com/RDStrauss/SEP_propagator}. The code is published under the Creative Commons license, but is not intended to be used for commercial applications. We ask anyone using this model to reference this paper in all research outputs and to contact the authors when used extensively.

\begin{figure*}[!t]
\centering
\includegraphics[width=0.49\textwidth]{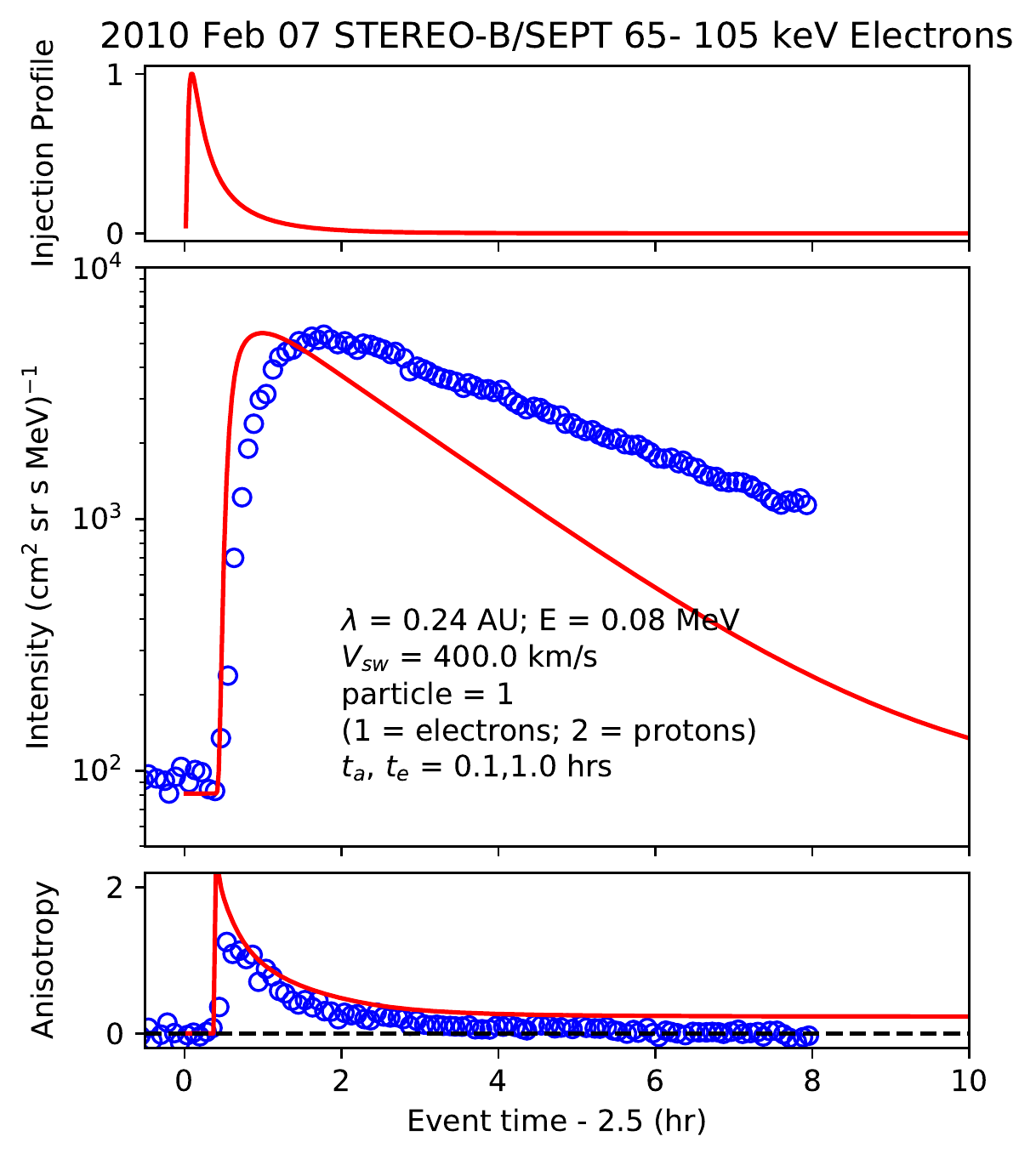}
\includegraphics[width=0.49\textwidth]{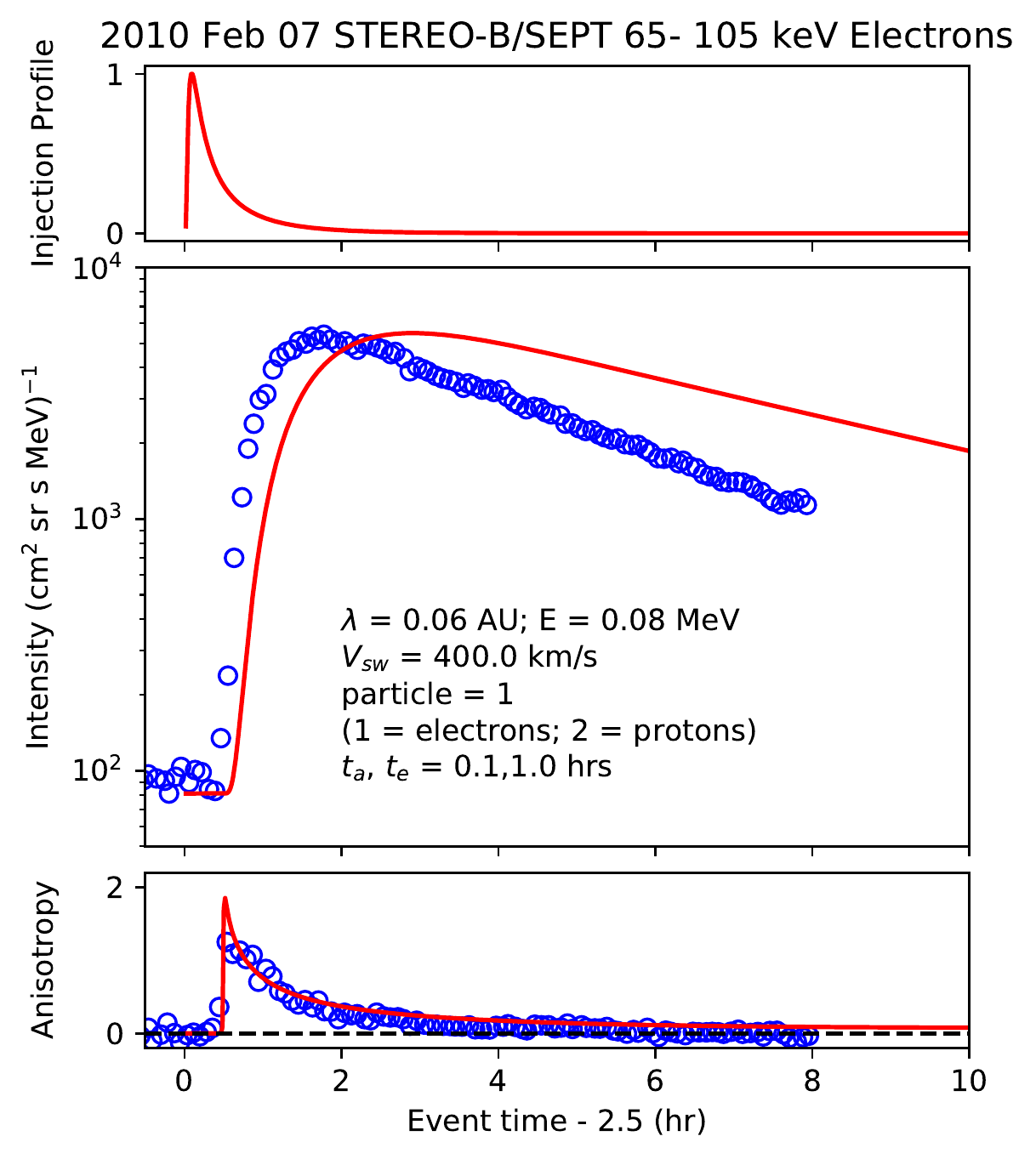}
\includegraphics[width=0.49\textwidth]{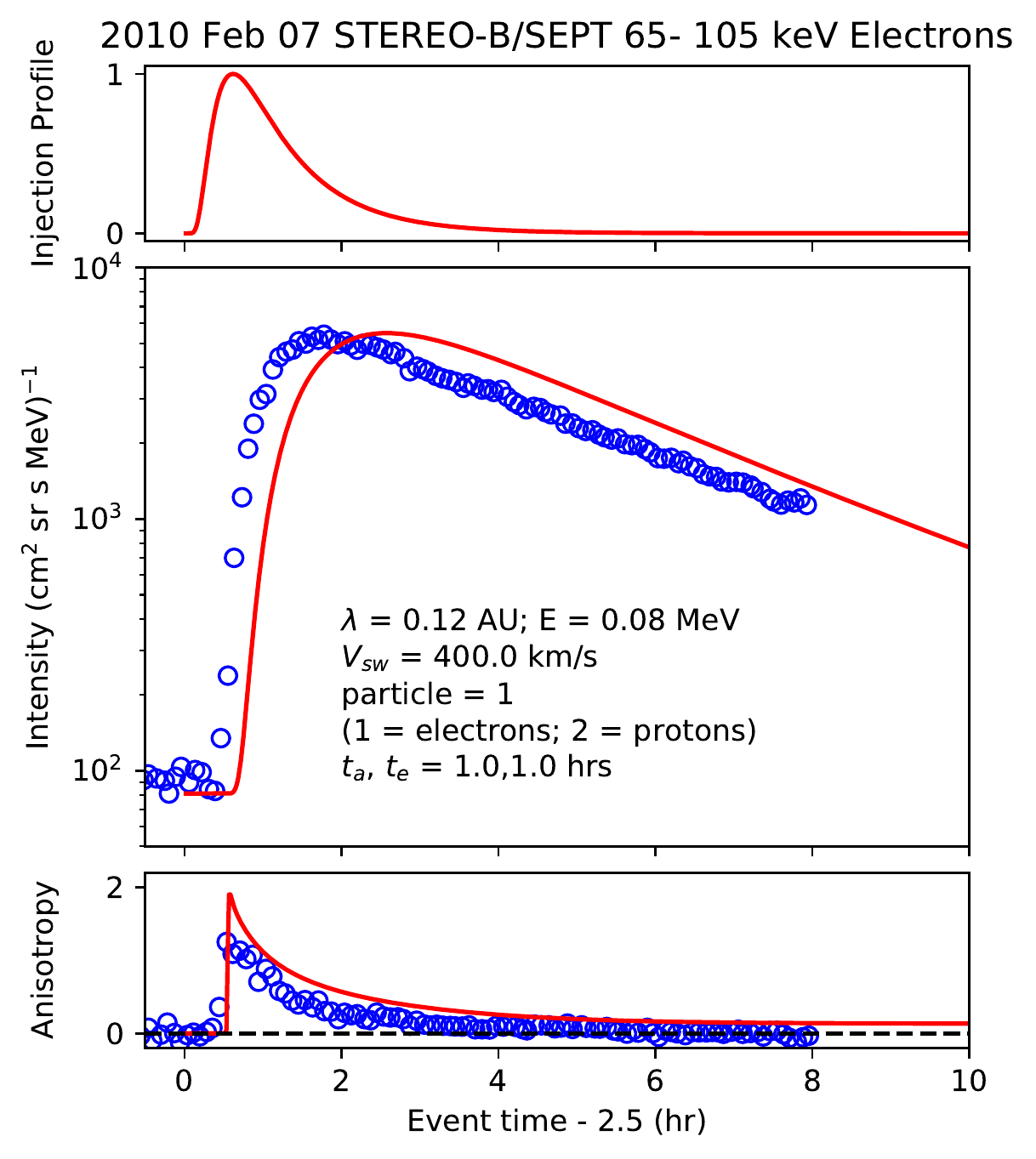}
\includegraphics[width=0.49\textwidth]{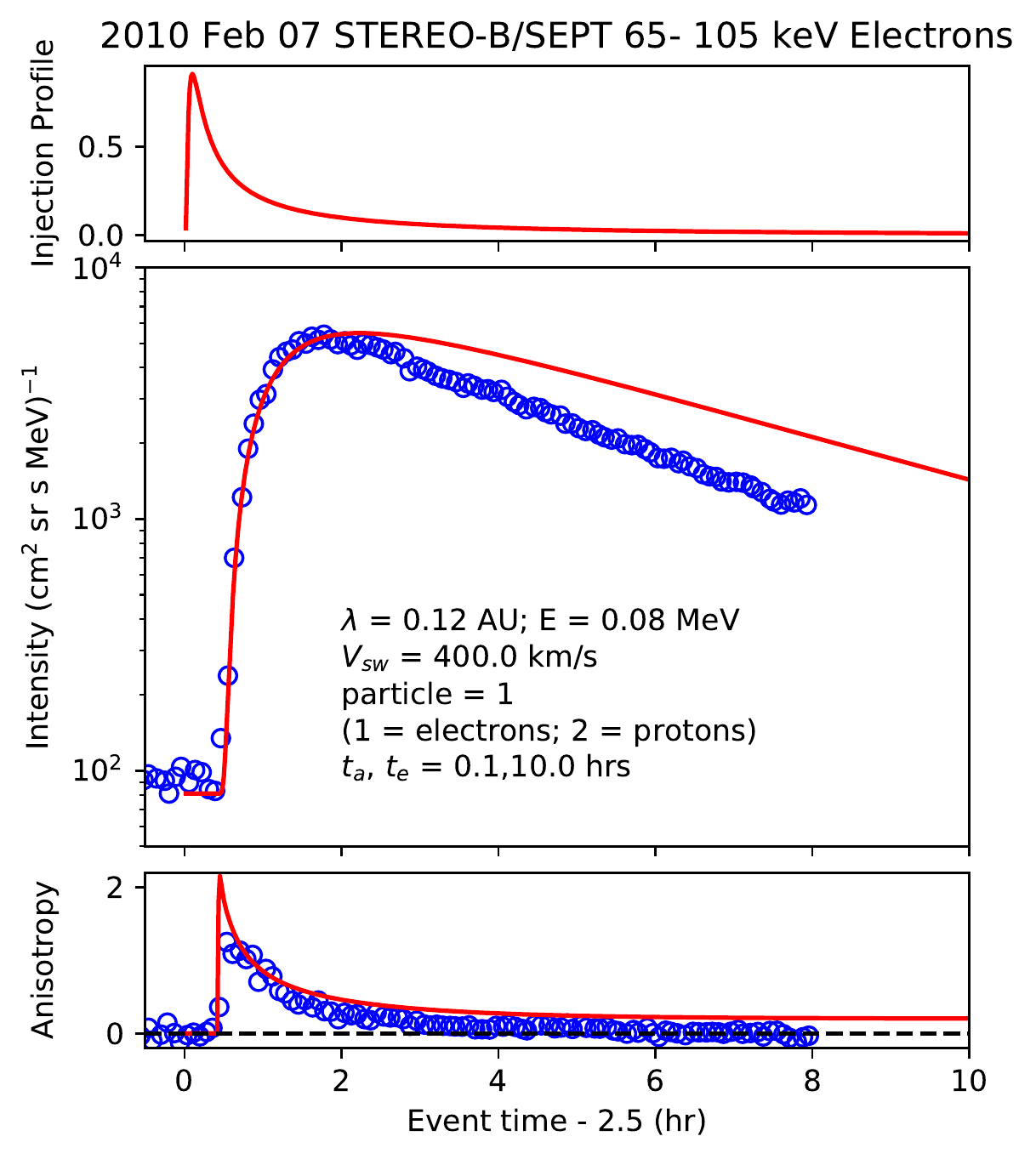}
\textbf{\caption[]{\label{fig:FD_param_variation}{\textnormal{Illustration of parameter sensitivity of the finite difference transport model in comparison to the 7 February 2010 electron event. \textit{Top left:} Smaller $\lambda_r$ of $0.06$~$\mathrm{AU}$. \textit{Top right:} Larger $\lambda_r$ of $0.24$~$\mathrm{AU}$. \textit{Bottom left:} Longer acceleration time of $1$ $\mathrm{h}$. \textit{Bottom right:} Longer escape time of $10$ $\mathrm{h}$.}}}}
\end{figure*}

The code contains a number of user-defined inputs, such as the particle species under consideration (i.e. electrons or protons), the effective radial MFP, the SW speed, the kinetic energy of the particles, and different options regarding the injected SEP distribution at the inner boundary condition. Details can be found in the comments section of the source-code. In Section~\ref{subsubsec:SEPevent}, this finite difference solver was applied to the 7 February 2010 electron event as observed by STEREO B. Fig.~\ref{fig:BestFitSEP} only showed a best fit scenario that can reproduce the observed particle intensity and anisotropy very well. Here, the sensitivity of the code to parameter variation is illustrated with four cases in Fig.~\ref{fig:FD_param_variation}. The top row shows the slower rise for a smaller MFP, in the left panel, and a quicker rise and quicker decay for a larger MFP, in the right panel. The bottom row shows a similar variation for a longer acceleration time, in the left panel, and a longer escape time, in the right panel, in the injection function. These example solutions are also included in the online repository.

\section{A Stochastic Differential Equation Solver}
\label{apndx:SDESolver}

Stochastic calculus is a study area with several works dealing with its mathematical formalism and application to a variety of problems, including \citet{gardiner1985}, \citet{vankampen1992}, \citet{kloedenplaten1995}, \citet{oksendal2000}, \citet{lemons2002}, and \citet{strausseffenberger2017}. Of special interest is \citet{gardiner1985}, \citet{kloedenplaten1995}, and \citet{strausseffenberger2017}, which gives an introduction of stochastic calculus specifically for the fields of natural sciences, an introduction focusing on numerical methods to solve stochastic differential equations (SDEs), and a review of the application of this to CR modelling with toy models to introduce the basic concepts, respectively. SDEs can be computationally expensive and these types of models did not become feasible until the dawn of parallel-processing. Nonetheless, \citet{mackinnoncraig1991} first applied SDEs in solving the FTE for binary collisions of particles with `cold' hydrogen atoms in the chromosphere and \citet{kocharovetal1998} first used them to solve the SEP model of \citet{ruffolo1995}. A three dimensional focused transport model for SEPs with and without energy losses are presented by \citet{qinetal2006} or \citet{zhangetal2009} and \citet{drogeetal2010}, respectively.

If $S$ and $M$ represents the stochastic variables corresponding to $s$ and $\mu$, respectively, then the two first order SDEs equivalent to the Roelof equation (Eq.~\ref{eq:FTPE}) are
\begin{align*}
{\rm d}S & = \mu v \, {\rm d}t \\
{\rm d}M & = \left[ \frac{(1 - \mu^2)v}{2 L(s)} + \frac{\partial D_{\mu \mu}}{\partial \mu} \right] {\rm d}t + \sqrt{2 D_{\mu \mu}} {\rm d}W_{\mu}(t) ,
\end{align*}
where ${\rm d}W_{\mu}(t)$ is a Wiener process. These SDEs are solved using the Euler-Maruyama scheme,
\begin{align*}
S(t + \Delta t) & = S(t) + M(t) v \Delta t \\
M(t + \Delta t) & = M(t) + \left[ \frac{(1 - M^2(t))v}{2 L(S(t))} + \left. \frac{\partial D_{\mu \mu}}{\partial \mu} \right|_{\mu = M(t)} \right] \Delta t + \sqrt{2 D_{\mu \mu}(M(t)) \Delta t} \Lambda ,
\end{align*}
from the initial values $S(t_0) = s(t_0)$ and $M(t_0) = \mu (t_0)$ at initial time $t_0$ ($= 0$ $\mathrm{h}$), where $\Delta t$ ($= 5 \times 10^{-5}$ $\mathrm{h}$) is the time step, and $\Lambda$ is a pseudo-random number which is Normally distributed with zero mean and unit variance \citep{kloedenplaten1995, strausseffenberger2017}.

A single solution of the SDEs represent only one possible realization of how a phase-space density element, or pseudo-particle in SDE nomenclature, would evolve. In order to calculate quantities of interest, the SDE is solved $10^6$ times. Temporal, spatial, and pitch-cosine bins are set up and the pseudo-particles are binned into the correct bin at each time step to create a phase-space density \citep{strausseffenberger2017}. To calculate, for example, the ODI at an observation point, only the spatial bin centred on the observation point is considered and for each temporal bin the pitch-angle bins are added together. The spatial bin surrounding the observer was chosen to have a volume of $\Delta s_{\rm obs} = v \Delta t$, since a pseudo-particle within this distance from the observer, would probably cross the observer within the next time step. The anisotropy, however, is simply calculated from the average pitch-cosine of each particle falling in the observer's spatial bin within a temporal bin. This approach of binning also allows the calculation of uncertainties through the standard deviation of each bin, although the uncertainties are mostly small due to the large number of pseudo-particles used.

The isotropic injection is realised by giving each pseudo-particle a random pitch-cosine which is uniformly distributed between -1 and 1. The inner reflecting boundary, in the case of a real SEP event, is handled similar to hard-sphere scattering of a planar surface, that is, if $S < 0$ $\mathrm{AU}$ then $S \rightarrow |S|$ and $M \rightarrow |M|$. An additional reflective boundary condition is imposed on the pitch-cosine to ensure that it says within its allowed range, that is, if $|M| > 1$ then $M \rightarrow \mathrm{sign}(M) 2 - M$ \citep{strausseffenberger2017}. The Reid-Axford injection is realised by a convolution of the delta injection solution with the Reid-Axford profile \citep[following the approach of][]{drogeetal2014}, since the transport coefficients are not time-dependent. Notice that the infinite derivatives of $D_{\mu \mu}$ in the anisotropic scattering case, is problematic. If the derivative around $\mu$ is too large (small), a dip (spike) will appear in the stationary PAD around $\mu = 0$, because pseudo-particles are `advected' away too efficiently (not `advected' away efficiently enough) from $\mu = 0$ in $\mu$-space by the derivative (\textit{N. Wijsen}, 2018, private communication). In order to avoid infinite derivatives, the derivative is limited to a maximum value \citep[see][for an evaluation of the validity of this approach]{vandenberg2018}, that is,
\begin{equation*}
\mathrm{if} \;\;\;\;\;\;\;\; \left| \frac{\partial D_{\mu \mu}}{\partial \mu} \right| > 2 \left| \frac{\partial D_{\mu \mu}}{\partial \mu} \right|_{\mu = 1} \;\;\;\;\;\;\;\; \mathrm{then} \;\;\;\;\;\;\;\; \frac{\partial D_{\mu \mu}}{\partial \mu} = \mathrm{sign} (\mu) 2 \left| \frac{\partial D_{\mu \mu}}{\partial \mu} \right|_{\mu = 1} .
\end{equation*}


\section{Model Slab Turbulence}
\label{apndx:ModelSlab}

Here a toy model for slab turbulence will be derived. It will be assumed that the total magnetic field can be written as the sum of a large-scale average/background magnetic field $\vec{B}_0$ and a fluctuating magnetic field $\delta \vec{B}$; that the fluctuations are perpendicular to the background magnetic field, such that $\vec{B}_0 \cdot \delta \vec{B} = 0$; that the fluctuations are random, such that $\langle \delta \vec{B} \rangle = \vec{0}$ and $\langle \vec{B} \rangle = \vec{B}_0$, where $\langle \cdots \rangle$ indicates a suitable average; that the fluctuations are due to a superposition of different types of small-amplitude waves of different wave numbers and gyrophases with frequencies which are deterministically governed by the dispersion relations of these waves, and that there are little to no interaction between the waves themselves (i.e. the wave viewpoint of turbulence); that only slab turbulence, which have wave vectors $k_{\parallel}$ parallel to the background magnetic field and is only dependent on the position along the background magnetic field, is the main contributor to pitch-angle scattering; that slab turbulence can be described as circularly polarised \citep[for how resonant wave-particle interactions can be described using circularly polarised waves, see e.g.][]{tsurutanilakhina1997, droge2000a, straussleroux2019}, non-dispersive Alfv\'{e}n waves, with angular frequency $\omega$ related to the wave number $k$ by $\omega/k = V_A$, where $V_A$ is the Alfv\'{e}n speed; and that the background magnetic field is in the $z$-direction of the Cartesian coordinate system, so that $\vec{B}_0 = B_0 \, \hat{\vec{z}}$ \citep{goldsteinetal1995, choudhuri1998, droge2000a, shalchi2009, brunocarbone2013}.

\begin{figure*}[!t]
\centering
\includegraphics[clip, trim=5mm 10mm 20mm 28mm, scale=0.4]{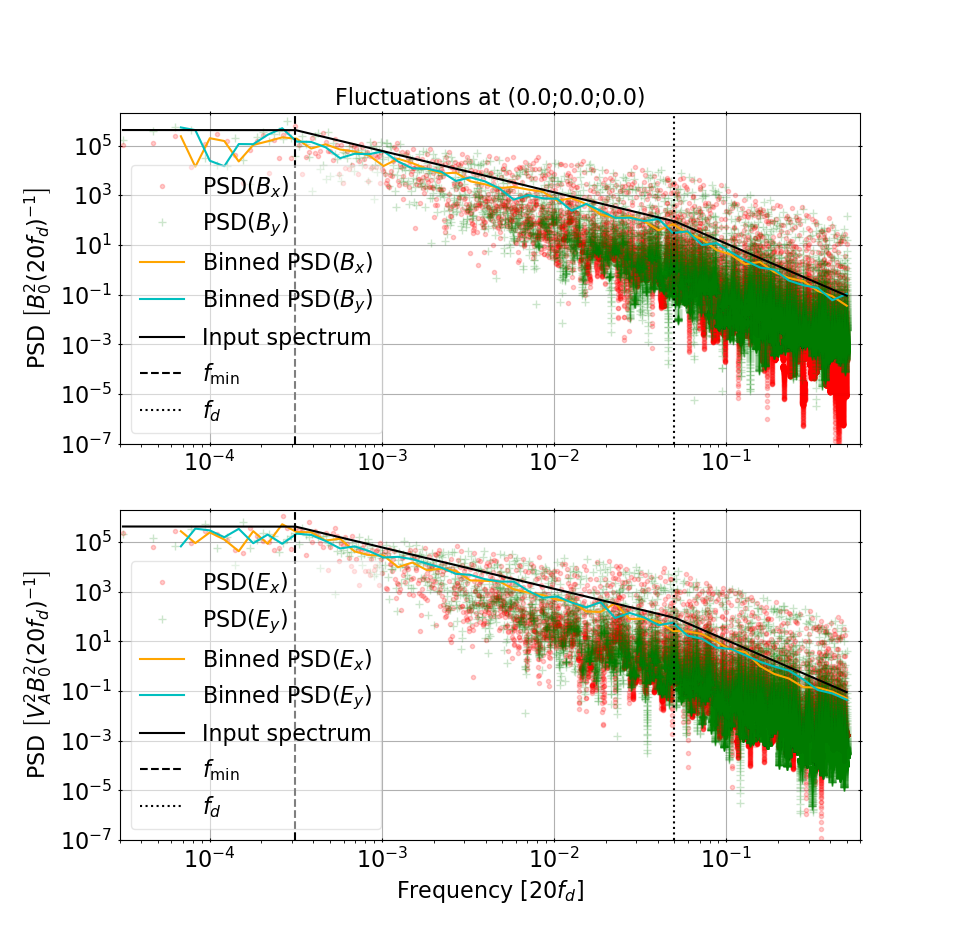}
\textbf{\caption[]{\label{fig:SlabTurbulence}{\textnormal{Spectra of the sampled magnetic (\emph{top panel}) and electric (\emph{bottom panel}) fluctuations of the toy slab turbulence model discussed in the text.}}}}
\end{figure*}

With these assumptions and waves propagating along the $z$-direction, the fluctuating magnetic field can have components
\begin{align*}
\delta B_x (z;t) & = \sum_i^{N_{\rm RH}^+} b_{0i} \cos \left[ k_{\parallel i} (z - V_A t) + \phi_i \right] + \sum_j^{N_{\rm RH}^-} b_{0j} \cos \left[ k_{\parallel j} (z + V_A t) + \phi_j \right] + \nonumber \\
 & \;\;\;\; \sum_m^{N_{\rm LH}^+} b_{0m} \sin \left[ k_{\parallel m} (z - V_A t) + \phi_m \right] + \sum_n^{N_{\rm LH}^-} b_{0n} \sin \left[ k_{\parallel n} (z + V_A t) + \phi_n \right] \\
\delta B_y (z;t) & = \sum_i^{N_{\rm RH}^+} b_{0i} \sin \left[ k_{\parallel i} (z - V_A t) + \phi_i \right] + \sum_j^{N_{\rm RH}^-} b_{0j} \sin \left[ k_{\parallel j} (z + V_A t) + \phi_j \right] + \nonumber \\
 & \;\;\;\; \sum_m^{N_{\rm LH}^+} b_{0m} \cos \left[ k_{\parallel m} (z - V_A t) + \phi_m \right] + \sum_n^{N_{\rm LH}^-} b_{0n} \cos \left[ k_{\parallel n} (z + V_A t) + \phi_n \right] ,
\end{align*}
where $b_{0l}$ are the amplitudes, $\phi_l$ are random phase differences which are uniformly distributed between $0$ and $2 \pi$, and $N_l$ is the number of waves of a particular type. This model considers four types of waves: right (RH) and left (LH) hand polarised waves propagating in the positive (+) and negative (-) $z$-direction. These fluctuating magnetic fields will induce fluctuating electric fields of the form
\begin{align*}
\delta E_x (z;t) & = V_A \left\lbrace \sum_i^{N_{\rm RH}^+} b_{0i} \sin \left[ k_{\parallel i} (z - V_A t) + \phi_i \right] - \sum_j^{N_{\rm RH}^-} b_{0j} \sin \left[ k_{\parallel j} (z + V_A t) + \phi_j \right] \right. + \nonumber \\
 & \;\;\;\;\; \left. \sum_m^{N_{\rm LH}^+} b_{0m} \cos \left[ k_{\parallel m} (z - V_A t) + \phi_m \right] - \sum_n^{N_{\rm LH}^-} b_{0n} \cos \left[ k_{\parallel n} (z + V_A t) + \phi_n \right] \right\rbrace \\
\delta E_y (z;t) & = V_A \left\lbrace - \sum_i^{N_{\rm RH}^+} b_{0i} \cos \left[ k_{\parallel i} (z - V_A t) + \phi_i \right] + \sum_j^{N_{\rm RH}^-} b_{0j} \cos \left[ k_{\parallel j} (z + V_A t) + \phi_j \right] \right. - \nonumber \\
 & \;\;\;\;\; \left. \sum_m^{N_{\rm LH}^+} b_{0m} \sin \left[ k_{\parallel m} (z - V_A t) + \phi_m \right] + \sum_n^{N_{\rm LH}^-} b_{0n} \sin \left[ k_{\parallel n} (z + V_A t) + \phi_n \right] \right\rbrace .
\end{align*}
The induced electric field is also circularly polarised and $90^{\circ}$ out of phase compared to the magnetic waves. With this form it can be verified that all of the Maxwell equations are satisfied.

The fluctuations should form a spectrum when sampled. A slab spectrum \citep[used for SEP modelling by][among others]{straussetal2017a} will be assumed to have the form
\begin{equation*}
g (k_{\parallel}) = g_0 \left\lbrace \begin{array}{lcl}
k_{\rm min}^{-s} & {\rm if} & \;\;\;\;\, 0 \le k_{\parallel} < k_{\rm min} \\
k_{\parallel}^{-s} & {\rm if} & k_{\rm min} \le k_{\parallel} \le k_d \\
k_d^{p-s} k_{\parallel}^{-p} & {\rm if} & \;\;\;\, k_d < k_{\parallel}
\end{array} \right. ,
\end{equation*}
with a flat energy range below $k_{\rm min}$, an inertial range with spectral index $s$ (assumed to be Kolmogorov, $s = 5/3$) between $k_{\rm min}$ and $k_d$, and a dissipation range with spectral index $p$ (assumed to be $p = 3$) above $k_d$. The total variance of the slab fluctuations is related to the spectra by \citep{shalchi2009, zank2014}
\begin{equation}
\label{eq:SlabVariance}
\delta B^2 = 8 \pi \int_0^{\infty} g (k_{\parallel}) \, {\rm d} k_{\parallel} ,
\end{equation}
from which the proportionality constant can be calculated as
\begin{equation}
\label{eq:ModelSlabG0}
g_0 = \frac{s - 1}{8 \pi} \delta B^2 k_{\rm min}^{s-1} \left[ s + \frac{s - p}{p - 1} \left( \frac{k_{\rm min}}{k_d} \right)^{s-1}  \right]^{-1} .
\end{equation}
The wave number at which the dissipation begins, follow a linear dependence on the proton cyclotron frequency in the SW \citep{duanetal2018, woodhametal2018}. For simplicity, it will therefore be assumed that
\begin{equation*}
k_d \approx \frac{|q_e| B_0}{m_p V_A} \;\;\;\;\;\;\;\; \mbox{and} \;\;\;\;\;\;\;\; k_{\rm min} \approx \frac{\pi k_d}{500} ,
\end{equation*}
where $q_e$ is the elementary charge of an electron, $m_p$ is the mass of a proton, and such that $k_{\rm min} \ll k_d$. Discrete wave numbers are chosen such that $\log k_{\parallel l}$ is equally spaced between $\log (k_{\rm min} / 10)$ and $\log (10 k_d)$.

By defining the ensemble average as
\begin{equation*}
\langle \cdots \rangle_{\theta} = \frac{1}{(2 \pi)^4} \int_0^{2 \pi} \!\!\! \int_0^{2 \pi} \!\!\! \int_0^{2 \pi} \!\!\! \int_0^{2 \pi} \!\!\! \cdots {\rm d}\theta_i \, {\rm d}\theta_j \, {\rm d}\theta_m \, {\rm d}\theta_n ,
\end{equation*}
where $\theta_l = k_{\parallel l} (z - V_A t) + \phi_l$ for $l = i,m$ and $\theta_l = k_{\parallel l} (z + V_A t) + \phi_l$ for $l = j,n$, it can be verified that $\langle \vec{B} \rangle_{\theta} = B_0 \, \hat{\vec{z}}$ and $\langle \vec{E} \rangle_{\theta} = \vec{0}$. The magnetic and electric variances can also be calculated as
\begin{subequations}
\begin{align}
\label{eq:ModelSlabBVariance}
\delta B^2 & = \sum_i^{N_{\rm RH}^+} b_{0i}^2 + \sum_j^{N_{\rm RH}^-} b_{0j}^2 + \sum_m^{N_{\rm LH}^+} b_{0m}^2 + \sum_n^{N_{\rm LH}^-} b_{0n}^2 \\
\label{eq:ModelSlabEVariance}
\delta E^2 & = V_A^2 \delta B^2 ,
\end{align}
\end{subequations}
respectively. The summations in Eq.~\ref{eq:ModelSlabBVariance} would approximate Eq.~\ref{eq:SlabVariance} if the amplitudes are chosen as
\begin{equation*}
b_{0l} = \sqrt{\frac{N_l}{N} 8 \pi g (k_{\parallel l}) \Delta k_{\parallel l}} ,
\end{equation*}
with $\delta B^2 = 0.1 \, B_0^2$ in $g_0$ (Eq.~\ref{eq:ModelSlabG0}) to reflect the fact that the variance is some fraction of the magnetic field strength, $\Delta k_{\parallel l}$ the difference between wave numbers centred on $k_{\parallel l}$, $N_l = N_{\rm RH}^+, N_{\rm RH}^-, N_{\rm LH}^+, N_{\rm LH}^-$ for $l=i,j,m,n$, respectively, and $N = N_{\rm RH}^+ + N_{\rm RH}^- + N_{\rm LH}^+ + N_{\rm LH}^-$ the total number of waves. It is assumed that each type of wave follows the same spectrum, but in reality the different types of waves might also have different spectra because the dissipation is set by different resonance conditions \citep[see e.g.][]{engelbrechtstrauss2018, straussleroux2019}.

This toy model for slab turbulence is verified in Fig.~\ref{fig:SlabTurbulence} where the spectra of both the electric and magnetic field are shown. The fluctuations, with $N_{\rm RH}^+ = N_{\rm RH}^- = N_{\rm LH}^+ = N_{\rm LH}^- = 500$ and $V_A = 10$ $\mathrm{m} \cdot \mathrm{s}^{-1}$, were sampled at the origin at a frequency of $20 f_d$ for a duration of $1/20f_{\rm min}$, where $f_d$ and $f_{\rm min}$ is the frequency corresponding to $k_d$ and $k_{\rm min}$, respectively. Notice that similar results can be found in the magnetostatic case ($V_A = 0$), without the electric field of course, if the fluctuations are sampled along the $z$-axis. It can be verified that the running average over time of the fluctuations go to zero (not shown), while the variance approach the correct values (keeping in mind that $\delta B^2 = \delta B_x^2 + \delta B_y^2$). The spectrum becomes clearer if the power spectral density is binned and it can be seen that it has the same form as the input spectrum. If the spectra of the two components are integrated and added together, a value close to the correct variance is found. If the number of waves are increased, the discrete wave numbers become less obvious in the spectra.


\section{Derivation of the Focusing Term and Steady State Pitch-angle Distribution}
\label{apndx:FocusPAD}

The pitch-angle transport term in the FTE,
\begin{equation*}
\frac{\partial}{\partial \mu} \left[ \frac{(1 - \mu^2) v}{2L(s)} f \right] ,
\end{equation*}
describes the mirroring or focusing of particles \citep{ruffolo1995, zank2014}. Following \citet{ruffolo1995}, the focusing term can be calculated directly from the mirroring condition (Eq.~\ref{eq:MirrorCondition}). The particles' pitch-angle change due to the movement of the particles into different regions of the magnetic field,
\begin{equation*}
\frac{{\rm d} \mu}{{\rm d} t} = \frac{{\rm d} \mu}{{\rm d} B} \frac{{\rm d} B}{{\rm d} s} \frac{{\rm d} s}{{\rm d} t} ,
\end{equation*}
where ${\rm d}s/{\rm d}t = v_{\parallel} = \mu v$. From the mirroring condition it follows that
\begin{equation*}
\frac{{\rm d} \mu}{{\rm d} B} = - \frac{1}{2 B_m \sqrt{1 - B / B_m}} = - \frac{1 - \mu^2}{2 \mu B} ,
\end{equation*}
where $B_m = B/(1 - \mu^2)$ was used. Hence, the change in pitch-angle becomes
\begin{equation*}
\frac{{\rm d} \mu}{{\rm d} t} = - \frac{1 - \mu^2}{2 \mu B} \frac{{\rm d} B}{{\rm d} s} \mu v = \frac{(1 - \mu^2) v}{2 L(s)} ,
\end{equation*}
where
\begin{equation}
\label{eq:FocusingLength}
\frac{1}{L(s)} = - \frac{1}{B(s)} \frac{{\rm d} B(s)}{{\rm d} s}
\end{equation}
relates the focusing length to the changing magnetic field.

An expression can be derived for the steady state PAD, $F(\mu)$, by neglecting any spatial dependences ($\partial f / \partial s = 0$ and $L$ constant), so that Eq.~\ref{eq:FTPE}, in the steady state ($\partial f / \partial t = 0$), reduces to
\begin{equation}
\label{eq:PADstationaryODE}
\frac{{\rm d}}{{\rm d} \mu} \left[ \frac{(1 - \mu^2) v}{2L} F(\mu) \right] = \frac{{\rm d}}{{\rm d} \mu} \left[ D_{\mu \mu} (\mu) \frac{{\rm d} F(\mu)}{{\rm d} \mu} \right] .
\end{equation}
Integrating this twice with respect to $\mu$ and applying the normalisation condition ($\int_{-1}^1 F {\rm d}\mu = 1$), yields \citep{earl1981, beeckwibberenz1986, heschlickeiser2014}
\begin{equation}
\label{eq:StationaryPAD}
F(\mu) = \frac{e^{G(\mu)}}{\int_{-1}^1 e^{G(\mu')} {\rm d}\mu'} ,
\end{equation}
where
\begin{equation}
\label{eq:PADg}
G(\mu) = \frac{v}{2L} \int_0^{\mu} \frac{1 - \mu'^2}{D_{\mu \mu} (\mu')} {\rm d}\mu' .
\end{equation}
The fact that the stationary PAD is some exponential function of the pitch-cosine, illustrates the fact that focusing causes the particles to be field aligned, with fewer particles moving opposite to the magnetic field. In the case of no focusing ($L \rightarrow \infty$), Eq.~\ref{eq:PADstationaryODE} reduces to ${\rm d}[ D_{\mu \mu} ({\rm d} F / {\rm d} \mu) ]/{\rm d}\mu = 0$, which yields $F(\mu) = 1/2$ upon integration and applying a reflective boundary (${\rm d}F/{\rm d}\mu = 0$) condition and the normalisation condition. This states that the global distribution relaxes to isotropy, as expected for pitch-angle scattering in the absence of focusing.


\section{The Diffusion-advection and Telegraph Approximation}
\label{apndx:DiffTel}

If one is only interested in the local properties over which $\lambda_{\parallel}^0 / L$ is approximately constant and if it is assumed that the distribution function can be written as $f = F_0 + F$ with $F \ll F_0$ (that is, small anisotropies), two analytical approximations are available for Eq.~\ref{eq:FTPE} with a delta injection of isotropic particles, $\delta (s - s_0) \delta (t)$, and a vanishing distribution function at infinity, $f(s \rightarrow \pm \infty; t) = 0$. Note that in what follows the given expressions differ from those given in some of the references due to the use of unitless variables in the references.


\subsection{The Diffusion-advection Approximation}
\label{subsec:Diffusion}

By assuming that the distribution is nearly isotropic, the evolution of the ODI is governed by a diffusion-advection equation \citep{litvinenkoschlickeiser2013, effenbergerlitvinenko2014}
\begin{equation}
\label{eq:DiffusionAdvectionEquation}
\frac{\partial F_0}{\partial t} = u \frac{\partial F_0}{\partial s} + \kappa_{\parallel} \frac{\partial^2 F_0}{\partial s^2} ,
\end{equation}
where $u = \kappa_{\parallel} / L$ is the coherent advection speed caused by focusing and $\kappa_{\parallel} = v \lambda_{\parallel} / 3$ is the spatial diffusion diffusion coefficient in the presence of focusing, with $\lambda_{\parallel}$ given by Eq.~\ref{eq:ParalMFPFocus}. The solution of Eq.~\ref{eq:DiffusionAdvectionEquation} is \citep{effenbergerlitvinenko2014}
\begin{equation}
\label{eq:DiffusionApprox}
F_0(s;t) = \frac{1}{\sqrt{4 \pi \kappa_{\parallel} t}} e^{- (s-s_0 - ut)^2 / 4 \kappa_{\parallel} t} ,
\end{equation}
and the anisotropy can be calculated from this using
\begin{equation}
\label{eq:DiffusionAnisotropy}
A(s;t) = \frac{3 \kappa_{\parallel}}{v} \left( \frac{1}{L} - \frac{1}{F_0} \frac{\partial F_0}{\partial s} \right) = \frac{3}{2v} \left( \frac{s-s_0}{t} + u \right) .
\end{equation}
A solution which might be more applicable to SEPs, is with a reflecting boundary at $s=0$ since SEPs are injected at the Sun and it can be assumed that the Sun's magnetic field would mirror particles away from the Sun. In this case the solution of Eq.~\ref{eq:DiffusionAdvectionEquation} is \citep{artmannetal2011},
\begin{equation*}
F_0(s>0;t) = \frac{1}{\sqrt{\pi \kappa_{\parallel} t}} \sinh \left( \frac{s s_0}{2 \kappa_{\parallel} t} \right) e^{- [(s-s_0 - ut)^2 + 2 s s_0] / 4 \kappa_{\parallel} t} ,
\end{equation*}
with an anisotropy given by
\begin{equation*}
A(s>0;t) = \frac{3}{2v} \left[ \frac{s-s_0 \coth (s s_0 / 2 \kappa_{\parallel} t)}{t} + u \right] .
\end{equation*}
according to Eq.~\ref{eq:DiffusionAnisotropy}. Notice that both expressions for the anisotropy has the unphysical prediction of infinite anisotropies as $t \rightarrow 0$ and that there is a persistent anisotropy at late times as $t \rightarrow \infty$ due to focusing. A known problem of the diffusion approximation is that it is too diffusive and violates causality, predicting that particles will arrive at a point before the particles could have physically propagated to that point \citep{litvinenkoschlickeiser2013, effenbergerlitvinenko2014}.


\subsection{The Telegraph Approximation}
\label{subsec:Telegraph}

In an attempt to preserve causality, the evolution of the ODI can also be described by the telegraph equation \citep{litvinenkoschlickeiser2013, effenbergerlitvinenko2014, litvinenkoetal2015}
\begin{equation}
\label{eq:TelegraphEquation}
\frac{\partial F_0}{\partial t} + \tau \frac{\partial^2 F_0}{\partial t^2} = u \frac{\partial F_0}{\partial s} + \kappa_{\parallel} \frac{\partial^2 F_0}{\partial s^2} ,
\end{equation}
where \citep{litvinenkonoble2013}
\begin{equation}
\label{eq:TelegraphTime}
\tau = \frac{\kappa_{\parallel} - \kappa_{\parallel}'}{u^2} ,
\end{equation}
with 
\begin{equation*}
\kappa_{\parallel}' = \frac{v^2}{4} \int_{-1}^1 Q(\mu') {\rm d}\mu' \; , \;\;\;\;\;\; 0 = \frac{{\rm d}}{{\rm d}\mu} \left[ \frac{2 L D_{\mu \mu}}{(1 - \mu^2) v} \left( \frac{{\rm d}}{{\rm d}\mu} \left[ \frac{D_{\mu \mu}}{1 - \mu^2} Q \right] + \mu \right) \right] - \frac{vQ}{2L} ,
\end{equation*}
and $Q(\mu = \pm 1) = 0$, which reduces to
\begin{equation}
\label{eq:ShalchiKappaParlWeakFocusing}
\kappa_{\parallel}' \approx \frac{v}{3} \left\lbrace \lambda_{\parallel}^0 + \lambda_{\parallel}^0 \left[ \frac{K(1)}{L} \right]^2 + \frac{6}{L^2} \int_{-1}^1 \mu' \left[ \frac{1}{3} K^3(\mu') - K(1) K^2(\mu') \right] {\rm d}\mu' \right\rbrace
\end{equation}
in the weak focusing limit ($\lambda_{\parallel}^0 / L \ll 1$), with $K(\mu) = (v/4) \int_{-1}^{\mu} (1 - \mu'^2) / D_{\mu \mu} (\mu') {\rm d}\mu'$ \citep{shalchi2011}. The solution of Eq.~\ref{eq:TelegraphEquation} is \citep{litvinenkoschlickeiser2013, effenbergerlitvinenko2014, litvinenkoetal2015}
\begin{equation}
\label{eq:TelegraphApprox}
F_0(s;t) = \frac{1}{2} e^{[(s-s_0)/L - t/\tau]/2}  \left\lbrace \begin{array}{ll} \frac{1}{2 \sqrt{\kappa_{\parallel} \tau}} \left[ I_0(z) + \left( 1 - \frac{\kappa_{\parallel} \tau}{L^2} \right) \frac{t}{2 \tau z} I_1(z) \right] & \mbox{if} \; |s-s_0| < t \sqrt{\frac{\kappa_{\parallel}}{\tau}} \\
1 & \mbox{if} \; s = s_0 \pm t \sqrt{\frac{\kappa_{\parallel}}{\tau}} \\
0 & \mbox{otherwise}
\end{array} \right. ,
\end{equation}
where $I_0$ and $I_1$ are modified Bessel functions of the first kind with argument
\begin{equation*}
z = \frac{1}{2} \sqrt{\left( 1 - \frac{\kappa_{\parallel} \tau}{L^2} \right) \left[ \left( \frac{t}{\tau} \right)^2 - \frac{(s-s_0)^2}{\kappa_{\parallel} \tau} \right]} .
\end{equation*}

For SEPs with an injection and reflecting boundary at $s=0$, the solution of Eq.~\ref{eq:TelegraphEquation} is roughly twice that of Eq.~\ref{eq:TelegraphApprox} \citep{litvinenkoetal2015},
\begin{equation*}
F_0(s>0;t) = e^{- t/2 \tau} \left\lbrace \begin{array}{ll} \frac{1}{2 \sqrt{\kappa_{\parallel} \tau}} \left[ I_0(z_0) + \frac{t}{2 \tau z_0} I_1(z_0) \right] & \mbox{if} \; s < t \sqrt{\frac{\kappa_{\parallel}}{\tau}} \\
1 & \mbox{if} \; s = t \sqrt{\frac{\kappa_{\parallel}}{\tau}} \\
0 & \mbox{otherwise}
\end{array} \right. ,
\end{equation*}
where
\begin{equation*}
z_0 = \frac{1}{2} \sqrt{\left( \frac{t}{\tau} \right)^2 - \frac{s^2}{\kappa_{\parallel} \tau}} .
\end{equation*}
The anisotropy for the telegraph equation can be calculated numerically, for simplicity, from
\begin{equation}
\label{eq:TelegraphAnisotropy}
A(s;t) = \frac{3 \kappa_{\parallel}}{v} \left[ \frac{1}{F_0} \left( \tau \frac{\partial^2 F_0}{\partial t \partial s} - \frac{\partial F_0}{\partial s} \right) + \frac{1}{L} \left( 1 - \frac{\tau}{F_0} \frac{\partial F_0}{\partial t} \right) \right] .
\end{equation}
The expressions for $\kappa_{\parallel}$ and $\tau$ in the absence of focusing ($L \rightarrow \infty$) can be found in \citet{litvinenkoschlickeiser2013}. \citet{earl1976, earl1981} and \citet{pauls1993} \citep[summarised by][]{paulsburger1991} derived and solved a modified telegraph equation. This solution yield the same ODI, but is dependent on coefficients which are more cumbersome to calculate. See \citet{malkovsagdeev2015} for a discussion on the validity of the telegraph equation.


\subsection{Transport Coefficients}
\label{subsec:Coefficients}

From all the equations introduced in the previous two paragraphs, it follows for isotropic scattering (Eq.~\ref{eq:IsotropicScattering}) that the various quantities are given by
\begin{align*}
\lambda_{\parallel}^0 & = \frac{v}{2D_0} \\
G(\mu) & = \mu \xi \\
F(\mu) & = \frac{\xi}{2} e^{\mu \xi} \mathrm{cosech} (\xi) \\
\kappa_{\parallel} & = L v \left[ \coth (\xi) - \frac{1}{\xi} \right] = \frac{\lambda_{\parallel}^0 v}{\xi} \left[\coth (\xi) - \frac{1}{\xi} \right] \\
\kappa_{\parallel}' & = \frac{L v}{\xi} \left[ 1 - \frac{\tanh (\xi)}{\xi} \right] \\
\tau & = \frac{L}{v} \tanh (\xi) = \frac{\lambda_{\parallel}^0}{v \xi} \tanh (\xi)
\end{align*}
where $\xi = \lambda_{\parallel}^0 / L$ is the focusing parameter \citep{roelof1969, beeckwibberenz1986, shalchi2011, litvinenkonoble2013, lasuiketal2017}.

For anisotropic scattering (Eq.~\ref{eq:QLT}) the various quantities are given by
\begin{align*}
\lambda_{\parallel}^0 & = \frac{3v}{2 D_0 (2-q) (4-q)} \\
G(\mu) & = \mathrm{sign}(\mu) \frac{4-q}{3} \xi |\mu|^{2-q} \\
F^{q=3/2}(\mu) & = e^{\mathrm{sign}(\mu) 5 \xi \sqrt{|\mu|} / 6} \div \left\lbrace \frac{24}{5 \xi} \sinh \left( \frac{5}{6} \xi \right) + \frac{144}{25 \xi^2} \left[ 1 - \cosh \left( \frac{5}{6} \xi \right) \right] \right\rbrace \\
\kappa_{\parallel}^{q=3/2} & = L v \left\lbrace \left[ \frac{5}{6} \xi + \frac{36}{5 \xi} \right] \cosh \left( \frac{5}{6} \xi \right) - \left[ 3 + \frac{216}{25 \xi^2} \right] \sinh \left( \frac{5}{6} \xi \right) \right\rbrace \nonumber \\
 & \;\;\;\;\; \div \left\lbrace \frac{5}{6} \xi \sinh \left( \frac{5}{6} \xi \right) + 1 - \cosh \left( \frac{5}{6} \xi \right) \right\rbrace \\
\kappa_{\parallel}'^{(\xi \ll 1)} & \approx \frac{v \lambda_{\parallel}^0}{3} \left[ 1 - \frac{2(5-q) (4-q)^2}{27 (8-3q)} \xi^2 \right]
\end{align*}
where $K(\mu) = (4-q) \lambda_{\parallel}^0 \left[ \mathrm{sign} (\mu) |\mu|^{2-q} + 1 \right] / 6$ in Eq.~\ref{eq:ShalchiKappaParlWeakFocusing} and $\tau$ is given by Eq.~\ref{eq:TelegraphTime} \citep{earl1981, beeckwibberenz1986, shalchi2011, litvinenkonoble2013}. Notice that analytical expressions for $F(\mu)$ and $\kappa_{\parallel}$ are only easily available for a Kraichnan spectrum with $q=3/2$. \citet{earl1981} gives some expressions in the limit of weak and strong focusing which can be used to find approximate expressions for these two quantities with an arbitrary $q$.


\section{The Heliospheric Magnetic Field and its Focusing Length}
\label{apndx:HMF}

The \citet{parker1958} HMF can be written as
\begin{equation*}
\label{eq:HMF}
\vec{B}_{\rm HMF} = A B_0 \left( \frac{r_0}{r} \right)^2 \left( \hat{\vec{r}} - \tan \psi \, \hat{\vec{\phi}} \right) ,
\end{equation*}
where $A$ is the polarity, $\hat{\vec{r}}$ and $\hat{\vec{\phi}}$ are unit vectors in the radial and azimuthal directions, respectively, and $B_0$ is a normalization value, usually related to the HMF magnitude as observed at Earth, $B_{\oplus} = 5$ $\mathrm{nT}$ (for solar minimum conditions) at $r_0 = 1$ $\mathrm{AU}$, such that
\begin{equation*}
\label{eq:HMFB0}
B_0 = \frac{B_{\oplus}}{\sqrt{1 + \left( \omega_{\odot} r_0/v_{\rm sw} \right)^2}} ,
\end{equation*}
with $\omega_{\odot} \approx 2 \pi / 25$ $\mathrm{days}$ $= 2.66 \times 10^{-6}$ $\mathrm{rad} \cdot \mathrm{s}^{-1}$ the solar rotation rate and $v_{\rm sw} = 400$ $\mathrm{km} \cdot \mathrm{s}^{-1}$ the radially directed SW speed. The HMF spiral angle $\psi$ is defined as the angle between the HMF line and the radial direction and is given by
\begin{equation}
\label{eq:SpiralAngle}
\tan \psi = \frac{\omega_{\odot} (r - r_{\odot})}{v_{\rm sw}} \sin \theta ,
\end{equation}
if it is assumed that the SW is immediately constant when leaving the solar surface, where $r_{\odot} \approx 0.005$ $\mathrm{AU}$ is the Sun's radius and $\sin \theta \approx 1$ for an observer in the ecliptic. The magnitude of the HMF is given by
\begin{equation*}
\label{eq:HMFmag}
B_{\rm HMF} =  B_0 \left( \frac{r_0}{r} \right)^2 \sqrt{1 + \tan^2 \psi} ,
\end{equation*}
from which it is evident that $B_{\rm HMF}$ decreases as $1/r$ in the equatorial regions \citep{parker1958, owensforsyth2013}.

The arc length of the \citet{parker1958} HMF line can be calculated by
\begin{equation*}
s = \int \!\! \sqrt{1 + \tan^2 \psi} \, {\rm d}r .
\end{equation*}
From the definition of the spiral angle (Eq.~\ref{eq:SpiralAngle}) it follows that ${\rm d}r = v_{\rm sw} {\rm d} (\tan \psi) / (\omega_{\odot} \sin \theta)$ and with this change in variables, the previous equation can be integrated analytically to give
\begin{equation}
\label{eq:ArcLength}
s = \frac{v_{\rm sw}}{2 \omega_{\odot} \sin \theta} \left[ \tan \psi \sec \psi + \mathrm{arcsinh} \left( \tan \psi \right) \right] ,
\end{equation}
where the integration constant must be zero to satisfy the condition $s(r_{\odot}) = 0$ \citep{lampa2011}. The focusing length (Eq.~\ref{eq:FocusingLength}) can be calculated from Eq.~\ref{eq:HMFmag} as
\begin{equation}
\label{eq:L(s)}
\frac{1}{L(s)} = \left( \frac{2}{r_{\odot} + v_{\rm sw} \tan \psi / \omega_{\odot} \sin \theta} - \frac{\omega_{\odot} \sin \theta}{v_{\rm sw}} \sin \psi \cos \psi \right) \cos \psi ,
\end{equation}
where Eq.~\ref{eq:SpiralAngle} was used to eliminate $r$. 


\bibliographystyle{aps-nameyear}     
\bibliography{References}  

\end{document}